\newif\ifMNRAS
\MNRASfalse 

\PassOptionsToPackage{pdfpagelabels=false}{hyperref} 

\ifMNRAS
    \documentclass[fleqn,usenatbib,useAMS]{mnras}
    \usepackage{graphicx}
\else

\documentclass[]{aastex63}
\fi

\usepackage{acronym}
\usepackage{hyperref}
\usepackage{amsmath,amsfonts,amssymb}
\usepackage{mathrsfs}
\usepackage{mathtools}
\usepackage[utf8]{inputenc}

\usepackage[normalem]{ulem}

\usepackage{pifont}

\newcommand{\david}[1]{{{#1}}}
\newcommand{\tianshu}[1]{{{#1}}}

\date{Received June 30, 2026; Accepted September 11, 2026}

\shorttitle{Rotation and CCSNe}

\begin{document}
\title{Effects of Rotation on 3D Core-Collapse Supernova Models for Low-Mass Progenitors} 
\correspondingauthor{Adam Burrows}
\author[0000-0002-3099-5024]{Adam Burrows}
\affiliation{Department of Astrophysical Sciences, Princeton University, NJ 08544, USA}
\email{burrows@astro.princeton.edu}
\author[0000-0002-0042-9873]{Tianshu Wang}
\affiliation{Department of Physics, University of California, Berkeley, CA, 94720-7300 USA}
\author[0000-0003-1938-9282]{David Vartanyan}
\affiliation{Department of Physics, University of Idaho, ID 83843, USA}

\begin{abstract}
We explore the dependence upon rotation rate alone of various supernova observables simulated to their saturation for the explosion of a 9-$M_{\odot}$ progenitor. We find that the explosion energy is non-monotonic with, and weakly dependent upon, spin across a broad range of initial spins. The asymmetries of the blast depend weakly on spin, with faster spins leading to only slightly greater asymmetries. There is little significant pole-equator neutrino heating asymmetry during explosion, even for rapid rotation, and only for the fastest rotator does the neutrino heating rate diminish noticeably. Hence, the effects of rotation alone
on all salient aspects of supernova dynamics are not large. We find that the recoil kick and spin are clearly aligned only for the most rapidly rotating model. Interestingly, for the fastest rotator, we witness a $T/|W|$ corotation instabilities near a value of 0.05 and spiral arm modes emerge. We find that the nucleosynthetic yields depend little upon the rotation rate and determine that the ratio of initial to final core spin period is near $\sim$4000, implying, given the modest inferred radio pulsar periods at birth, that the initial spin periods of most supernova cores are likely quite long. However, we focus on only one progenitor and do not include magnetic fields. Nevertheless, at least for low-mass progenitors which explode early, we find muted consequences of rotation in most major particulars across a wide range of initial spins.
\end{abstract} 

\ifMNRAS
    \begin{keywords}
    stars - supernovae - general    
    \end{keywords}   
\else 
    \keywords{
    stars - supernovae - general }
\fi

\section{Introduction}
\label{sec:int}  
\twocolumngrid

At their terminal stages, stars more massive than $\sim$8 $M_{\odot}$ create a dense Chandrasekhar core which becomes unstable to dynamical implosion. Theory suggests that this core collapses within hundreds of milliseconds (ms) to nuclear densities, at which point its inner fraction rebounds
into its outer infalling mantle, generating a spherical shock wave. Matter compression and heating by that shock wave as it moves outward liberates a burst of predominantly electron neutrinos ($\nu_e$). This burst undermines the shock's vigor and it stalls into accretion. The lowering of the effective gamma of the matter by photodissociation, thereby decreasing the efficiency of converting thermal energy into pressure, further facilitates this initial fizzle.  

However, as originally suggested by \citet{wilson1985} and further articulated by \citet{1985ApJ...295...14B}, prolonged energy deposition between the core and stalled shock wave (in the so-called ``gain region") over hundreds of milliseconds of a fraction of the neutrino power radiated from the residual protoneutron star (PNS) can reenergize the shock into explosion. Nevertheless, the original suggestion by \citet{wilson1985} that ``neutron-finger" (doubly-diffusive) convection boosted the core neutrino luminosities to the levels necessary to ignite explosion proved incorrect \citep{bruenn_dineva}, though lepton-gradient-driven convection in the PNS residue does boost the neutrino luminosities usefully. It is now thought that this boost, along with the Reynold's (and perhaps Maxwell's \citep{varma_muller2023}) stress generated behind the stalled shock wave by vigorous neutrino-driven convection, together relaunch the explosion after a delay of tens to hundreds of milliseconds. However, this is a fundamentally multidimensional problem and the computational capturing of this complex phenomenon with physical fidelity has engaged the cognizant theory community for decades.  Through this international effort, a compelling case has recently emerged that this delayed neutrino-heating, turbulence-aided, and PNS-convection-boosted mechanism is indeed the primary agency of explosion \citep{janka2012,burrows2013,lentz:15,2015ApJ...799....5C,2015MNRAS.453..287M,Muller2019,stockinger2020,bollig2021,sandoval2021,nakamura2022,wang,vartanyan2023,burrows_correlations_2024,janka2025}. 

To date, the primary focus of the effects on CCSN explosions of initial core rotation has been on those cores that may be rotating at high spin rates with millisecond periods at bounce. Such models have been explored with large initial magnetic fields which might act as transducers to transfer the high rotational kinetic energy of the core to matter ejected in jets \citep{leblanc1970,burrows2007_mag,Kuroda2014,2020ApJ...896..102K,Mosta2014,mosta2015,shankar_mosta2025,Obergaulinger2018,Obergaulinger2020,Obergaulinger2021,Aloy2021,varma_muller_rot_2023,bugli2023,powell2023,kovalenko2026}. A cold neutron star with a spin period of 5 milliseconds has a rotational kinetic energy of $\sim$10$^{51}$ ergs ($\equiv$ one Bethe) and a more rapidly rotating PNS would provide an even greater energy reservoir that might explain hypernovae and the most energetic Type Ic (broad-line) supernovae \citep{woosley1999,inserra2013}. However, such rapid initial core rotation is likely quite rare, thought to constitute but $\sim$1\% of all supernovae \citep{modjaz2008}. This perspective is informed by the inferred dynamically slow mean spin rates ($\sim$300 $\pm$ 200 milliseconds) of radio pulsars at birth \citep{emmering1989,faucher_kaspi,popov2012,noutsos2013,Igoshev_2022} and currently \citep{popov2012,Igoshev2013}\footnote{Observed magnetars \citep{Kaspi2016,Kaspi2017} on average are even more slowly rotating, with periods \citep{yavuz2026} from 0.33 seconds \citep{Livingstone2011} to $\sim$12 seconds \citep{dib2014}.}. We note that the spin rates induced during collapse and explosion for initially non-rotating cores can be of this magnitude \citet{blondin_shaw,rantsiou,coleman,burrows_kick_2024}, but that this mechanism likely falls far short of generically explaining measured neutron star spins\footnote{For instance, the Crab pulsar has a current spin period of 33 milliseconds and an inferred birth spin period of $\sim$15$-$19 milliseconds \citep{Kou2015}. This is far faster than the calculated possible induced spins \citep{burrows_kick_2024} for its inferred $\sim$10 $M_{\odot}$ progenitor.}. 

For relative simplicity, most massive star progenitor evolutionary models have been conducted without rotation \citep[e.g.,][]{swbj16,sukhbold2018}, but some massive star modelers have attempted to incorporate rotation and torques. These models employed prescriptions for angular-momentum transfer, either by magnetic field torques due to a Taylor-Spruit dynamo \citep[leaving a core period of 10's of seconds;][]{Heger2003,2005ApJ...626..350H} or by fluid instabilities
\citep[leaving a core period of seconds;][]{Hirschi2004,Hirschi2005a,Hirschi2005b,hirschi2017}. These recipes have been criticized \citep{fuller2019,fuller2022}, in part, for their inconsistency with asteroseismological data for red giants \citep{cantiello2014,deheuvels2012,beck2013}, which indicate that angular momentum transport in post-main-sequence stellar models is more efficient than assumed. In addition, the slow measured periods of white dwarfs \citep{suijs2008} suggest that they too are subject to efficient angular-momentum loss in the core of their birth star and, perhaps, that most white-dwarf-like cores in massive stars too are rotating slowly.

These observations have motivated the use generically of non-rotating progenitor cores for detailed CCSN supernova simulations. To date, this has been our group's approach using our code F{\sc{ornax}} (see \S\ref{method}). However, what in fact is (or would be) the hydrodynamic effect of initial progenitor core rotation? The distortional and centrifugal effects of rotation on the neutrino fields and hydrodynamics must, if significant, translate into altered outcomes and asymptotic observables. The rotation axis sets a direction, but at what spin rate does this affect the eject distributions? Rotation could affect the speed of explosive disassembly, itself a major determinant of nucleosynthesis
during $\alpha$-rich freeze-out \citep{wang_nucleo_2024}. In addition, rotation could determine the direction of the recoil kick to the residual neutron star and, hence, explain in part the spin-kick correlation seen for some radio pulsars \citep{Ng2007,noutsos2013,yao2021,biryukov2025}.     

It is these and related questions that motive this work. In this paper, we conduct a sequence of state-of-the-art 3D simulations along a rotational continuum from slow to fast, ignoring magnetic effects \footnote{This does not mean magnetic effects would not be important (particularly for rapid rotators), merely that this paper focuses on rotational effects alone.}. There is an established tradition for studying rotating supernova progenitors without magnetic fields. \citet{fryer2000}, \citet{2002ApJ...574L..65F}, and \citet{2004ApJ...601..391F} concluded that rotation hurts explosion by lowering the $\bar{\nu}_e$ flux and its associated heating \citep[see also][]{Ott2008}. \citet{suwa2010}  and \citet{nakamura2014} found that rotation aided explosion, mostly by rotationally expanding the size of the gain region and increasing the mass it contained \citep[see also][]{summa2018} in their 3D context).
\citet{iwakami2014} and \citet{TaKoSu16} highlight the rotational excitation of $m=1$ spiral-arm modes at modest $T$/$|W|$\footnote{the ratio of the rotational kinetic energy to the absolute value of the gravitational energy} and find a role for non-axisymmetric rotational instabilities in explosions of rapidly rotating progenitor cores. \citet{nakamura2014} and \citet{TaKoSu16} observed equatorial explosions. \citet{ott_2005_onearmed} also witnessed the excitation of such spiral modes, but not in the context of explodability.      

Most of the modern simulations that explore the potential effects of rotation \citep[e.g.,][]{Kuroda2014,TaKoSu16,summa2018,andresen2019,Shibagaki2021,fujibayashi2021} address only rapid rotation.
What distinguishes our investigation here is that, whereas previous simulations focused on explodability and the hydrodynamic issues early after core bounce, we endeavor to carry the simulations to late times when the explosion energy, nucleosynthesis, recoil kick speeds, gravitational-wave signal, and neutron star gravitational mass have stop changing. In this way, we seek to determine the potential role of rotation on observables. 

In \S\ref{method}, we summarize our computational approach and its parameters. We then in \S\ref{hydro} proceed to discuss the explosion characteristics and hydrodynamic development of the various models. For the $\Omega_0$ = 1.0 radian s$^{-1}$ model, we see the emergence of a $T/|W|$ instability and the generation of vigorous spiral waves.  Too late to affect the explosion itself, this mode has direct consequences for the gravitational wave (GW) emission discussed in \S\ref{GW}. In \S\ref{lum}, we present the radiation quantities and heating rates for the span of simulated initial rotation rates. Then, we focus in \S\ref{kicks} on the recoil kicks and the potential for spin-kick correlation. The differences in the character of PNS convection due to rotation are described in \S\ref{pns}, and then in \S\ref{nucleo} we summarize the salient features of the nucleosynthesis  and its variation with rotation. The dependence of the gravitational wave emissions as a function of initial rotation rate is provided in \S\ref{GW}, and we wrap up in \S\ref{conclusions} with summary thoughts concerning the systematic behavior with rotation, but without magnetic fields, of 3D CCSN models and various associated observables and provide reflections on remaining outstanding questions.

 

\section{Method}
\label{method}

We use for this exploration of the effects of initial rotation on 3D core-collapse supernova (CCSN) models our workhorse multi-group radiation hydrodynamics code F{\sc{ornax}} \citep{skinner2019,vartanyan2018b}. Over the last ten years, we have employed this code to address a broad spectrum of topics in CCSN theory \citep[e.g.,   ][]{radice2017b,burrows2018,vartanyan2018a,vartanyan2018b,vsg2018,burrows_2019,radice2019,hiroki_2019,vartanyan2019,Nagakura2020,vartanyan2020,burrows_2020,nagakura2021,2021Natur.589...29B,2022MNRAS.510.4689V,coleman,burrows_40,burrows_correlations_2024,burrows_40,Channels2025,Rusakov2026}. To date, F{\sc{ornax}} has conducted more than forty long-term ($\ge$ 2$-$9 seconds post-bounce) 3D simulations on both GPU and CPU machines (e.g., NERSC/Perlmutter, ALCF/Polaris, ALCF/Aurora, and TACC/Frontera). This is the majority of the model-seconds simulated by the international theory community, with the major focus on determining the systematic dependence of supernova observables on progenitor and progenitor core structure at the time of Chandrasekhar collapse \citep{burrows_correlations_2024}. Most models explode naturally without artifice, though no aspect of the input physics (e.g., the neutrino-matter couplings, treatment of multi-D and general relativistic gravity), massive-star progenitor structures at collapse (e.g.,
density profiles, convective seed perturbations, B-fields, and rotation rates), or numerical techniques should be considered converged or retired. 

For this study, we use the SFHo nuclear equation of state \citep{Steiner2013} and focus on the solar-metallicity 9-$M_{\odot}$ ZAMS \footnote{Zero-age main-sequence} mass model of \citet{swbj16}, acknowledging that a future study encompassing a spectrum of massive star progenitors would certainly be desired. We have chosen this progenitor in part because its low compactness\footnote{The ``compactness" approximately characterizes the shallowness of the mass density profile of the core of the progenitor massive star and is defined as
\begin{equation}     
\xi_M= \frac{M/M_{\odot}}{R(M)/1000\, \mathrm{km}}\,,
\end{equation}
where the subscript $M$ denotes the interior mass coordinate at which it is evaluated. We choose to evaluate it at $M$ = 1.75 M$_{\odot}$. For the 9-$M_{\odot}$ model of \citet{swbj16} its value is $6.7\times 10^{-5}$.}
results in simulations that asymptote in its final observable properties (e.g., neutron-star mass, explosion energy, kicks, nucleosynthesis, gravitational radiation emission) relatively quickly \citep{burrows_2020,burrows_correlations_2024}, thereby saving on computational resources. 

The spatial grid for these 3D simulations is $1024\times128\times256$ ($r\times\theta\times\phi$), extending from zero to 100,000 kilometers in spherical radius.  The angular gridding in the inner radial region is progressively deresolved as the center is approached to mitigate the effects of the Courant timestep limitation in the angular directions. This creates a so-called static ``dendritic" grid \citep{skinner2019} which helps speed up the code considerably, while still allowing one to conduct fully 3D simulations to the very center. \tianshu{The dendritic grid ensures that the angular cell sizes never exceed 1 km at small radii and maintains a 1.4-degree angular resolution at larger radii. At 20 km, which is a representative radius for proto-neutron star scale, the radial cell size is about 0.5 km and there are 64$\times$128 cells on the sphere.} Moreover, for the one non-rotating run and the three rotating runs of this paper, the initial pre-bounce collapses are performed in 3D. The rotating runs span initial angular frequencies ($\Omega_0$) of 0.01, 0.1, and 1.0 rad s$^{-1}$ (models 9-rot-0.01, 9-rot-0.1, and 9-rot-1.0), all with a cylindrical radius scale $A$ of 1000 kilometers (km) using the functional form of \citet{1985A&A...146..260E}, i.e., $v_{\phi}=\frac{\Omega_0R\sin(\theta)}{1+(R\sin(\theta)/A)^2}$, where the axis of rotation is the $z$ direction and $R$ is the spherical radius. The corresponding initial central spin periods are 628, 63, and 6.28 seconds, from slow to faster initial rotation. This parametrization is a standard in the field for which the true initial spin profiles are not in general available. Its virtue is that it allows models easily to span a broad range of initial periods to explore the overall behavior and trends with spin rate.  However, this profile form should not be consider definitive and a more motivated profile is keenly desired. We note that the few 1D massive-star progenitor evolution models conducted with approximate rotation and angular momentum transfer, with magnetic torques; \citet{Heger2003,2005ApJ...626..350H} and without; \citet{hirschi2017}, have used algorithms and initial spin profiles that are not currently compelling \citep{cantiello2014,fuller2019,fuller2022}. Indeed, the subject of the spin structures and associated B-fields in the core of massive stars, in and out of binaries, deserves renewed attention.  Absent this, we have opted for this simple, though common, spin parametrization. 

We use twelve neutrino energy groups for each of three species ($\nu_e$, $\bar{\nu}_e$, and ``$\nu_{\mu}$" [$\equiv$ $\nu_{\mu}$, $\bar{\nu}_{\mu}$, $\nu_{\tau}$, $\bar{\nu}_{\tau}$]). The $\nu_e$ neutrinos are logarithmically spaced from 1 to 300 MeV and the $\bar{\nu}_e$ and ``$\nu_{\mu}$" species are logarithmically spaced from 1 to 100 MeV. 

\tianshu{We calculate the nucleosynthetic yields in the same way as in \citet{wang_nucleo_2024,wang_ti44,wang_low_2024}. About 320,000 tracers are injected into each model and are evolved backward in time in a post-processing manner. The tracers are placed logarithmically along the r-direction above 500 km and uniformly along the $\theta$- and $\phi$-directions at the end of each simulation. A more detailed description of the tracer method can be found in \citet{wang_nucleo_2024}. The tracer thermal trajectories (density and temperature) are fed into SkyNet \citep{lippuner2017} together with neutrino field information, and a 1530-isotope network including elements up to $A=100$ is used. The reaction rates are taken from the JINA Reaclib \citep{cyburt2010} database, and we include neutrino interactions with protons and neutrons. However, reactions for the $\nu-$process \citep{woosley1990} are not included. The NSE (nuclear statistical equilibrium) criterion is set at 0.6 MeV ($\sim$7 GK). A nucleosynthesis calculation starts from the point after which the tracers never reach NSE again. The initial abundance of the tracer is set either by NSE with the electron fraction read from F{\sc ornax} or by the abundances read from the progenitor model, depending upon whether this tracer has reached NSE during its evolution. We use the \citet{lodders2021} values whenever solar abundances are referred to in this paper.}

\section{Hydrodynamics as a Function of Initial Rotation Rate}
\label{hydro}

\begin{figure*}[htbp!]
    \includegraphics[width=0.45\textwidth]{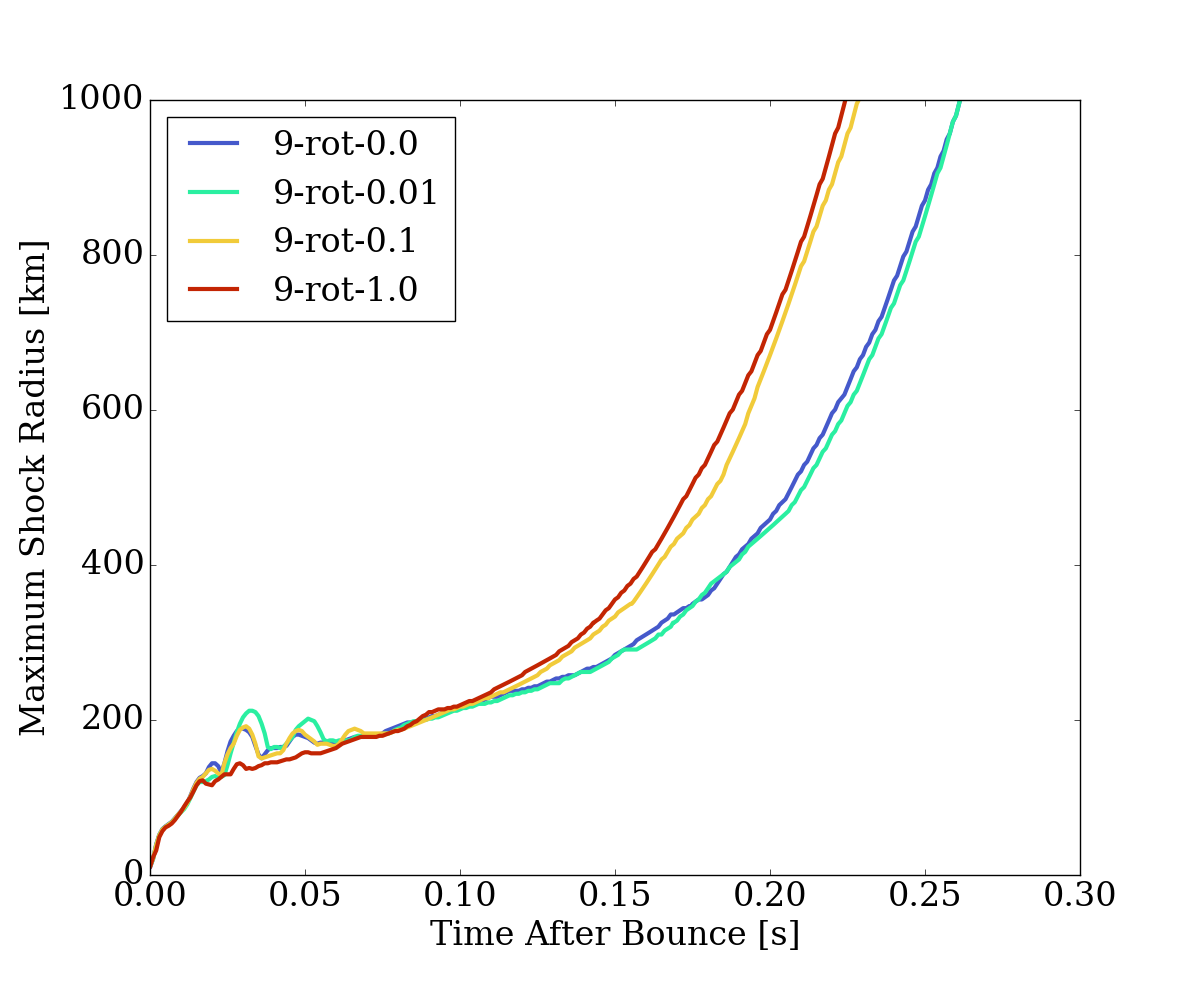}
    \includegraphics[width=0.45\textwidth]{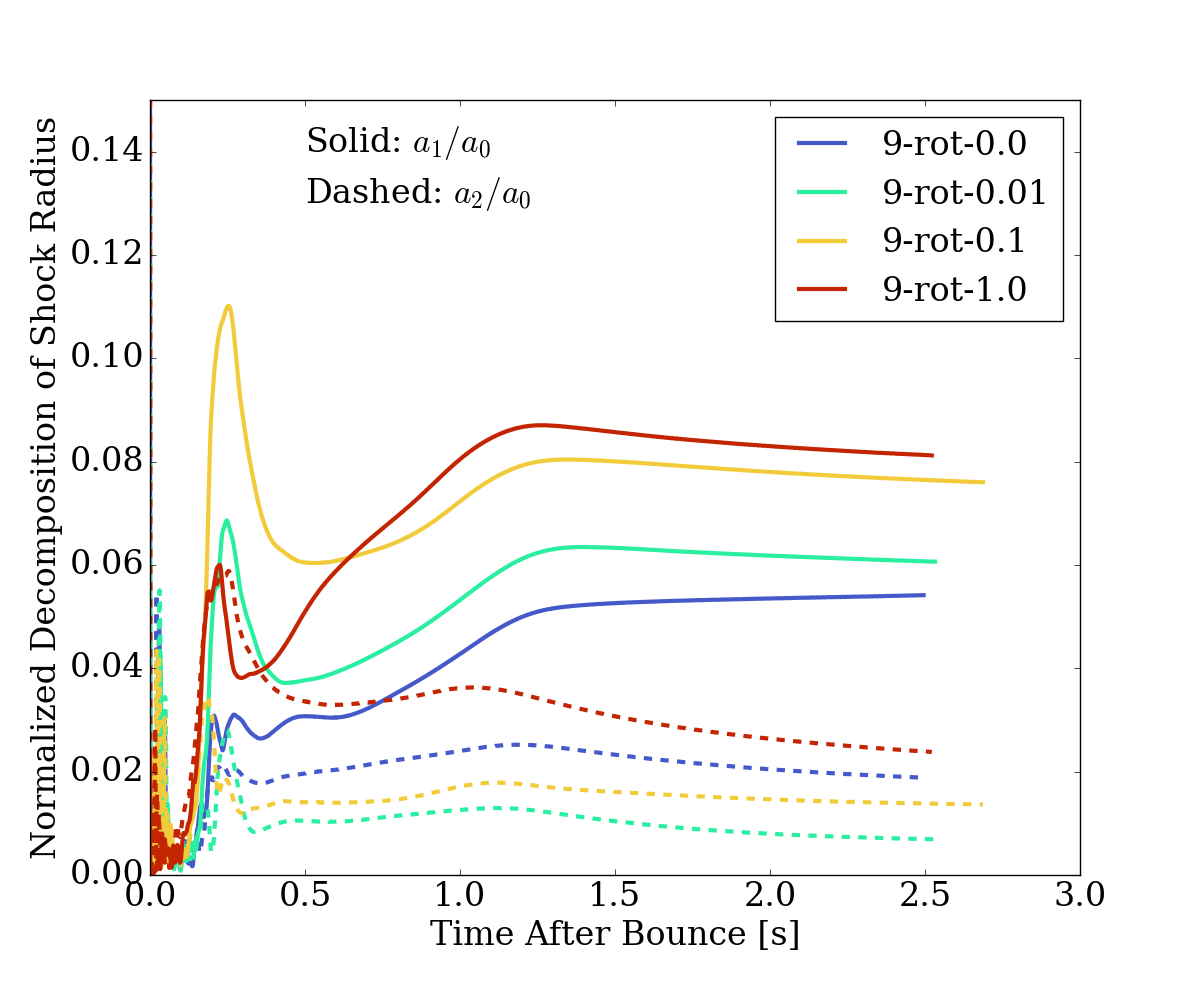}
\caption{Maximum shock radius (left) and normalized spherical harmonic decomposition of the shock radius (right). The decomposition is done using $a_l=\sqrt{\sum_{m=-l}^{m=l}|\int f(\theta,\phi)Y_{lm}(\theta,\phi)d\Omega|^2/(4\pi(2l+1))}$. We note that the mean shock radius and the time of explosion are weak functions of initial spin rate.  It is only at and faster than $\Omega_0$ $\approx$ 0.1 rad s$^{-1}$ that we see the shock radius peel off from the other models, and this only slightly.} 
    \label{fig:rshock}      
\end{figure*}

As Table \ref{tab:main} indicates, the initial central     rotation periods range from $\sim$630 to $\sim$6.3 seconds. The final angular momenta left in the nascent neutron stars by the end of calculations, divided by the estimated moment of inertia \citep{lattimer2005,breu2016}, yield final spin periods of $\sim$132 to 1.7 milliseconds\footnote{The final period of the initially non-rotating model reflects induced rotation \citep{rantsiou,burrows_kick_2024} and is interesting, if slow.}. The latter butts up against that of the fastest spinning pulsar known, while the former is significantly faster than the mean spin period of measured pulsars \citep{emmering1989,faucher_kaspi,Kaspi2016}. Yet, as we see in the left panel of Figure \ref{fig:rshock}, it is only for our fastest two models that we see much of an effect of the initial spin on the maximum shock radius as the explosion gets underway. This is consonant with what we witness throughout $-$ it is only for rapid initial rotation rates that rotational effects are much in evidence in the hydrodynamics.

The right panel of Figure \ref{fig:rshock} depicts the mean-normalized dipole and quadrupole moments
of the shock wave as a function of time after bounce. We do see a weak monotonic increase in the dipole distortion with increasing spin, but only over a range of $\sim$30\%. This is not what might be expected if magnetic fields were included in the simulations, particularly for our 9-rot-1.0 model, but isolating the effect of rotation alone on the various behaviors and observables is our goal here.

In Table \ref{tab:main}, we also see the variation in the explosion energy and the final gravitational mass with initial spin. The explosion energy spans about $\sim$20\% and is not monotonic with initial spin. We suggest this non-monotonicity is a consequence of the competition between two effects of rotation: 1) the decrease in the neutrino emissions and heating rates in the gain region \citep{1985ApJ...295...14B} with increasing spin (\S\ref{lum}); 2) the decrease in the effective gravitational potential against which driving by neutrino heating is working; and 3) the reservoir of rotational kinetic energy in the exploding mantle. The former lowers the explosion energy, while the latter two raise it, but, as we see in Table \ref{tab:main}, neither makes a huge difference. Note that in this paper we are exploring the rotational effect of a model, the 9-$M_{\odot}$ of \citet{swbj16}, which explodes easily. Whether rotation alone can make a qualitative difference for some more massive models for which explosion might be more sensitive to details remains to be seen.

\begin{figure*}[htbp!]
    \centering
    \includegraphics[width=0.45\textwidth]{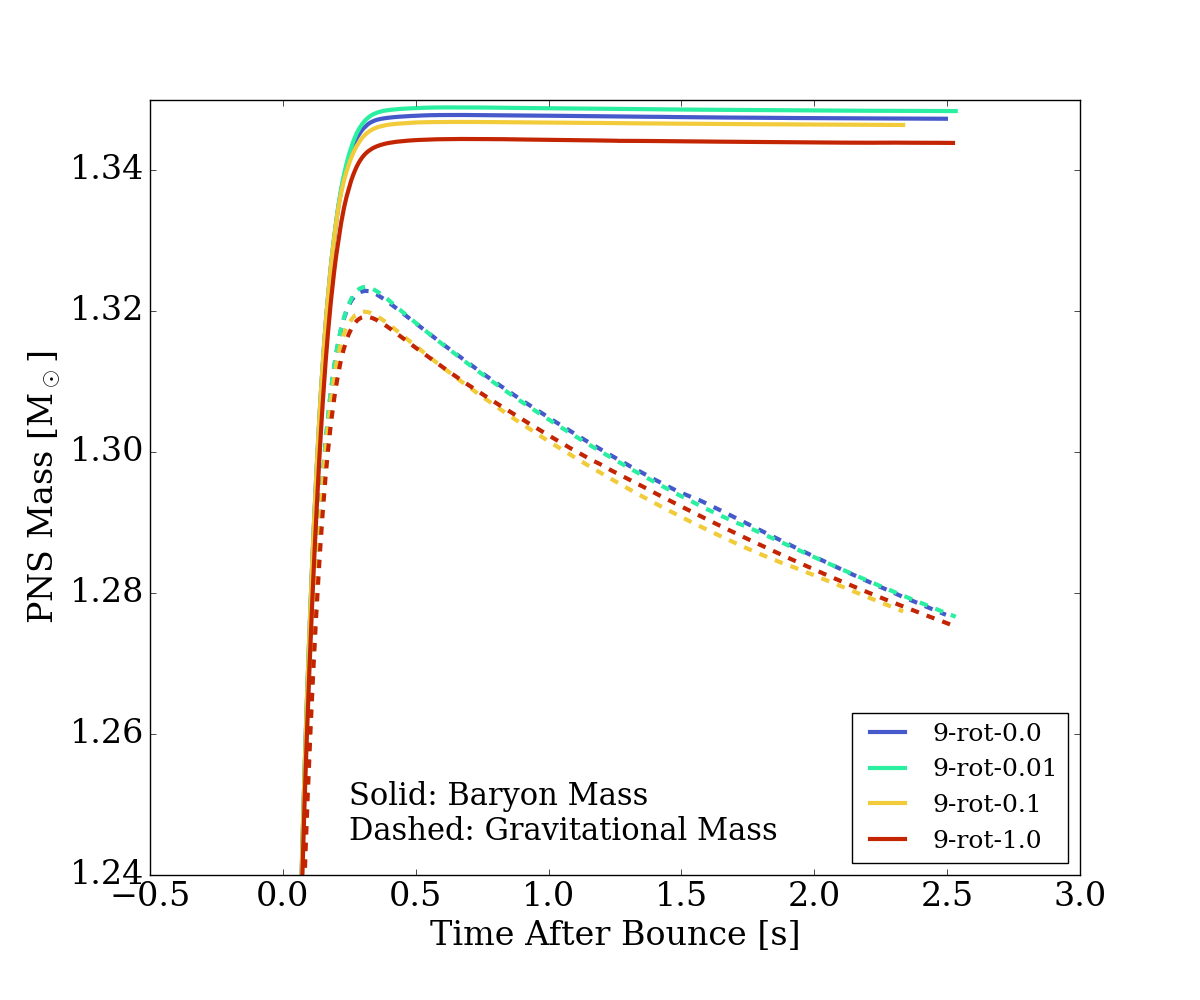}
    \includegraphics[width=0.45\textwidth]{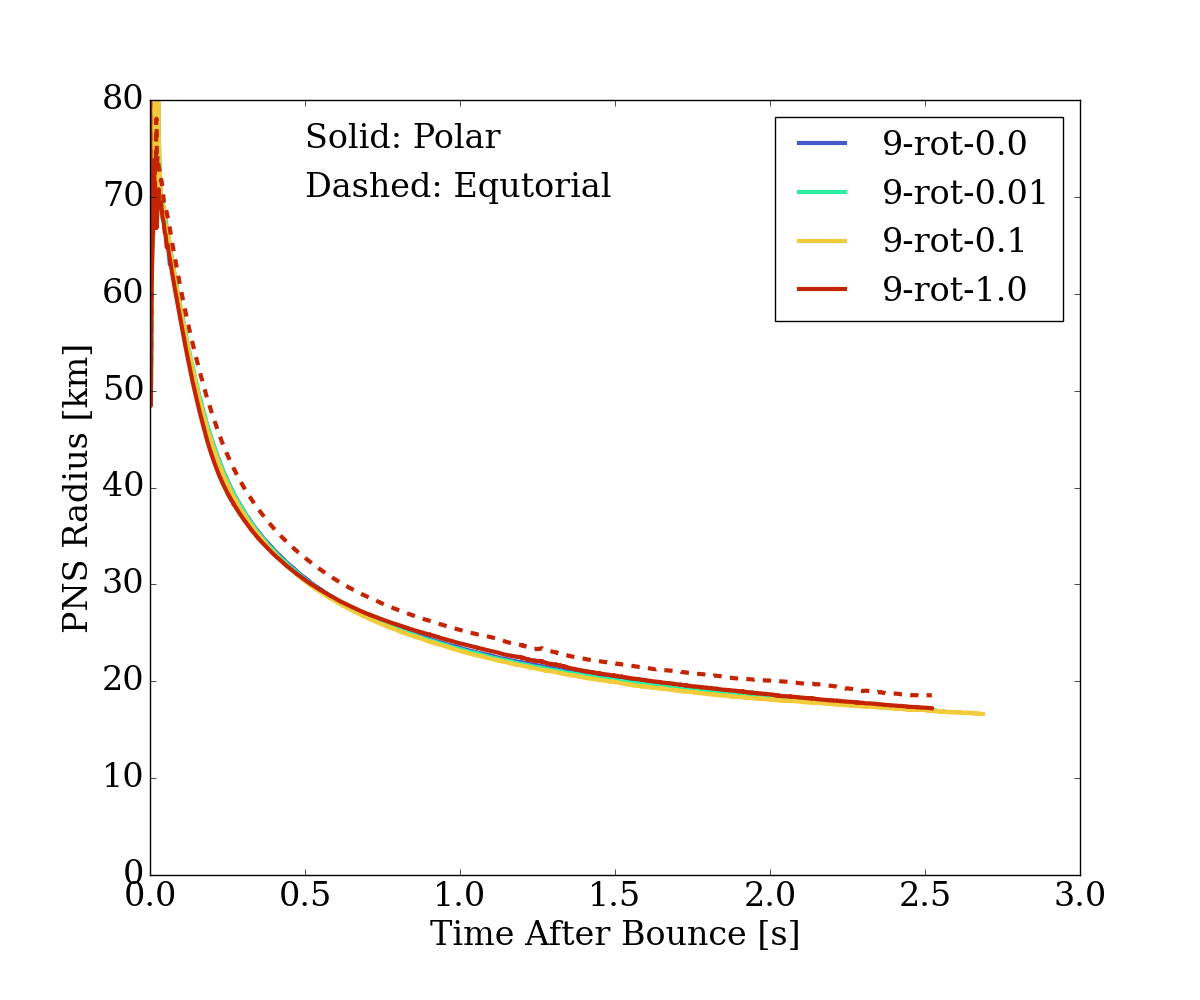}
    \caption{Proto-neutron star baryon mass and gravitational mass (left) and PNS radius (right, defined as the mean radius at which the mass density is $10^{11}$ g cm $^{-3}$). Note that the evolution of the gravitational mass tracks the integrated neutrino energy losses. By the end of the simulations $\sim$0.04 $M_{\odot}$c$^2$ has radiated. Note also that the final baryon mass of the residues changes little after $\sim$0.5 seconds post-bounce, decreasing only very slightly due to winds from their surfaces \citep{wang_wind}.}
    \label{fig:mpns}      
\end{figure*}

We see in the left panel of Figure \ref{fig:mpns} and in Table \ref{tab:main} that the variation with rotation rate in the mass of the newly-born neutron star is quite small. The two more rapidly rotating models have slightly lower residual masses due to their earlier and marginally more energetic explosions and the slightly smaller associated post-explosion mass accretion rates. Not easily seen in Figure \ref{fig:mpns} is the signature of winds that emerge from these cores which shave mass ever so fractionally off the object left behind. 

\begin{figure*}[htbp!]
    \centering
    \includegraphics[width=0.45\textwidth]{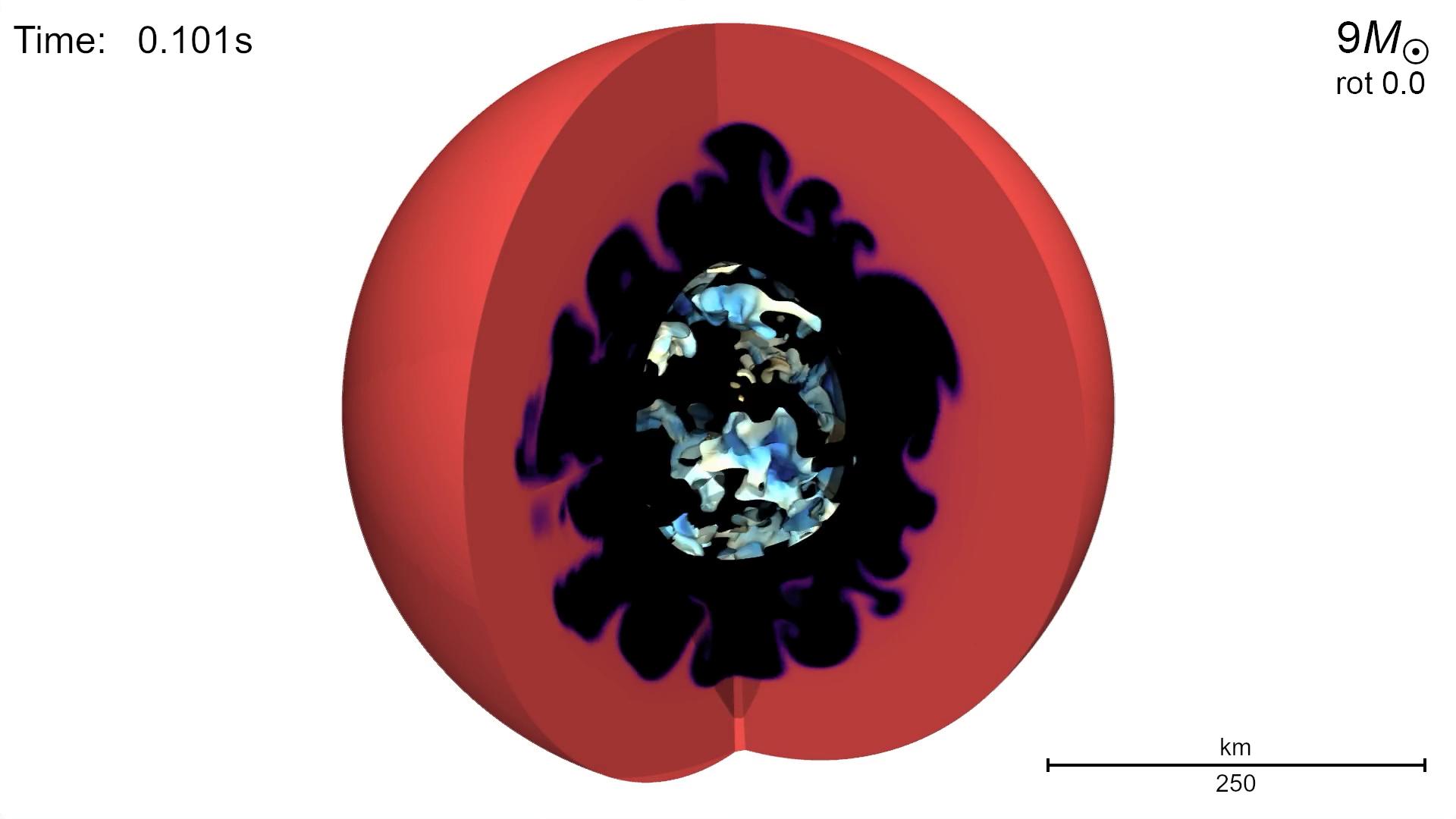}
    \includegraphics[width=0.45\textwidth]{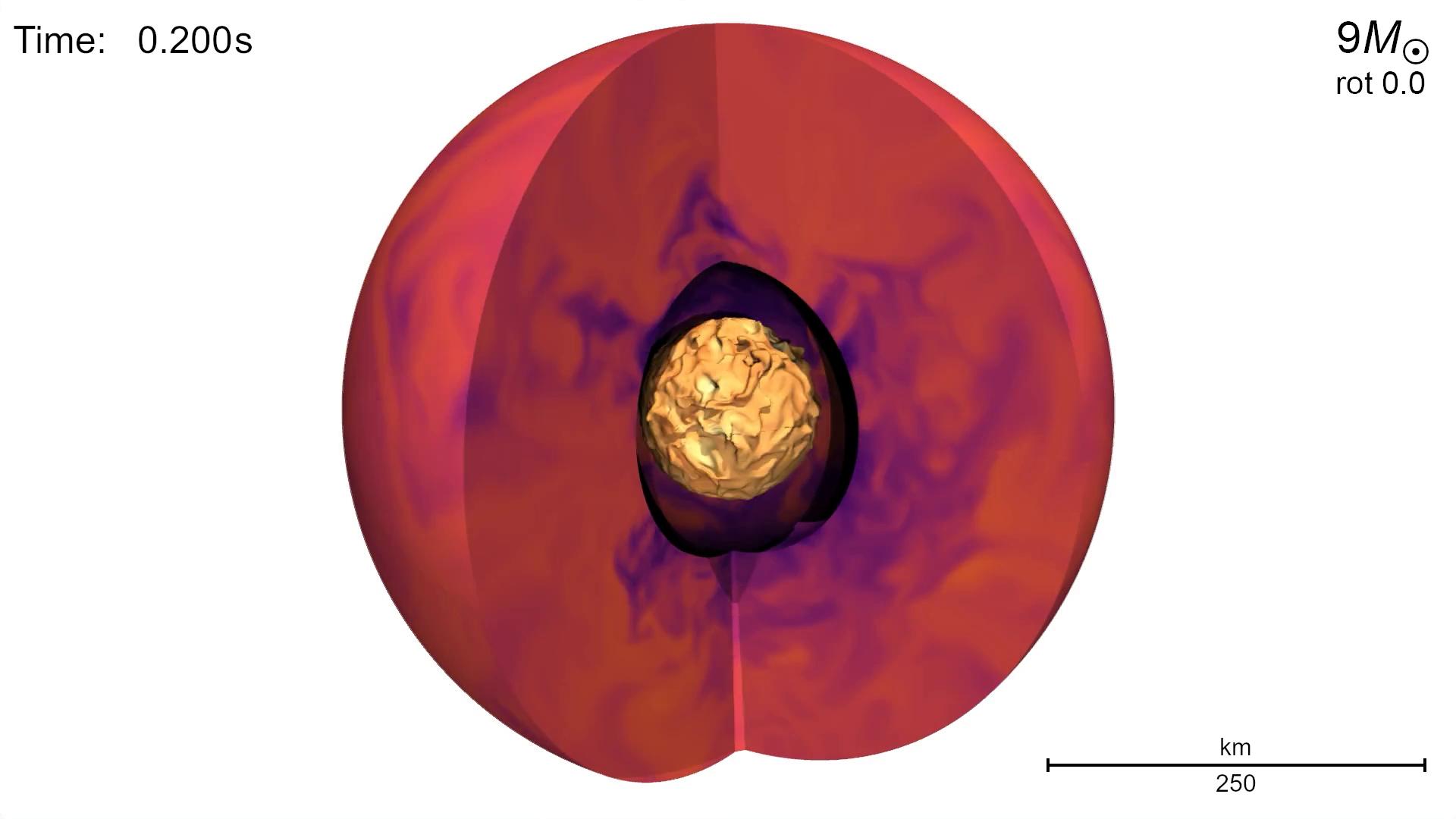}
    \includegraphics[width=0.45\textwidth]{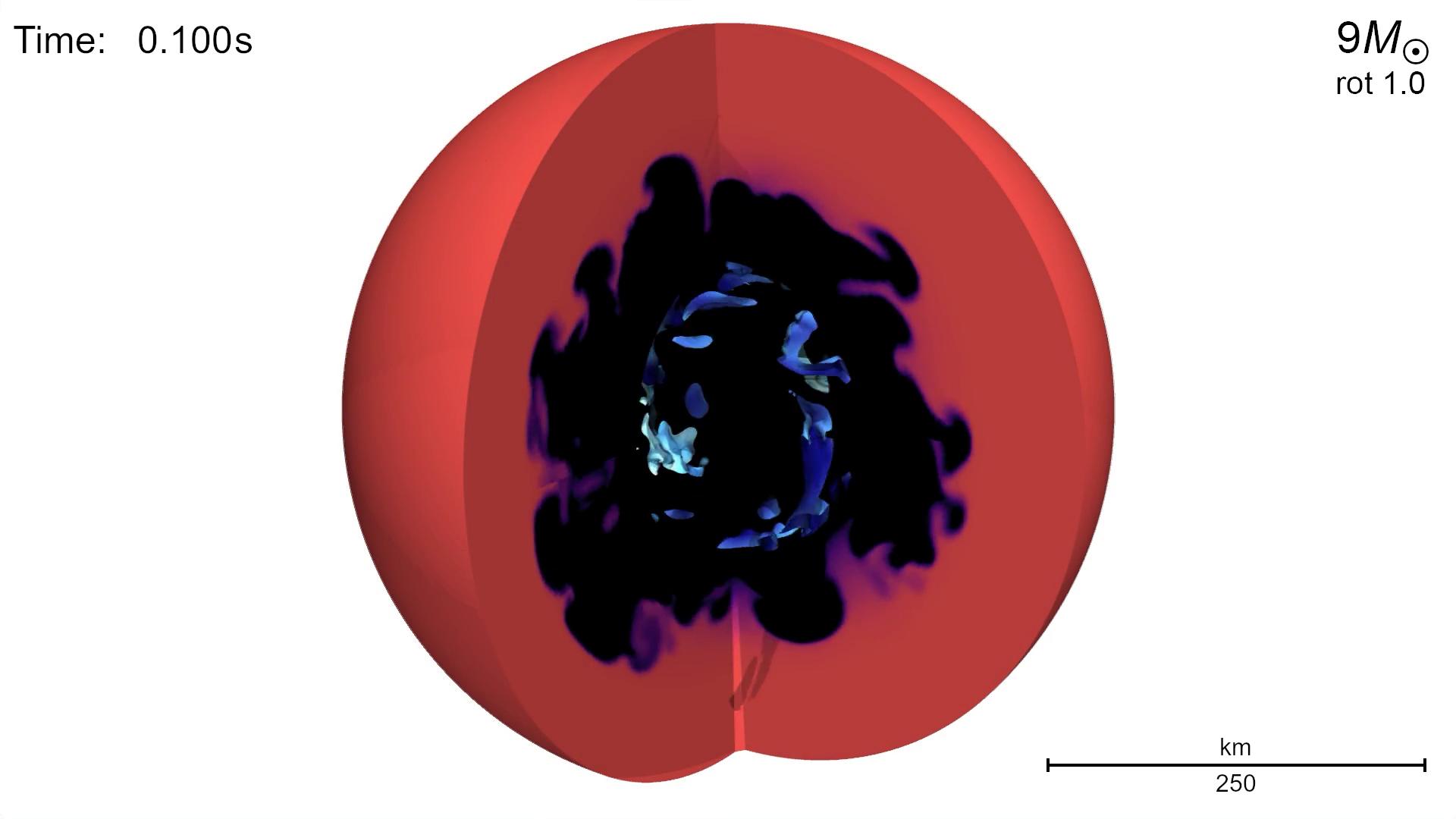}
    \includegraphics[width=0.45\textwidth]{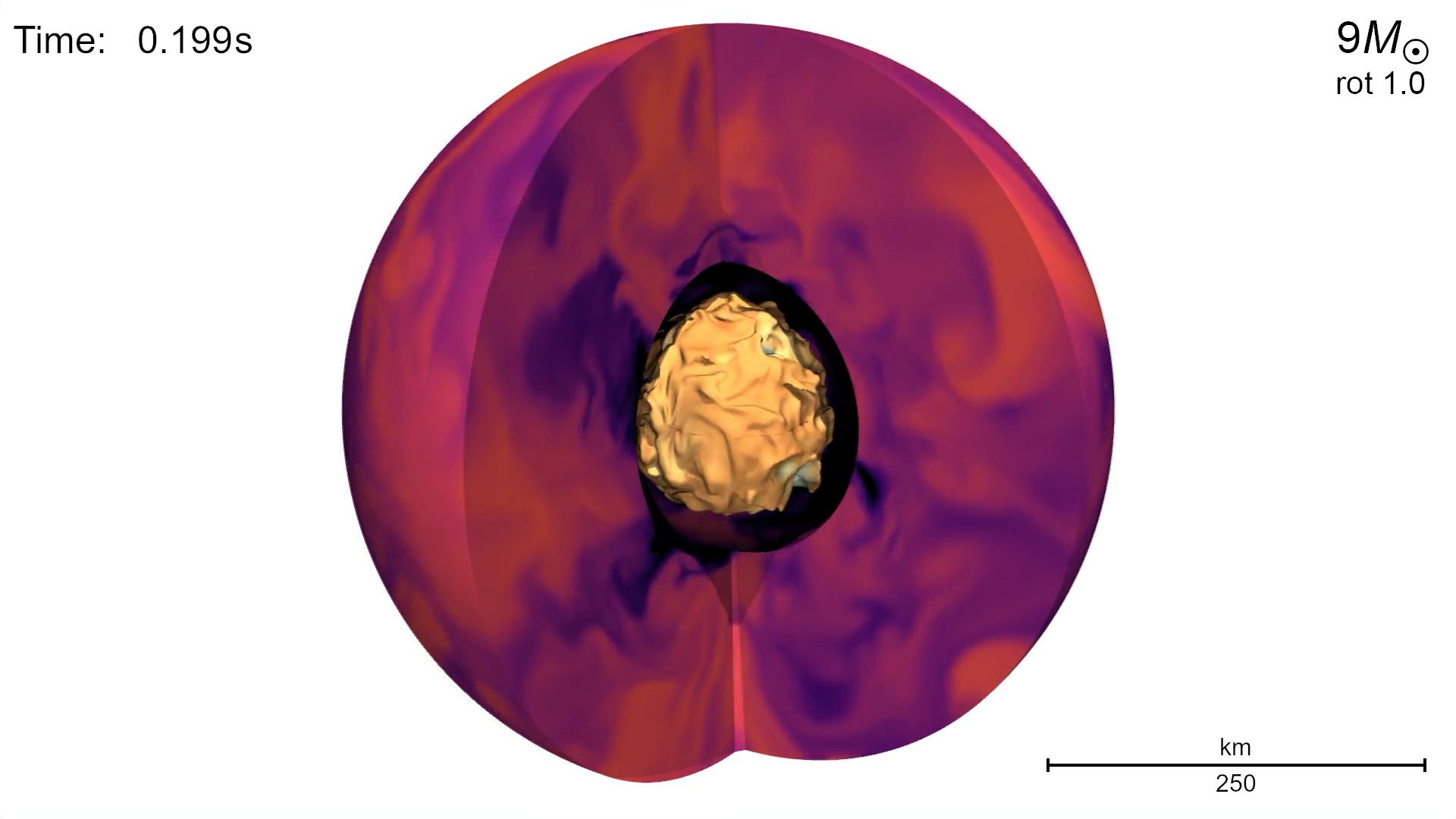}
    \caption{Cutaway slices depicting electron fraction ($Y_e$; blue low and red higher) on the surfaces, with an inner embedded isodensity surface (at $10^{9.5}$ g cm$^{-3}$, colored by $Y_e$), for the non-rotating and most rapidly rotating (9-rot-1.0) models at 100 (left) and $\sim$200 milliseconds (right) after bounce. On the left plots, the darker regions depict those in which there has been significant electron capture behind the shock that bounds it. On the right plots, the shock has left the inner domain depicted. Note that the spinning inner sphere at later times is larger (due to centrifugal support). This is one of the few salient differences between most of the models and the 9-rot-1.0 model that we see.}
    \label{fig:peach}      
\end{figure*}

In Figure \ref{fig:peach}, we provide cutaway slices depicting electron fraction ($Y_e$) on the surfaces, with an inner embedded isodensity surface (at $10^{9.5}$ g cm$^{-3}$, colored by $Y_e$), for the non-rotating (9-rot-0.0) and most rapidly rotating (9-rot-1.0) models at two different times after bounce. On the left panels, the darker regions depict where significant electron capture interior to the shock has deleptonized the region around the inner core  during the early post-bounce phase. On the right panels, the shock has left the inner domain, which is now experiencing net outflow. Note that due to the large centrifugal support the spinning inner sphere at later times is larger. This is one of the few salient differences between most of the models we have simulated in this paper and the 9-rot-1.0 model.

\begin{figure*}[htbp!]
    \centering                
    \includegraphics[width=0.45\textwidth]{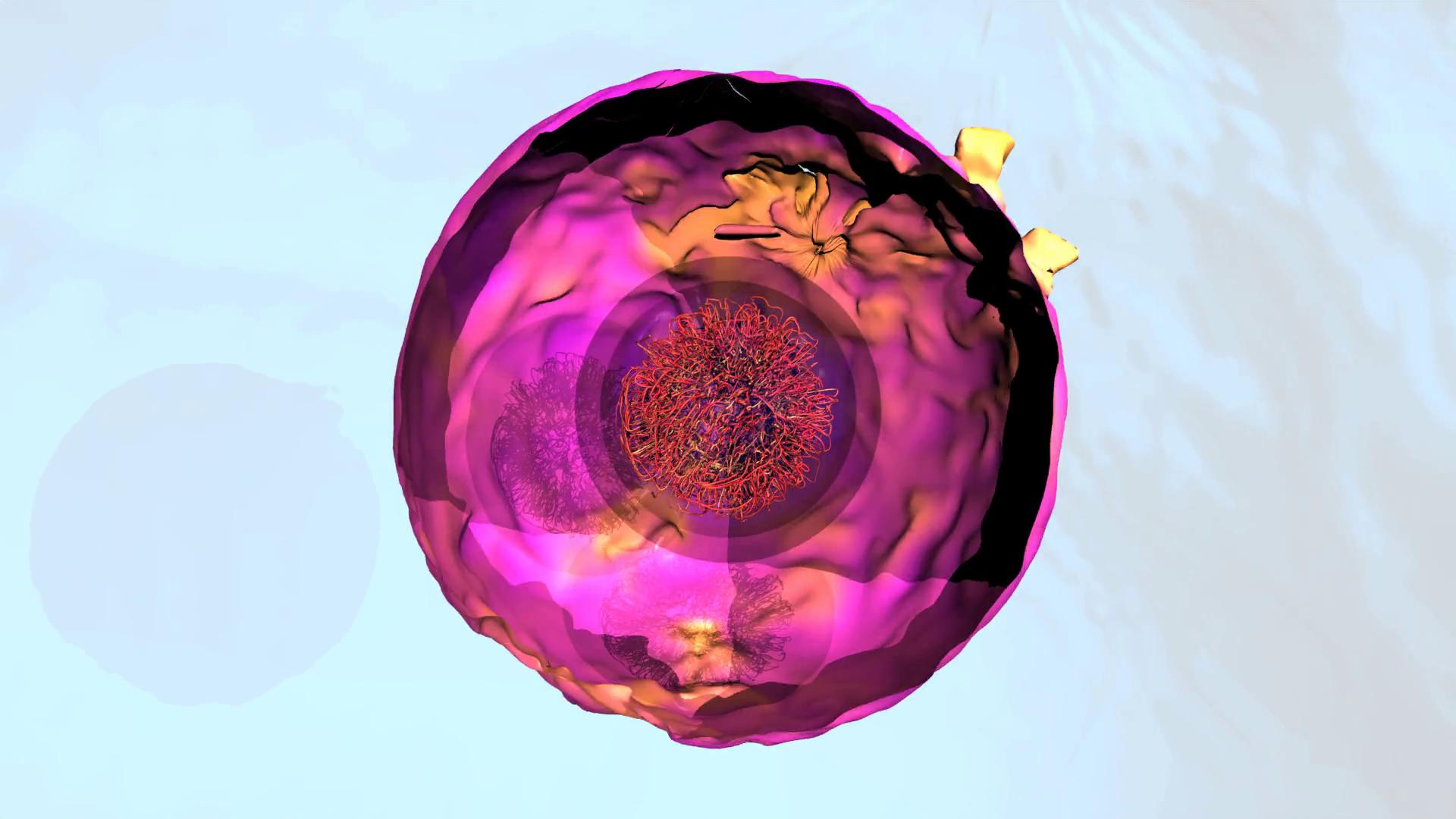}
    \includegraphics[width=0.45\textwidth]{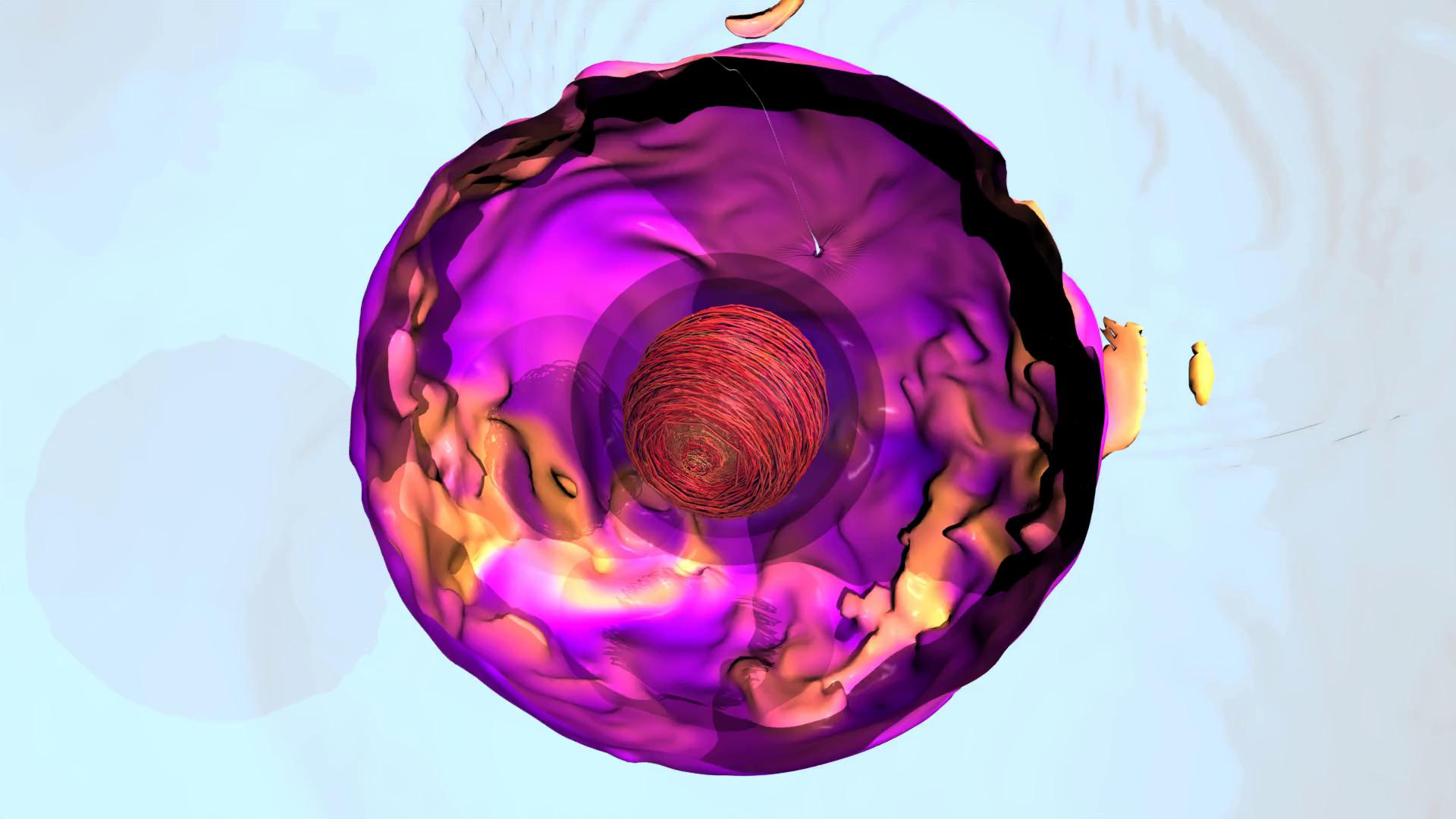}
    \caption{Nested isodensity spheres (at $10^{9.5}$ and $10^{11}$ g cm$^{-3}$), colored by $Y_e$, bounding the inner protoneutron star convective region indicated by particle trajectories, for the non-rotating (left) and $\Omega_0 = 1.0$ radian per second (right) models. Noted that rotation for the initially rotating model clearly dominates the motion of its inner core and that its isosurfaces are at larger radii. Chaotic convective motion is clearly in evidence in the core of the initially non-rotating model. Both snapshots are at $\sim$200 milliseconds after bounce. As Figures \ref{fig:peach} and \ref{fig:sauron} demonstrate, the effect of rotation is the slight expansion of the inner structure of the PNS and its environs.}
    \label{fig:sauron}      
\end{figure*}

In Figure \ref{fig:sauron}, we portray nested isodensity spheres 200 milliseconds after bounce, colored by $Y_e$. These bound the PNS and the inner convective region (traced with particle trajectories) for the 9-rot-0.0 and 9-rot-1.0 models. Note that rotation for the 9-rot-1.0 model clearly supports its structure. In addition, chaotic convective motion is clearly in evidence in the core of both models. As Figures \ref{fig:peach} and \ref{fig:sauron} demonstrate, the early effect of rapid rotation is the slight expansion of the inner structures left behind. However, initially the effect is not dramatic, and for the other more slowly rotating models (9-rot-0.01 and 9-rot0.1) is even weaker.

\begin{figure*}[htbp!]
    \centering
    \includegraphics[width=0.45\textwidth]{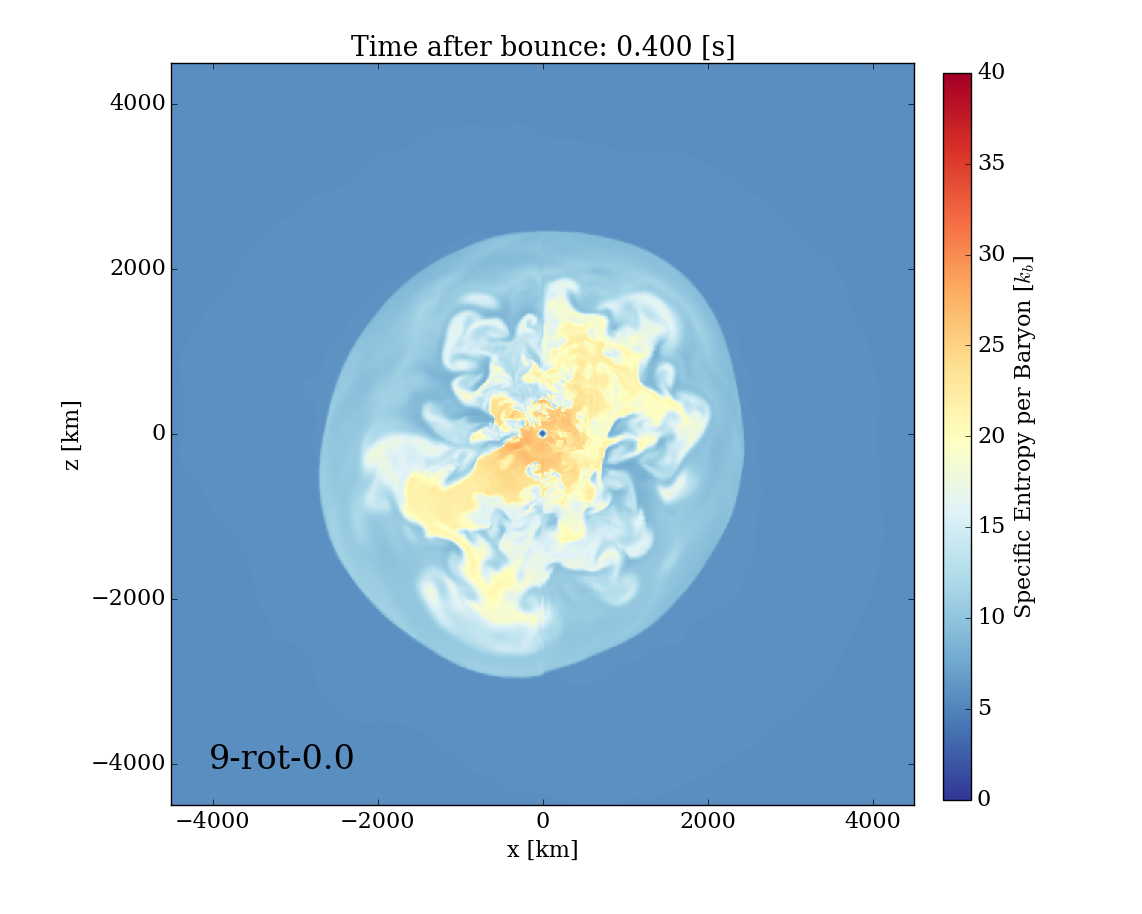}
    \includegraphics[width=0.45\textwidth]{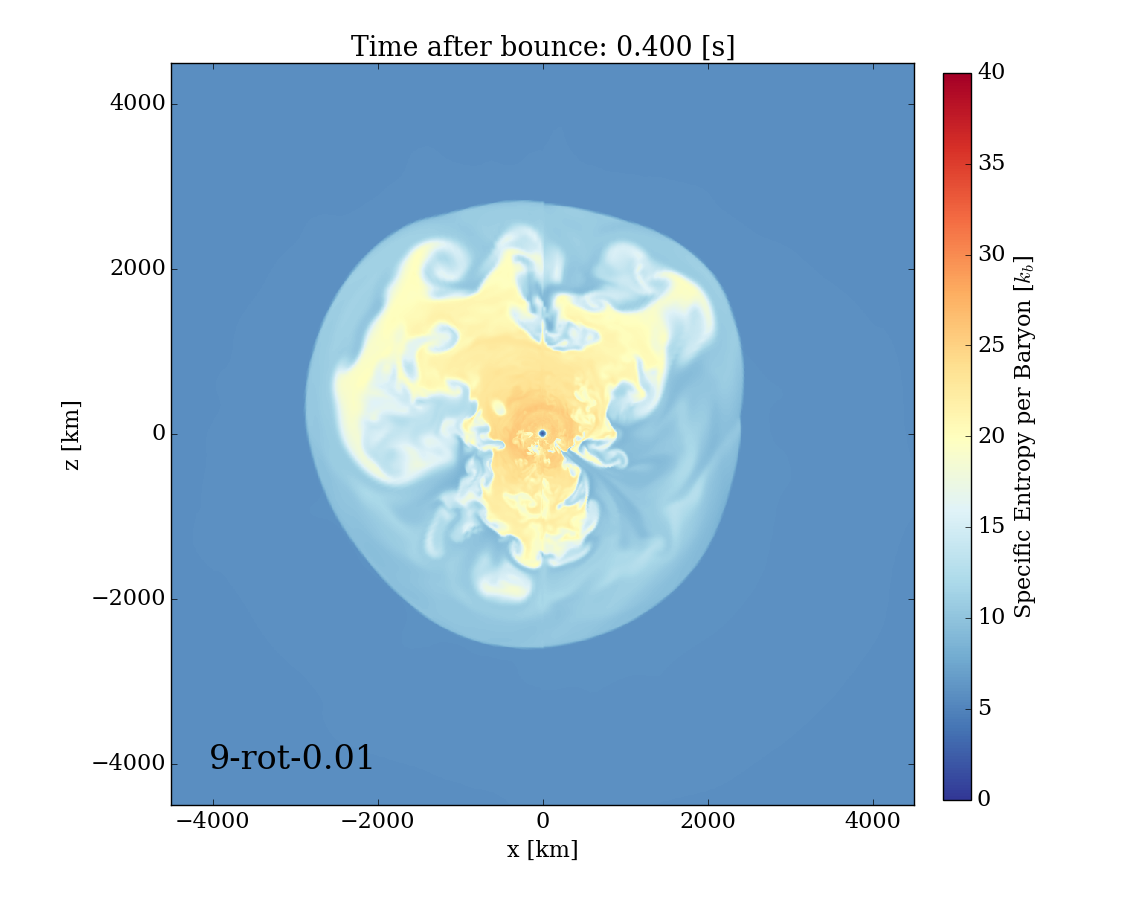}
    \includegraphics[width=0.45\textwidth]{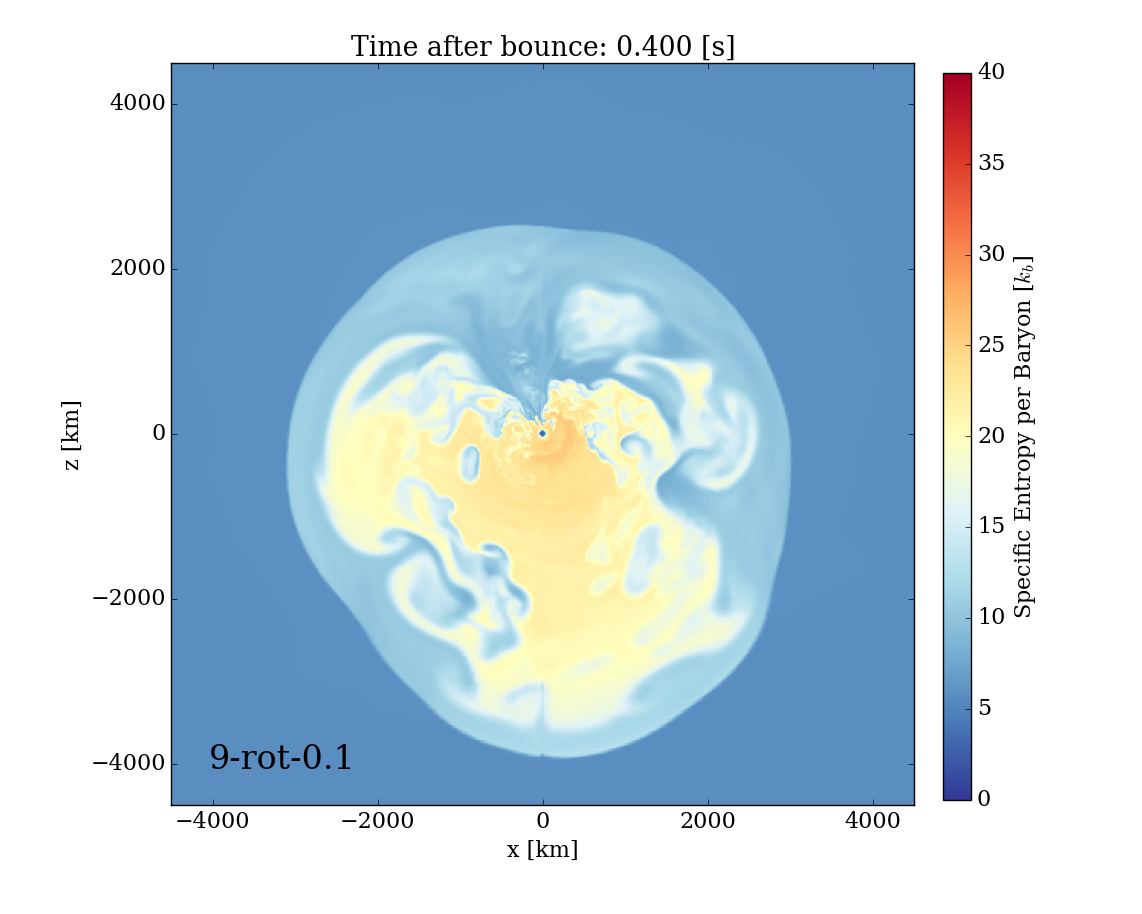}
    \includegraphics[width=0.45\textwidth]{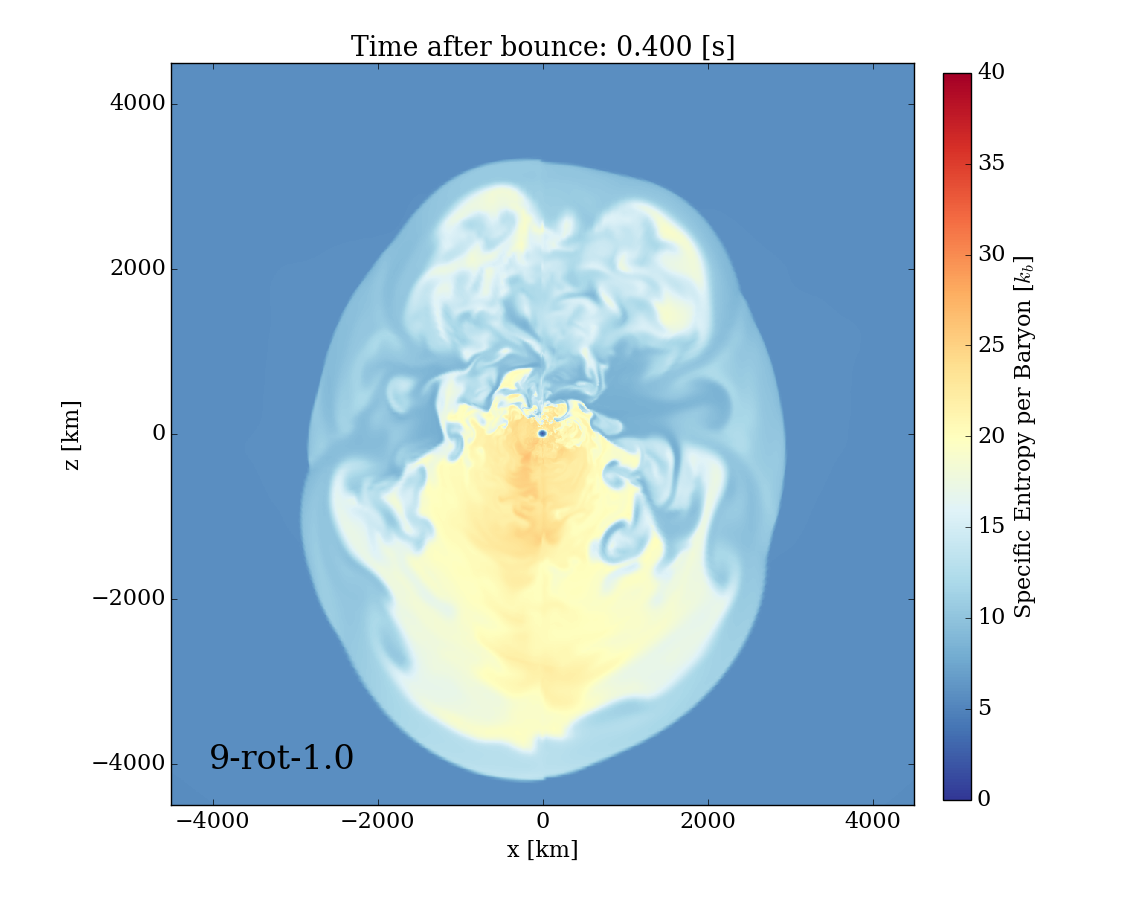}
    \caption{Entropy slices in the x-z plane of all models at 0.4 seconds post-bounce. The $\Omega_0=0.0$ and $0.01$ rad s$^{-1}$ models show more spherical explosions, while modest pole-equator structures develop in the faster rotating 9-rot-1.0 model. This behavior is consistent with the shock radius decomposition shown in the right panel of Figure \ref{fig:rshock}.}
    \label{fig:morphology}      
\end{figure*}

Figure \ref{fig:morphology} renders entropy slices in the x-z plane of all models at 0.4 seconds after bounce. The 9-rot-0.0 and 9-rot-0.01 models explode more spherically, while a modest pole-equator contrast is in evidence only in the fastest rotating 9-rot-1.0 model. This behavior is consistent with the shock radius decompositions shown in the right panel of Figure \ref{fig:rshock} and emphasizes the fact that without magnetic fields even models with rapid rotation do not explode with a jet-like morphology.

\begin{figure*}[htbp!]
    \centering
    \includegraphics[width=0.45\textwidth]{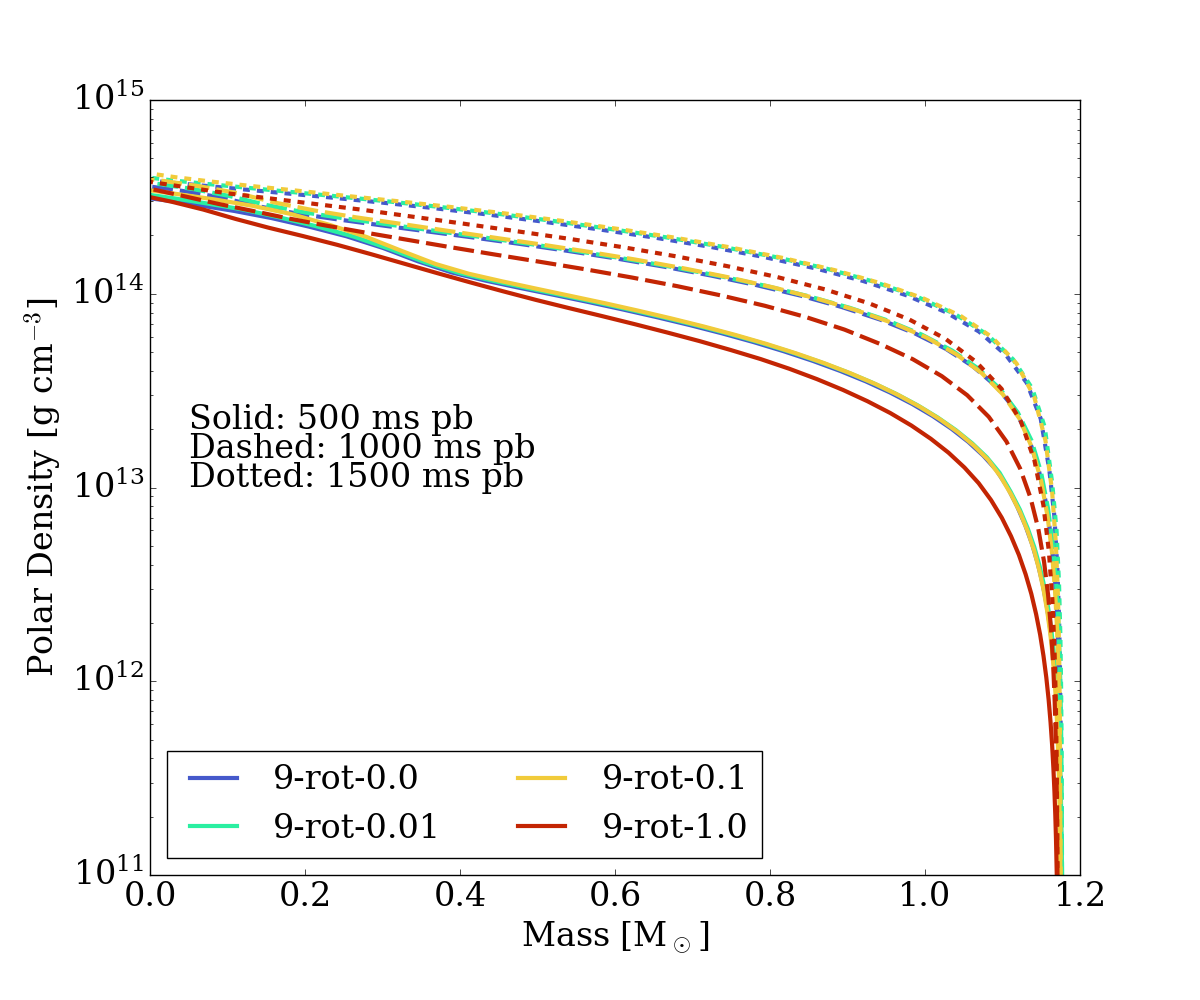}
    \includegraphics[width=0.45\textwidth]{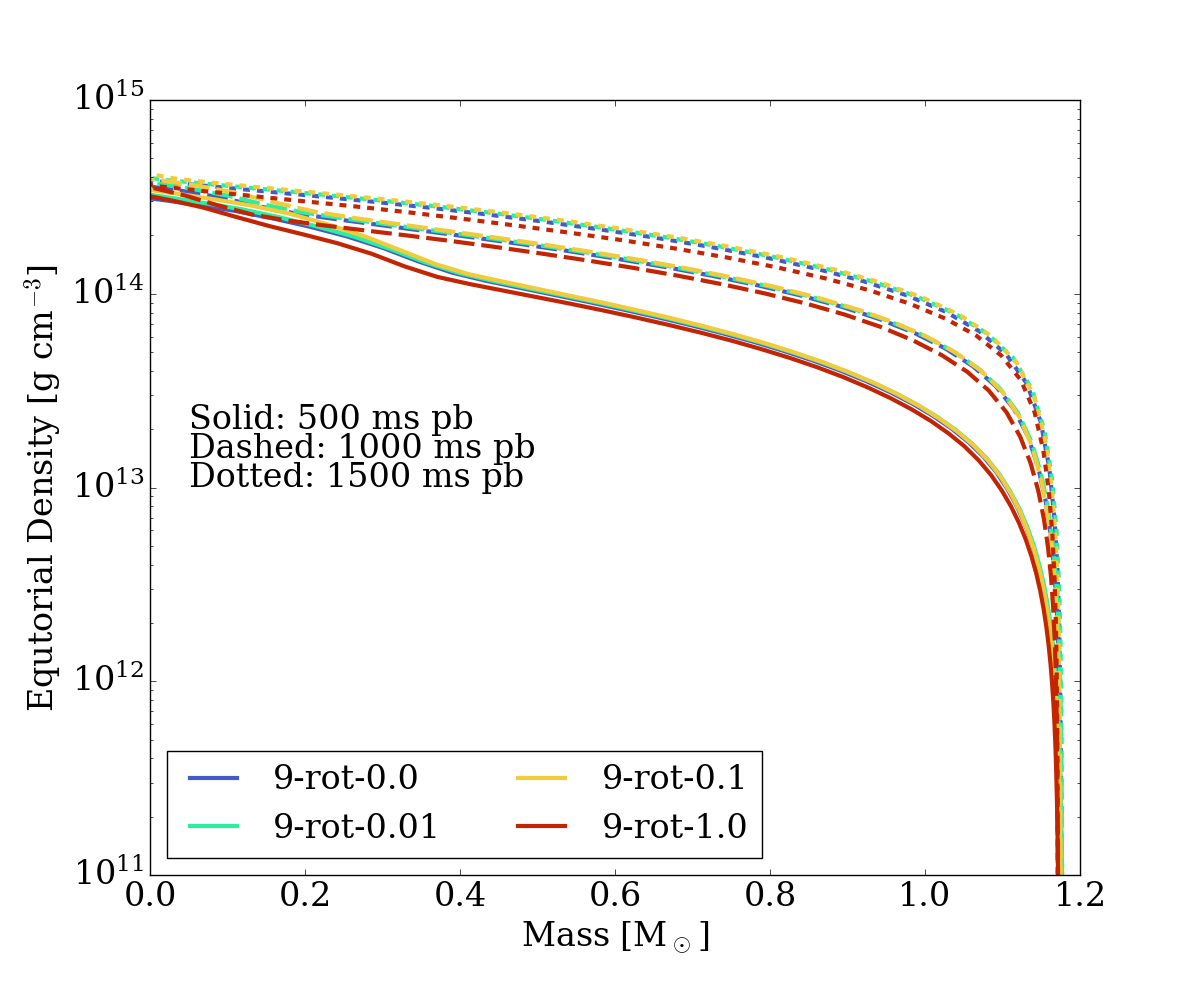}
    \includegraphics[width=0.45\textwidth]{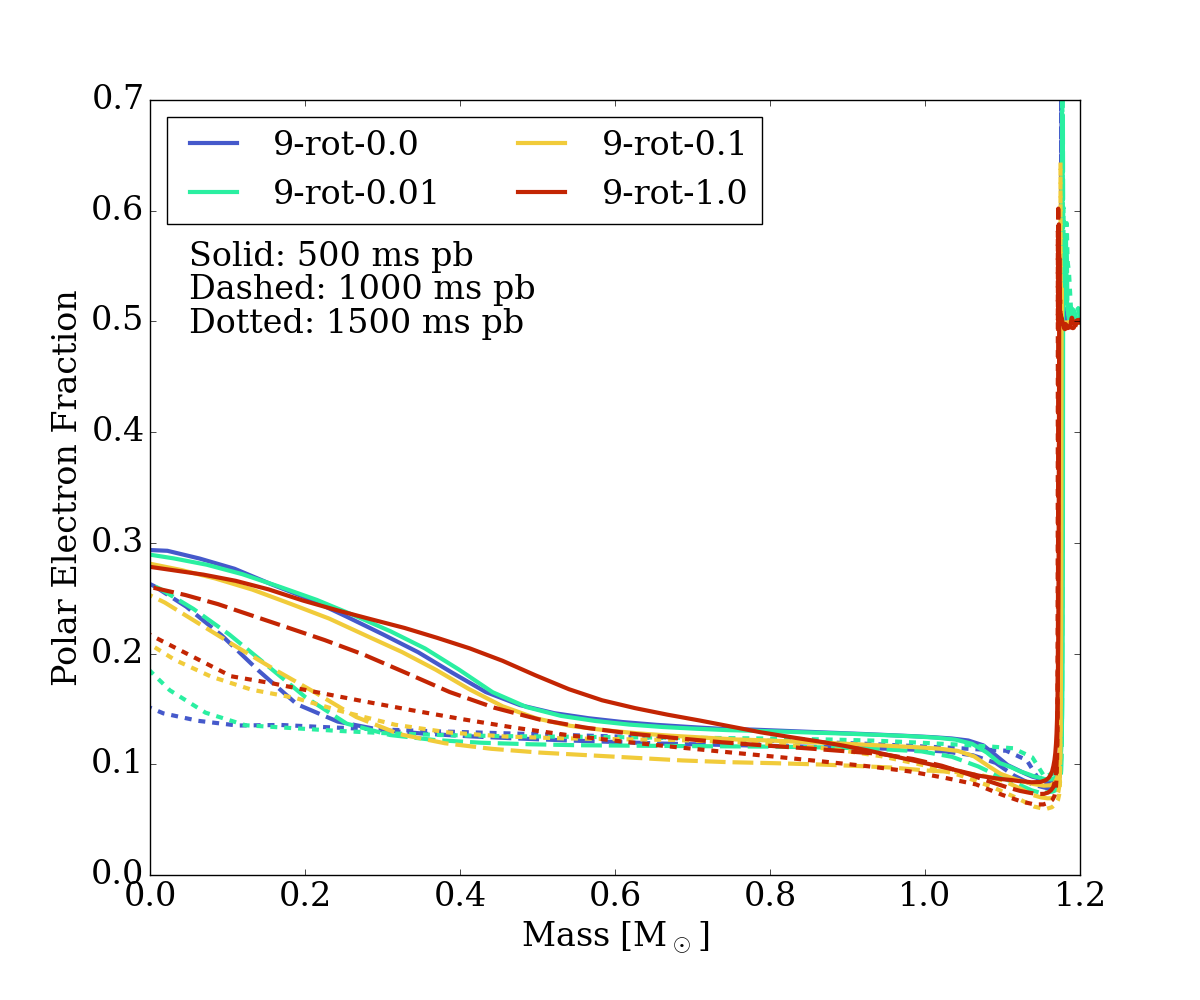}
    \includegraphics[width=0.45\textwidth]{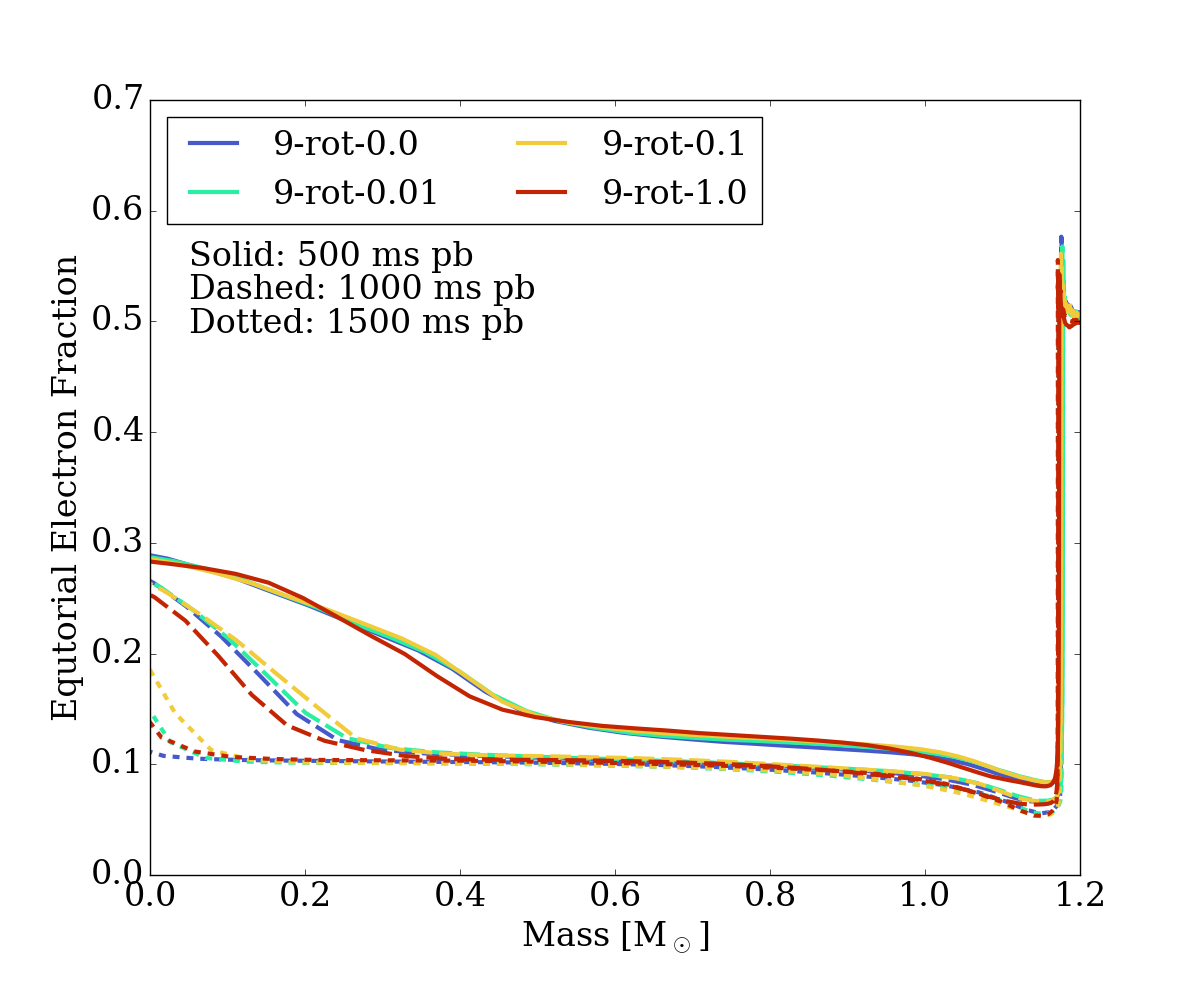}    
    \caption{Density (top) and electron fraction (bottom) profiles at 500, 1000, and 1500 ms post-bounce along polar (left) and equatorial (right) directions. Compactification with time is manifest for all models. It is only the fastest rotating model (red) that clearly deviates from the others. One sees the same trend on the bottom plots, with the polar profiles of the 9-rot-1.0 model deviating most. These plots again suggest, however surprisingly, that rotation comes into its own only for high initial spins.}
    \label{fig:profile}      
\end{figure*}

\begin{figure*}[htbp!]
    \centering
    \includegraphics[width=0.45\textwidth]{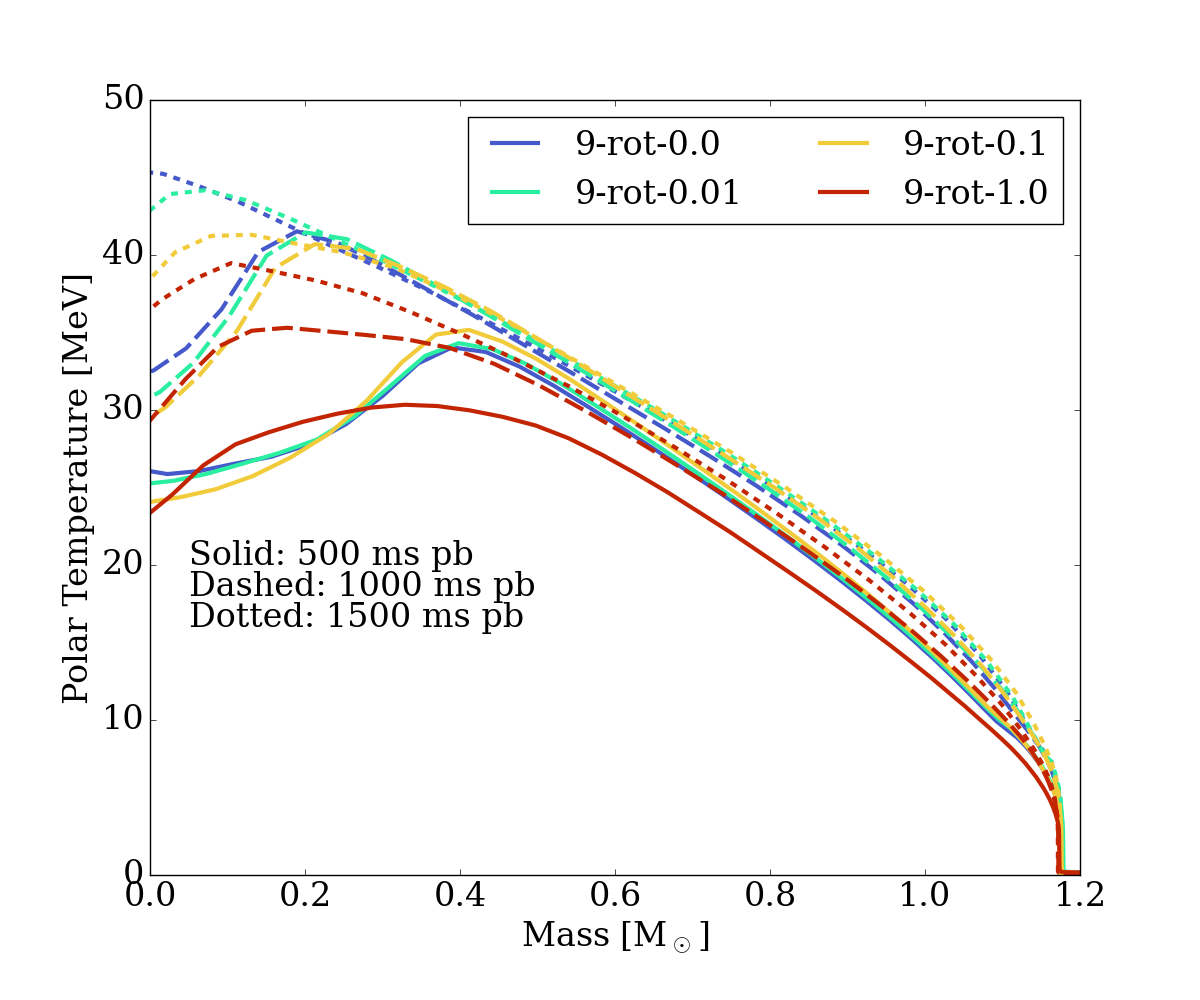}
    \includegraphics[width=0.45\textwidth]{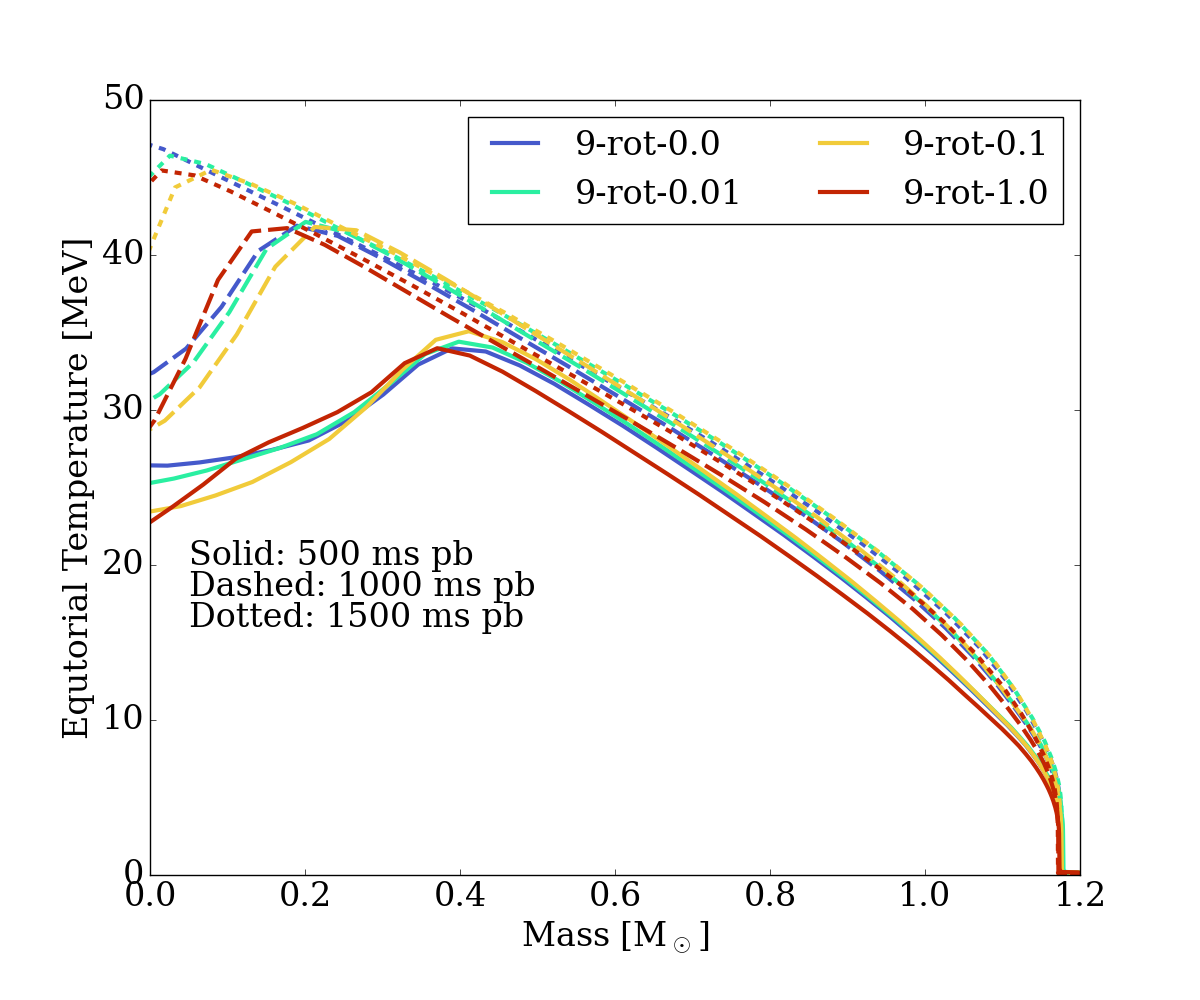}
    \includegraphics[width=0.45\textwidth]{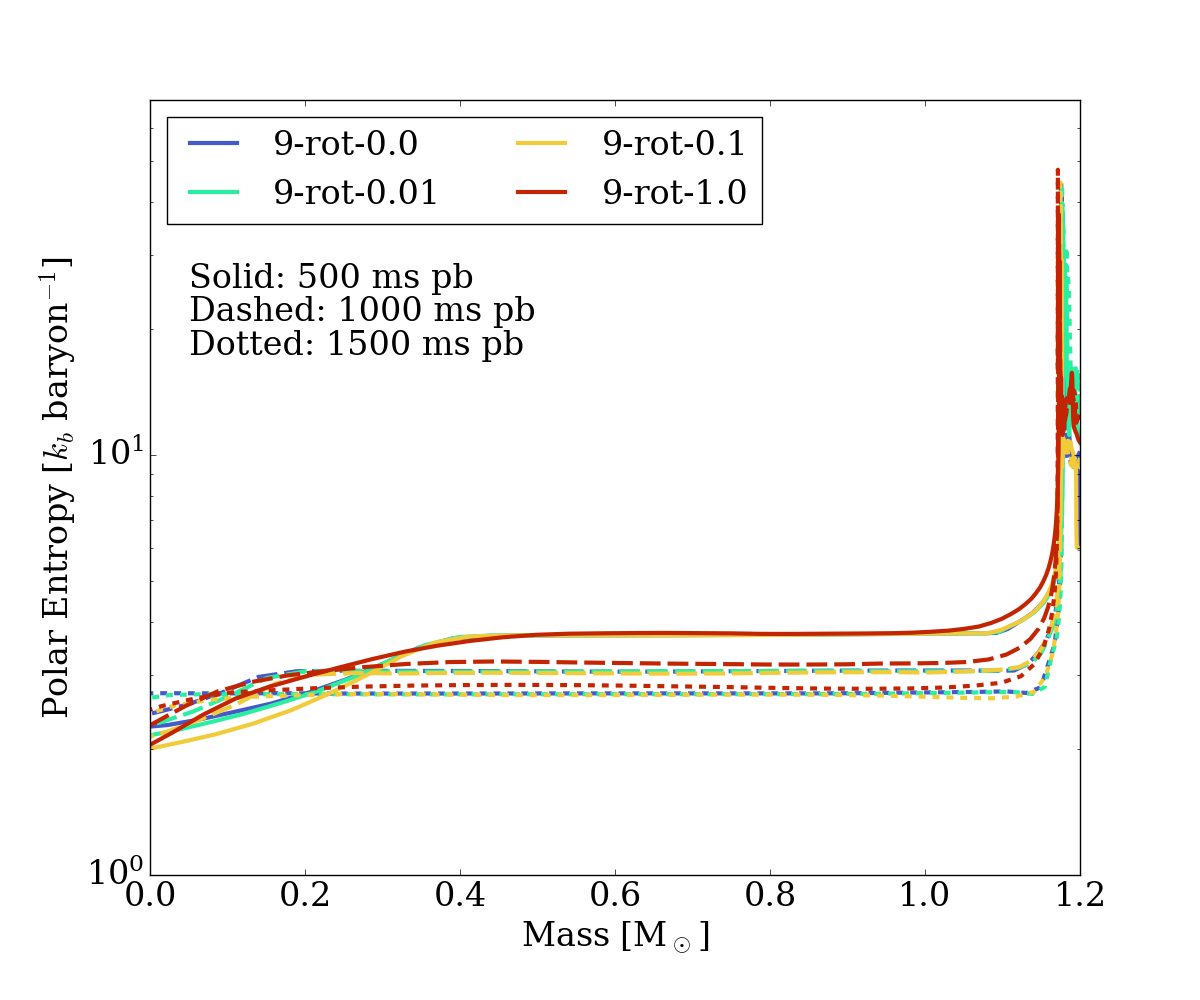}
    \includegraphics[width=0.45\textwidth]{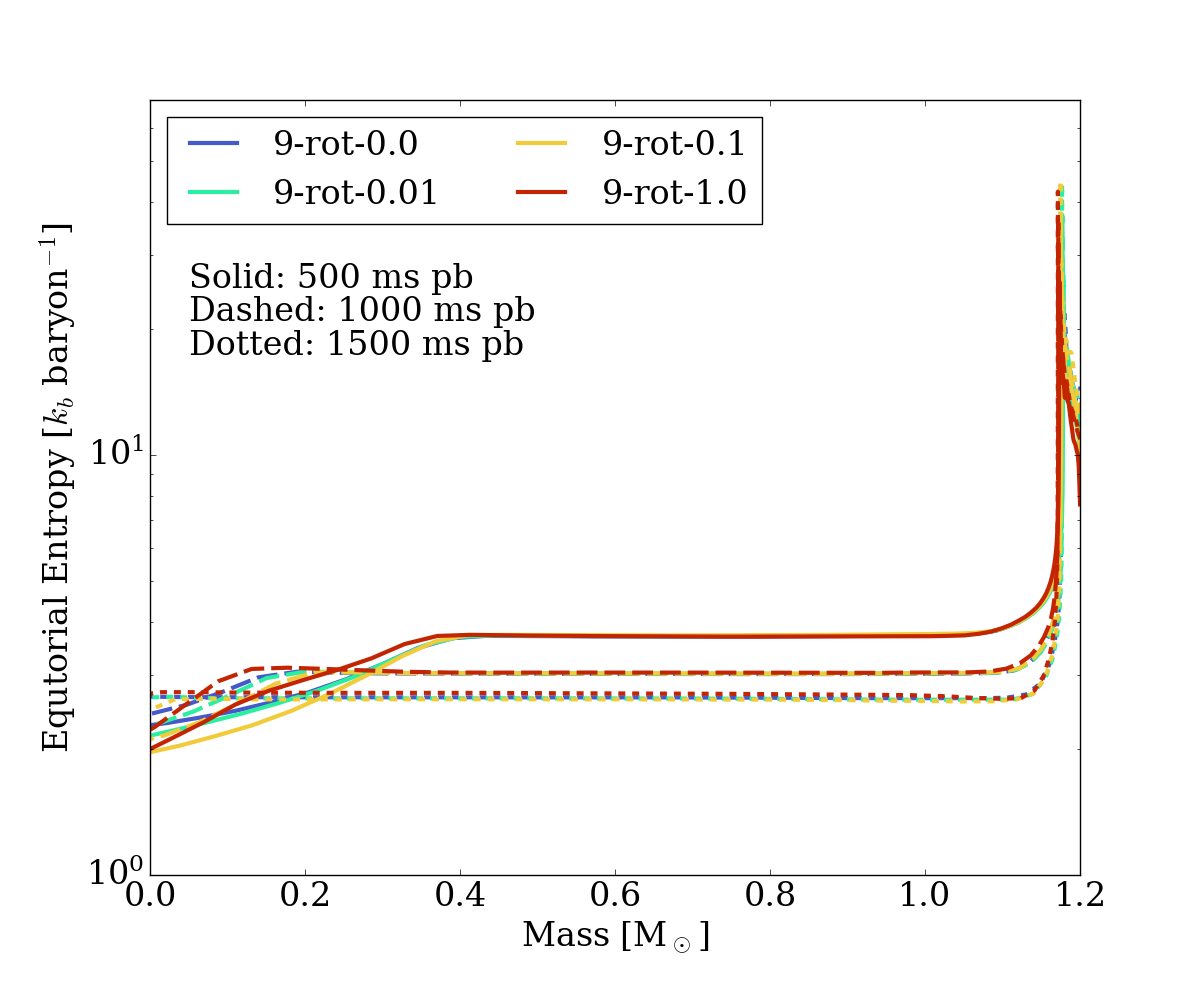}
    \caption{Same as Figure \ref{fig:profile}, but for temperature (top) and entropy (bottom) profiles in the polar (left) and equatorial (right) directions.}
    \label{fig:profile-2}      
\end{figure*}

The fact that the fast rotating model 9-rot-1.0
is the singular outlier is emphasized with Figures \ref{fig:profile} and \ref{fig:profile-2}. There we plot the polar (left) and equatorial (right) profiles at various times after bounce of the baryon mass density and electron fraction ($Y_e$) (Figure \ref{fig:profile}) and the temperature and entropy (Figure \ref{fig:profile-2}). It is only model 9-rot-1.0 which is much distinguished from the others..

Table \ref{tab:main} lists the initial and final (after full deleptonization and cooling) spin periods of the models included in this paper.  We see that even the 9-rot-0.01 model results in a pulsar spin period of $\sim$132 ms, faster than the current mean pulsar periods observed \citep{emmering1989,faucher_kaspi,Kaspi2016} and according to our analysis (\S\ref{hydro}) is not hydrodynamically of much consequence. The kinetic energy of rotation at two seconds after bounce for this model is but $\sim$1.7$\times$10$^{49}$ ergs. 

Heger et al. \citep{Heger2003,2005ApJ...626..350H} calculated massive star evolutionary models with spin and magnetic fields using a Taylor-Spruit dynamo prescription for both field generation and angular momentum transport by magnetic torques. They concluded that the core spin periods were 
$\sim$30$-$60 seconds at the time of collapse, the faster for the more massive progenitors. These numbers are close to that of our 9-rot-0.1 model, from which we might determine that pulsars would be born with spin periods near $\sim$15 milliseconds. However, this is near that inferred for the Crab at birth \citep{Kou2015} and is $\sim$20 times faster than the inferred pulsar mean. Interestingly, asteroseismological measurements of the spins in the cores of stars generally yield spin rates $\sim$10 times lower than would be expected from the standard Taylor-Spruit dynamo \citep{cantiello2014,fuller2019,fuller2022}. The upshot is that the cores of singlet massive stars are now expected to be very slowly rotating. Given this and the results communicated in \S\ref{hydro} concerning the very modest effects on the explosion of even our 9-rot-0.01 and 9-rot-0.1 models, one might conclude that rotation would have little conceptual or physical consequence. 

\begin{figure*}[htbp!]
    \centering
    \includegraphics[width=0.45\textwidth]{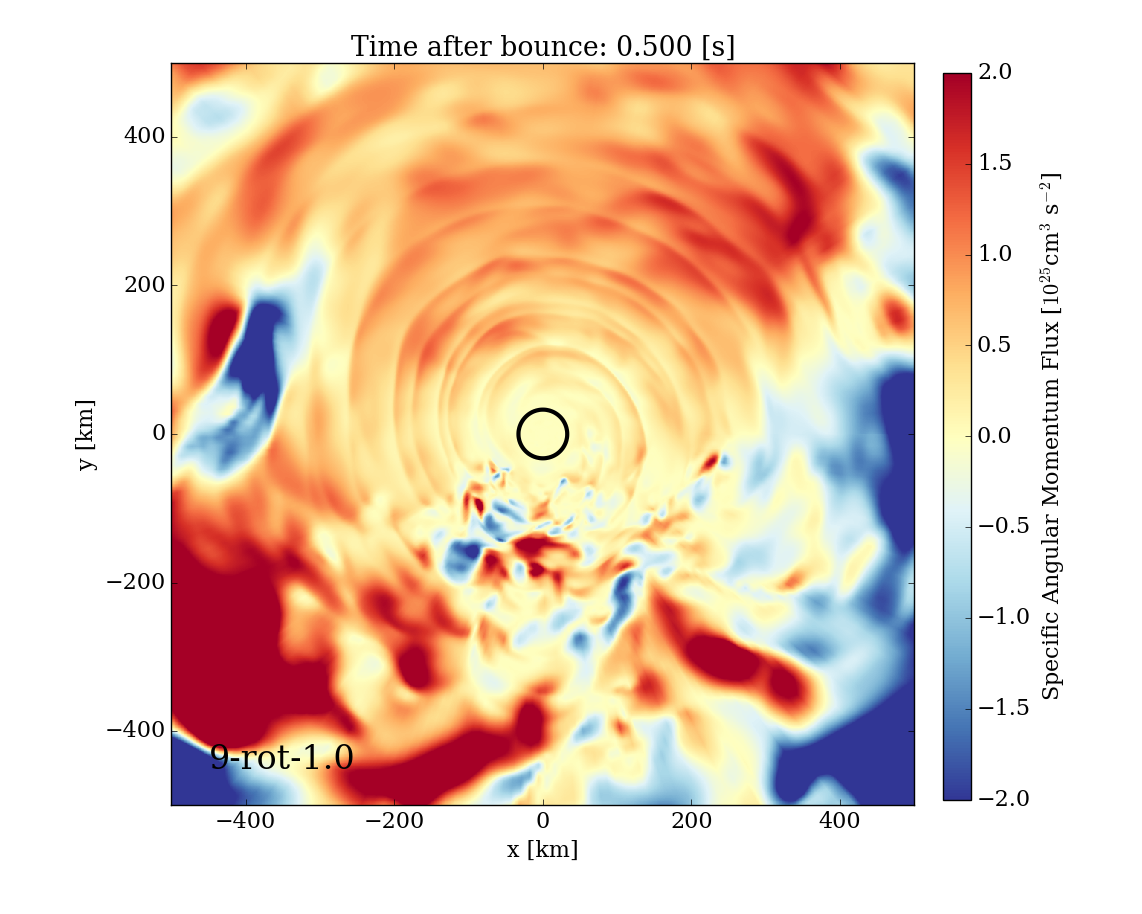}
    \includegraphics[width=0.45\textwidth]{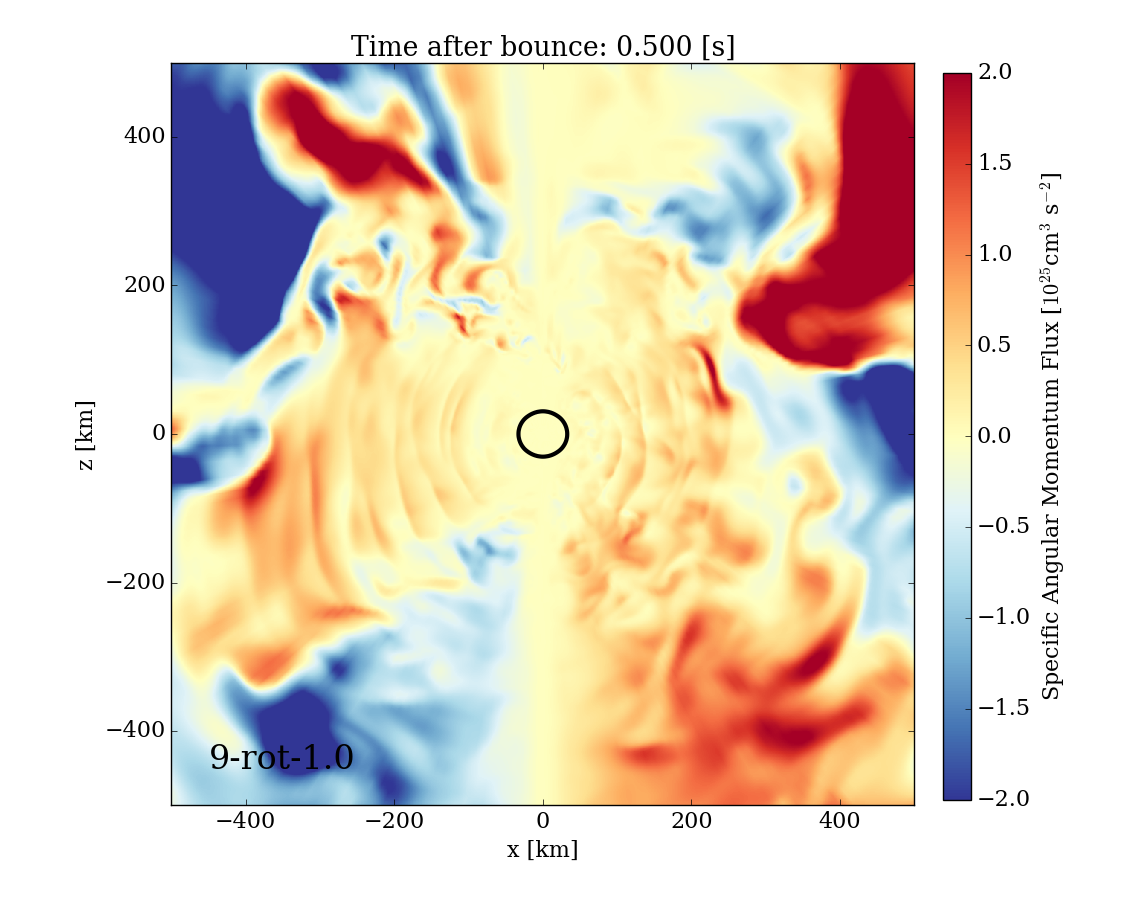}
    \includegraphics[width=0.45\textwidth]{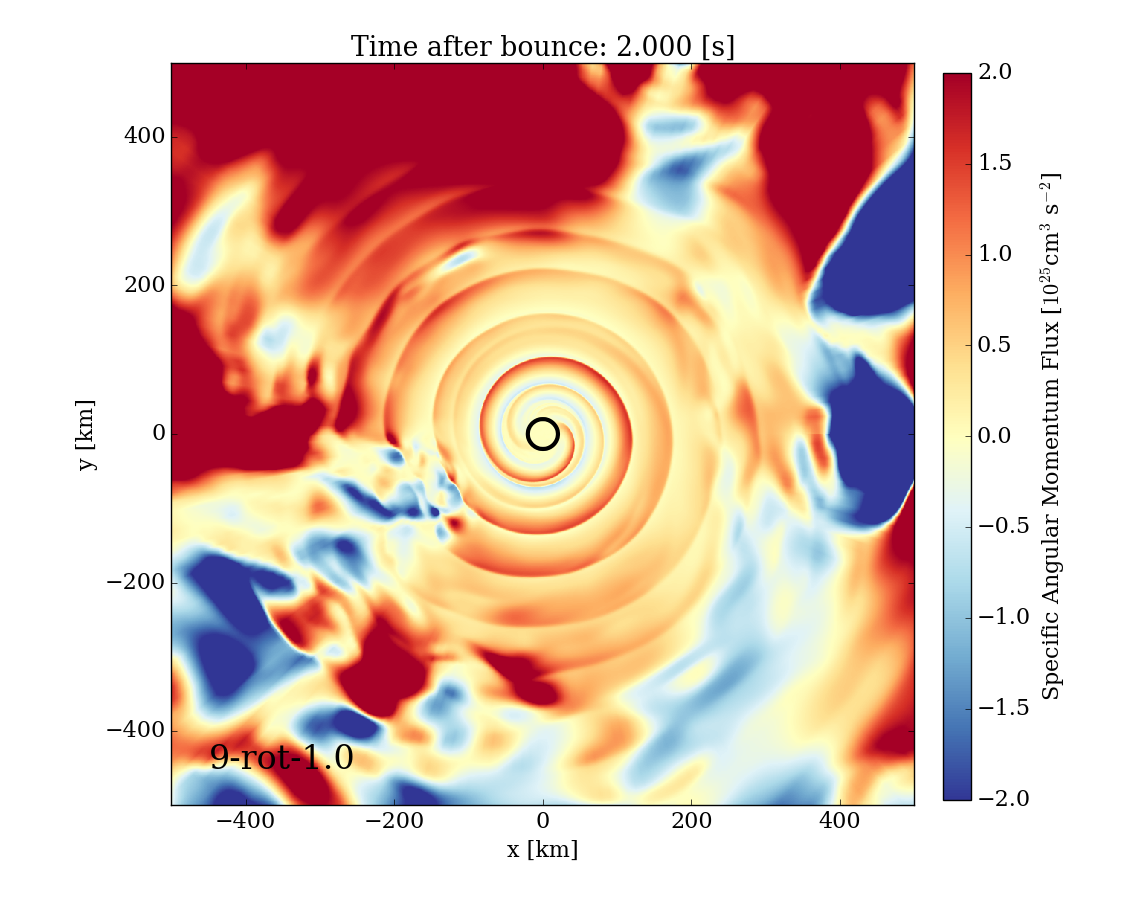}
    \includegraphics[width=0.45\textwidth]{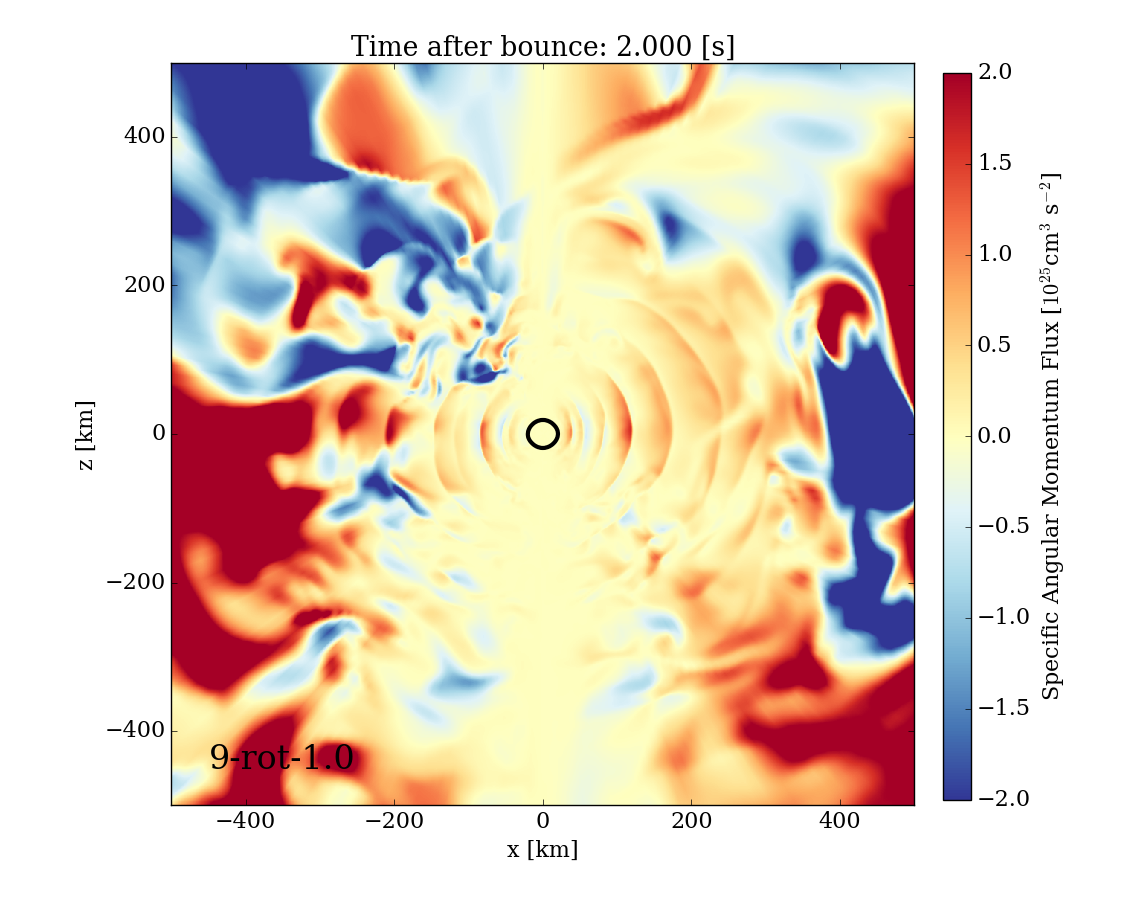}
    \caption{Specific angular momentum flux in the y-x (left) and z-x (right) planes at 0.5 (top) and 2 (bottom) seconds post-bounce for the 9-rot-1.0 model. Within $\sim$300 ms of bounce, the core that is shrinking quasi-statically due to neutrino losses during its Kelvin-Helmholtz cooling phase and correspondingly spinning up seems to hit various co-rotation resonances. This is more clearly seen in Figure \ref{fig:egw2}. The result is the generation and emission of spiral waves/arms with various and changing mixes in angular quantum number $m$. See text for a discussion.}
    \label{fig:Lflux}      
\end{figure*}

\begin{figure*}[htbp!]
    \centering
    \includegraphics[width=0.45\textwidth]{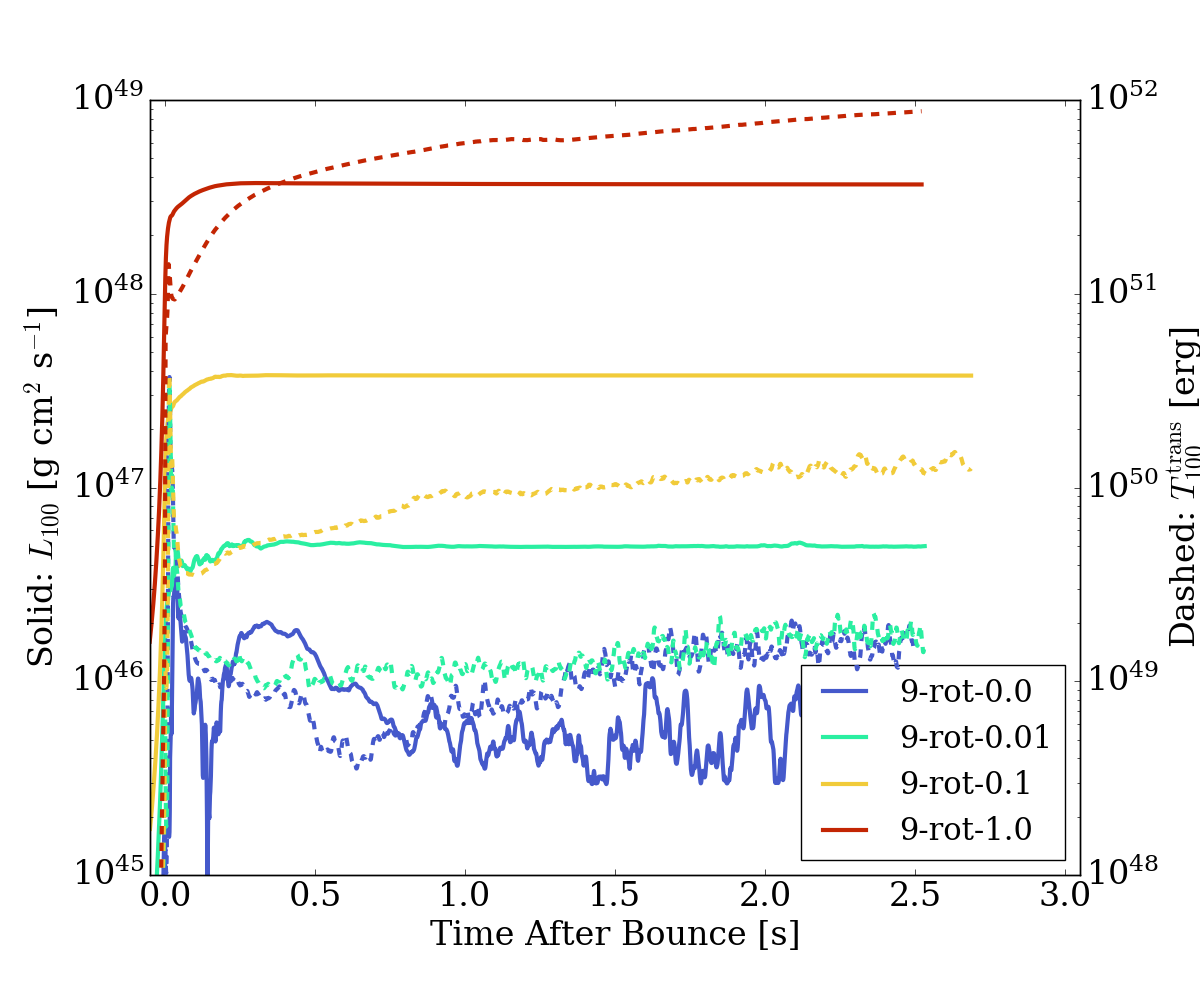}
    \includegraphics[width=0.45\textwidth]{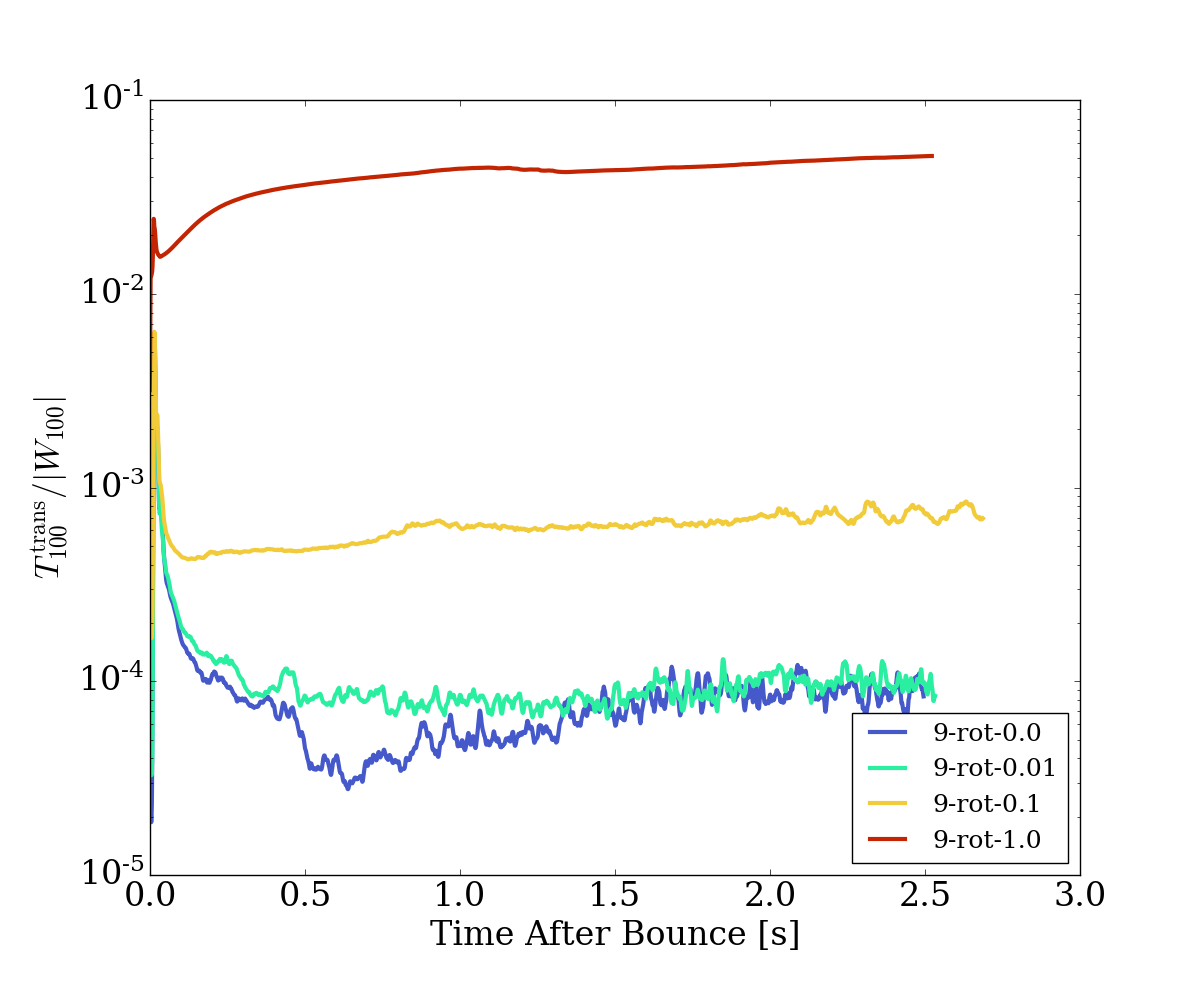}
    \caption{\tianshu{Temporal evolution of the angular momentum (solid lines in the left panel), transverse kinetic energy (dashed lines in the left panel), and $T/|W|$ (right panel) of all models. The quantities are measured for the inner 100 km; we have tested that changing this radius leads to negligible differences as long as the PNS is enclosed. The transverse kinetic energy shown here includes all transverse motions. Thus, convection and turbulence are included as well. However, for models with initial spin greater than ``0.01," the transverse kinetic energy is dominated by the bulk rotation of the PNS and averaging out the non-bulk motions doesn't change the plot in a visible way. As shown by the right panel, because the accretion weakens the angular momentum no longer changes after about 0.4 seconds post-bounce (the fluctuations in the non-rotating model are due to convection). This demonstrates angular momentum is well conserved. The increase of the transverse kinetic energy and $T/|W|$ is due to the contraction of the PNS. In the fastest-spinning model, the $T/|W|$ reaches about 0.03 at 300 ms, and it gradually increases to above 0.05 after 2.3 seconds post-bounce.}}
    \label{fig:t_over_w}      
\end{figure*}

However, our model 9-rot-1.0, with $\sim$60 times the final spin kinetic energy of model 9-rot-0.1, manifests some very interesting behavior. In Figure \ref{fig:Lflux}, we plot the specific angular momentum flux in two slices at two times post-bounce for our 9-rot-1.0 model. After early explosion, for which spin contributed only modest modifications even for this fastest-rotating model, the protoneutron star continues to cool, deleptonize, and shrink. The shrinkage spins the PNS up secularly, which increases the $T$/$|W|$ ratio, and the degree of differential rotation evolves. 
\tianshu{Figure \ref{fig:t_over_w} shows the temporal evolution of angular momentum, transverse kinetic energy, and $T/|W|$ of all models. Within $\sim$300 ms of bounce, $T/|W|$ of the fastest spinning model approaches $\sim$0.03 and then gradually grows to above $\sim$0.05 at about 2.3 seconds post-bounce as the PNS radius shrinks. During this process, the core hits various co-rotation resonances between the spin and core pulsation modes.} This phenomenon has been seen and diagnosed in detail for rapidly rotating models by numerous authors, including \citet{shibata2002}, \citet{watts2005}, \citet{shibagaki2020_2}, and \citet{takiwaki2021MNRAS}. For our most rapidly-rotating model, we see the emergence of numerous spiral waves/arms with various and changing mixes in angular quantum number $m$. We see not only an $m=1$ arm, first seen by \citet{iwakami2014}, \citet{TaKoSu16}, and  \citet{ott_2005_onearmed,ott2012_rapid}, but multiple modes\footnote{Unlike what \citet{iwakami2014} and \citet{TaKoSu16} concluded, for this model the spiral arm energy flux is not an agency of explosion, nor does it contribute significantly to the explosion energetics.}. The response is particularly strong around $\sim$1.0 second after bounce for a duration of $\sim$300 ms, after which spiral waves of various winding numbers (at least $m =1,2$) persist, but with diminished strength.

\citet{takiwaki2021MNRAS} associate such modes with shear instabilities and the destabilization of 
Rossby waves near the PNS boundary. A comprehensive recent discussion of a similar $T/|W|$ instability for a rapidly rotating 35-$M_{\odot}$ progenitor is given in \citet{cusinato2026}. These authors associate the excitation of such modes with the overlap of the corotation radius with the PNS convective region and strong gradients in angular speeds. Unfortunately, our data dump cadence for these models was 1000 Hz, so we are Nyquist limited to 500 Hz. Hence, a detailed modal analysis of the spiral arm structures and angular momentum and energy fluxes witnessed issuing from the core has proven difficult. Given this, we defer a more detailed analysis of these $T$/$|W|$ instabilities due to rapid rotation to a later work and when our I/O cadence is elevated appropriately. Though these modes have little effect on the overall supernova explosion of model 9-rot-1.0, they have a significant effect on its gravitational-wave signature. Fortunately, our dump cadence for the gravitational-wave data was at 25 kHz to 50 kHz and we discuss the manifestations of the low-$T/|W|$ instability in the gravitational-wave signal of model 9-rot-1.0 in \S\ref{GW}. We note from Figure \ref{fig:Lflux} that the wave emissions are predominantly in the equatorial directions, with the y-x panels showing no angle dependence, while the z-x panels show an emission preference away from the z-direction. This is as expected \citep{ott_2005_onearmed,ott2012_rapid,Takiwaki2014,TaKoSu16,cusinato2026}. 
   
\section{Neutrino Luminosity and Flux Behavior as a Function of Rotation Rate}
\label{lum}

\begin{figure*}[htbp!]
    \centering
    \includegraphics[width=0.45\textwidth]{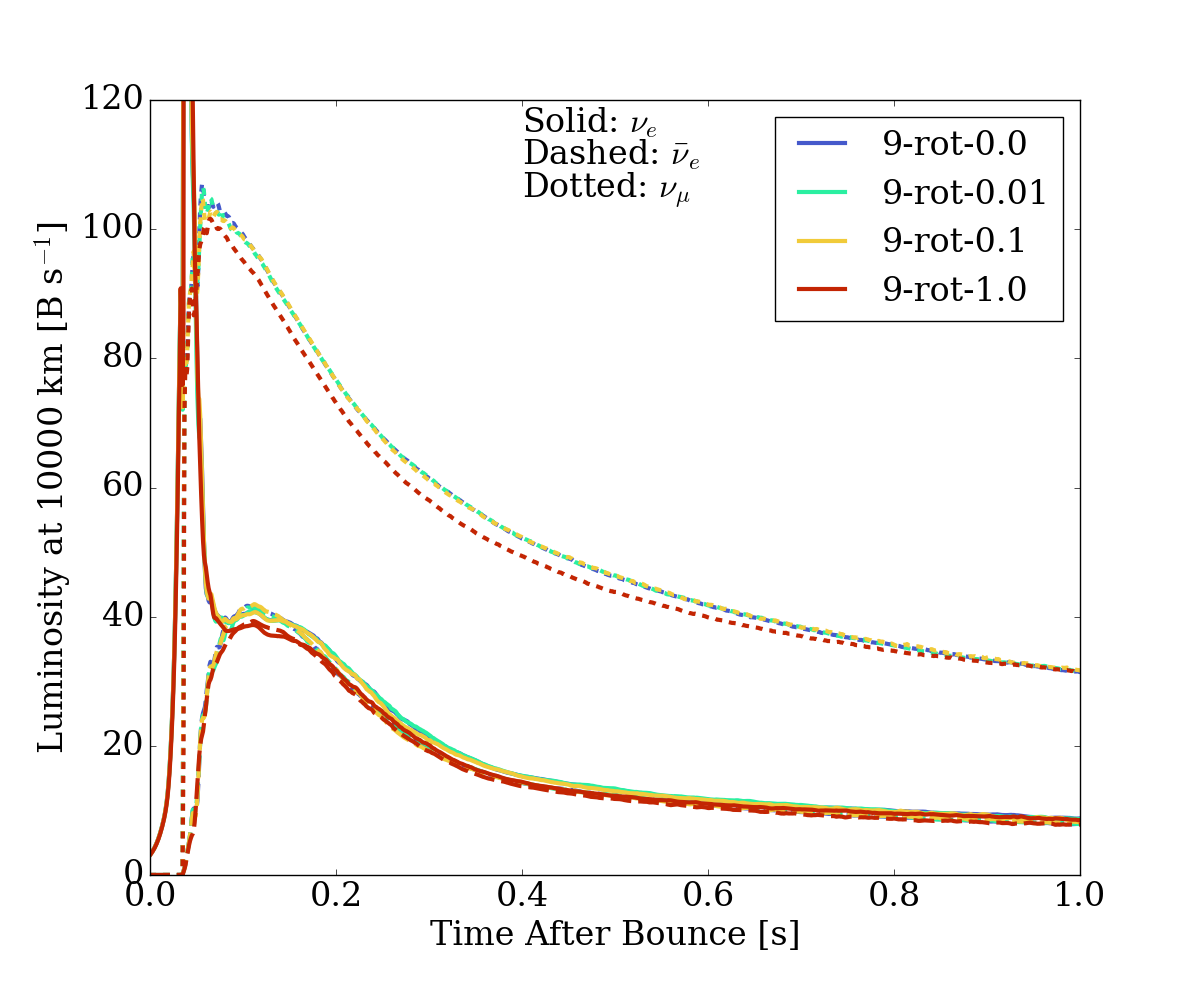}
    \includegraphics[width=0.45\textwidth]{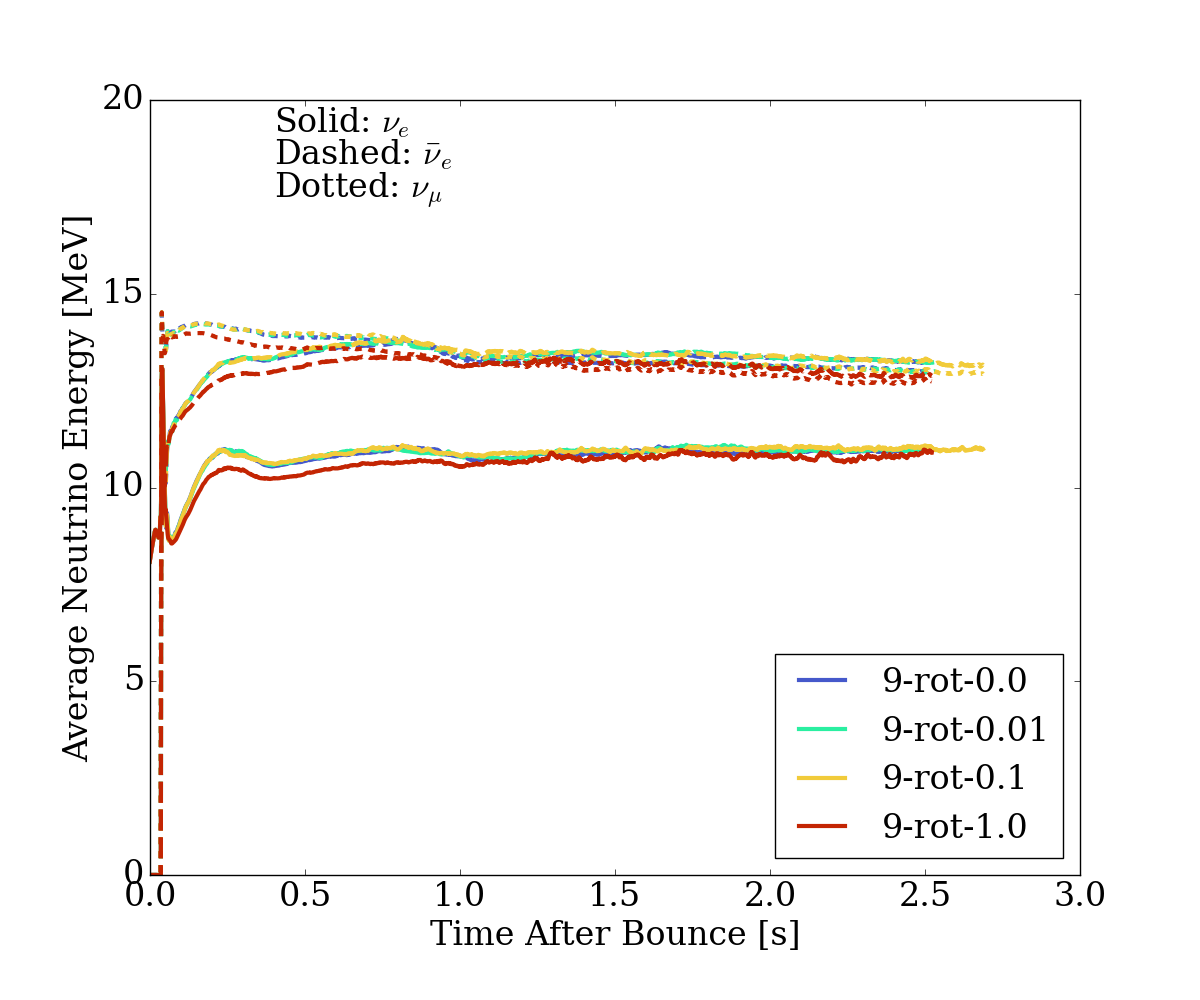}
    \caption{Angle-integrated neutrino energy luminosity (left, in Bethes per second) and average energy (right, in MeV) for the various neutrino species in the laboratory frame at 10000 km. One Bethe $\equiv$ 10$^{51}$ ergs. The dotted lines are for the $\nu_{\mu}$, $\nu_{\tau}$, $\bar{\nu}_{\mu}$, and $\bar{\nu}_{\tau}$ neutrinos collectively. The 9-rot-1.0 model is the only one that deviates much from the others, with its luminosities and mean neutrino  energies slightly lower before $\sim$300 milliseconds after bounce than the corresponding values for the other models.}
    \label{fig:Lnu}      
\end{figure*}

Figure \ref{fig:Lnu} depicts the evolution of the various angle-integrated neutrino luminosities (left) and mean neutrino energies (right) for all four models explored in this paper. The first thing to note is that there is little difference between the models with the various initial spin rates, despite the fact there is a two order of magnitude difference between the slowest rotator and the fastest. The 9-rot0.0, 9-rot-0.01, and 9-rot-0.1 models are quite close.  It is only the 9-rot-1.0 model that deviates from the pack, but in luminosity by perhaps only $\sim$10\% and in mean energy by no more than $\sim$0.5 MeV.  Moreover, the difference between the 9-rot-1.0 model and the others is most obvious only during the first $\sim$0.3 seconds for the luminosity and $\sim$1.0 second for the mean energies. This is due to the greater centrifugal support experienced by the 9-rot-1.0 model in the early post-bounce phases. 

\begin{figure*}[htbp!]
    \centering
    \includegraphics[width=0.95\textwidth]{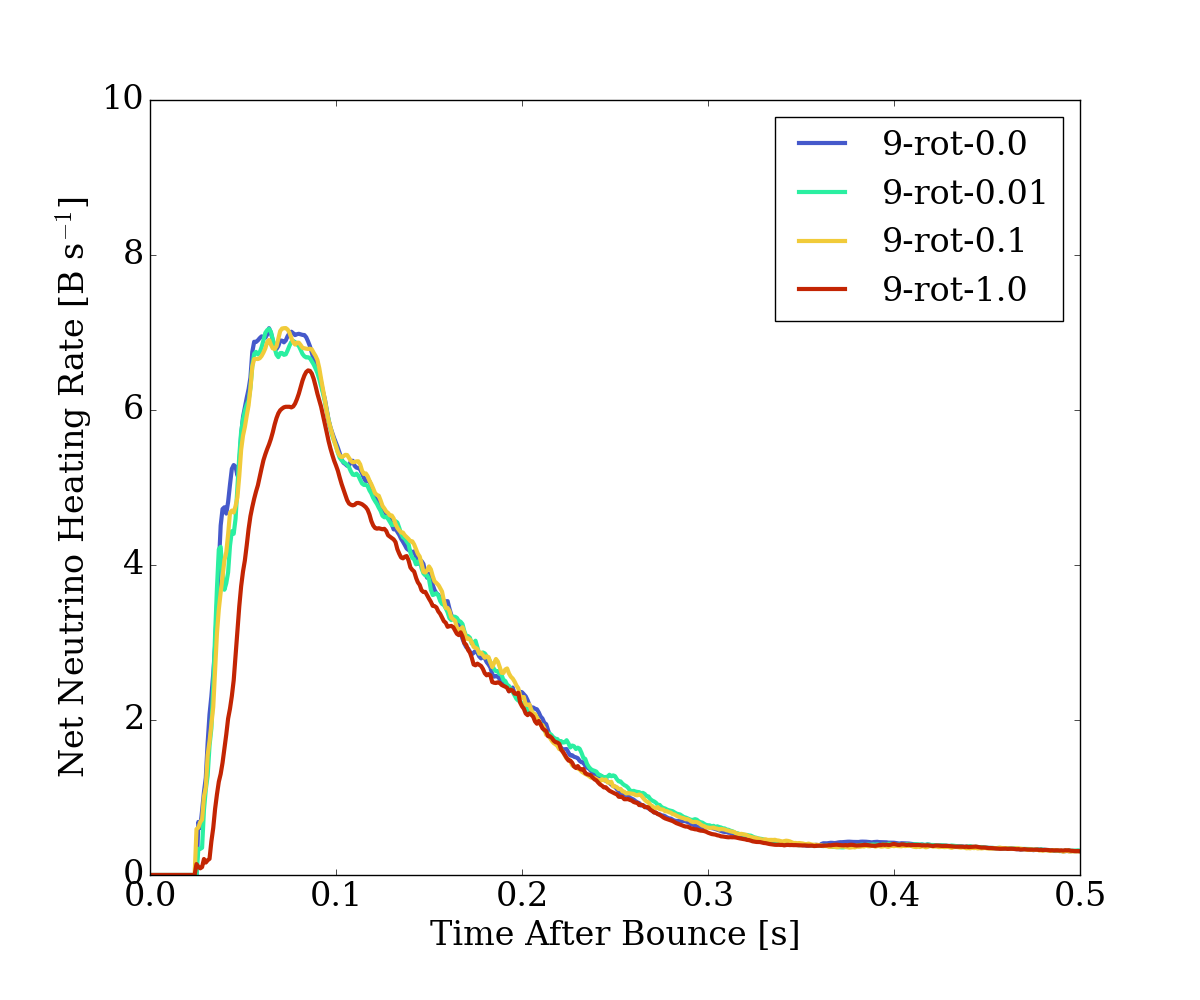}
    \caption{Net heating rate in the gain region in Bethes per second as a function of time (in seconds) for all new models of this paper. Centrifugal support for model 9-rot-1.0 lowers its mean neutrino energies and luminosities in the gain region behind the shock during the first $\sim$200 milliseconds after bounce to a significant degree. See text for a discussion.}
    \label{fig:Qdot}      
\end{figure*}

\begin{figure*}[htbp!]
    \centering
    \includegraphics[width=0.95\textwidth]{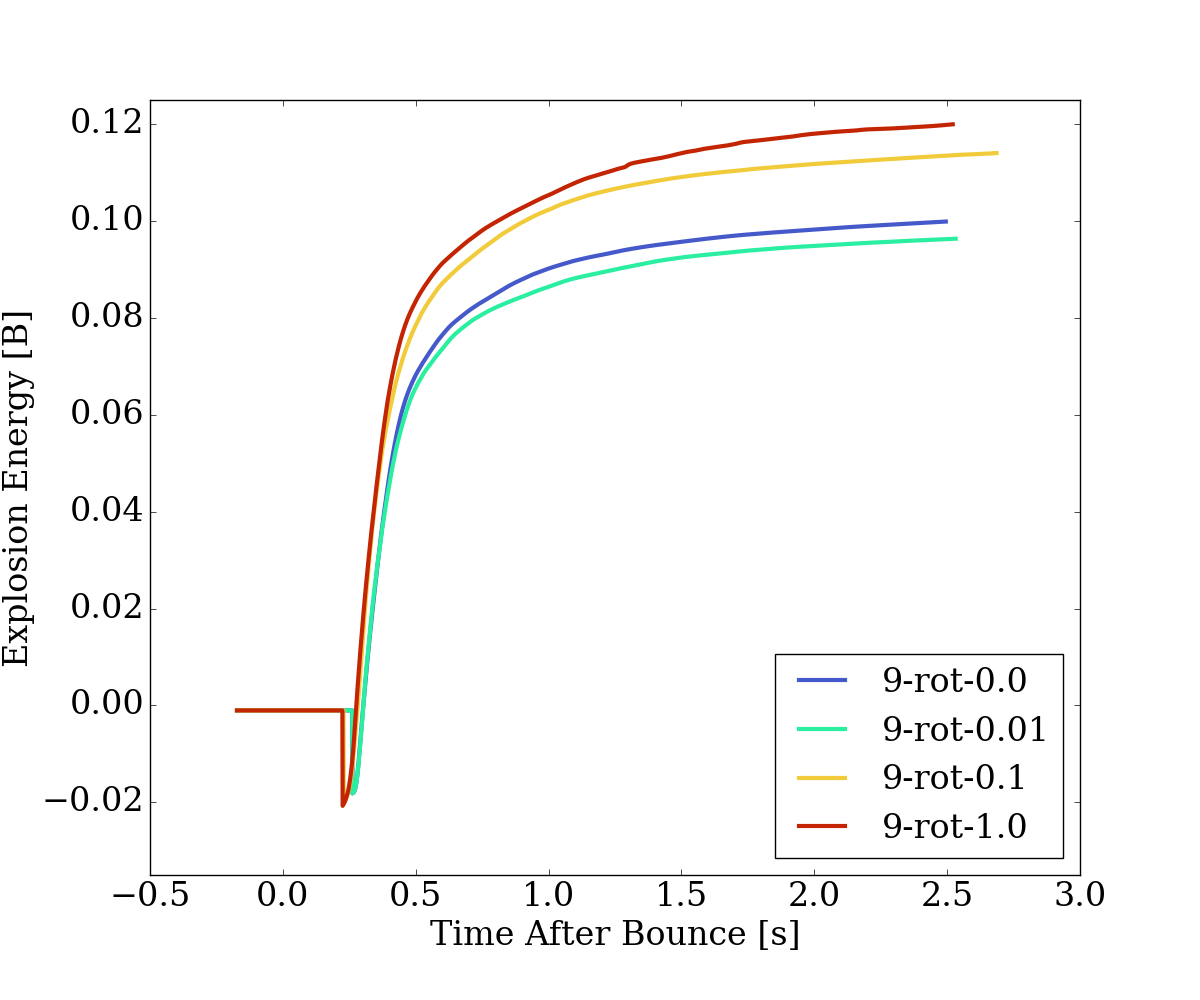}
    \caption{Explosion energy (in Bethes, B) versus time after bounce (in seconds) for all four models of this study. The blue curve depicts the baseline behavior of the initially non-rotating model. The behavior is not monotonic, with the 9-rot-0.1 and 9-rot-1.0 models separating slightly from the others. The magnitude of the increase in energy of the former vis \`a vis the latter is as much as $\sim$20\%. Note that the overall effect is not large, and that it requires a modest to rapid rate of initial rotation to show a measurable effect. Note also that for a small degree of initial rotation the explosion energy decreases slightly relative to the non-rotating baseline. The explosion energy is a sum of the kinetic, internal, gravitational, recombination, and overburden energies for zones with outward velocities exterior to 1000 kilometers.}
    \label{fig:explene}      
\end{figure*}

The differences described above translate into differences in the integrated neutrino heating rates behind the shock in the gain region. Figure \ref{fig:Qdot} portrays these differences as a function of time after bounce for our four models. As seen, it is again only the 9-rot-1.0 model which departs significantly from the others, but only during the first $\sim$0.2 seconds. However, this is the key launch phase of the explosion. The upshot is that all else being equal we would expect the explosion energy of the 9-rot-1.0 model to be lower.  However, as Figure \ref{fig:explene} clearly indicates the explosion energy of the 9-rot-1.0 model is actually higher than for the others. This is a consequence of the fact that the more rapidly rotating models experience more centrifugal support in the regions exterior to the PNS core, with the result that the effective potential against which the neutrino heating is working is diminished. In addition, more rapidly rotating models start with a bit more kinetic energy in and around the gain region and this contributes to the final tally. Hence, the diminished heating rates due to the rotation of the regions around the inner neutrinospheres counteract the decrease in the effective potential energy well depth and the associated initial transverse kinetic energy in the mantle behind and around the shock. As Figure \ref{fig:explene} indicates, as the rotation rate increases slightly the diminished heating rate ``wins," and the lower explosion energy curve for the 9-rot-0.01 model (teal) versus the initially non-rotating model 9-rot-0.0 (blue) so indicates. However, the two more rapidly rotating models (9-rot-0.1 and 9-rot-1.0) explode more energetically than the initially non-rotating model 9-rot-0.0, by $\sim$10\% to $\sim$20\%, respectively, indicating that sufficient rotation is a boon to explosion energy.

\begin{figure*}[htbp!]
    \centering
    \includegraphics[width=0.45\textwidth]{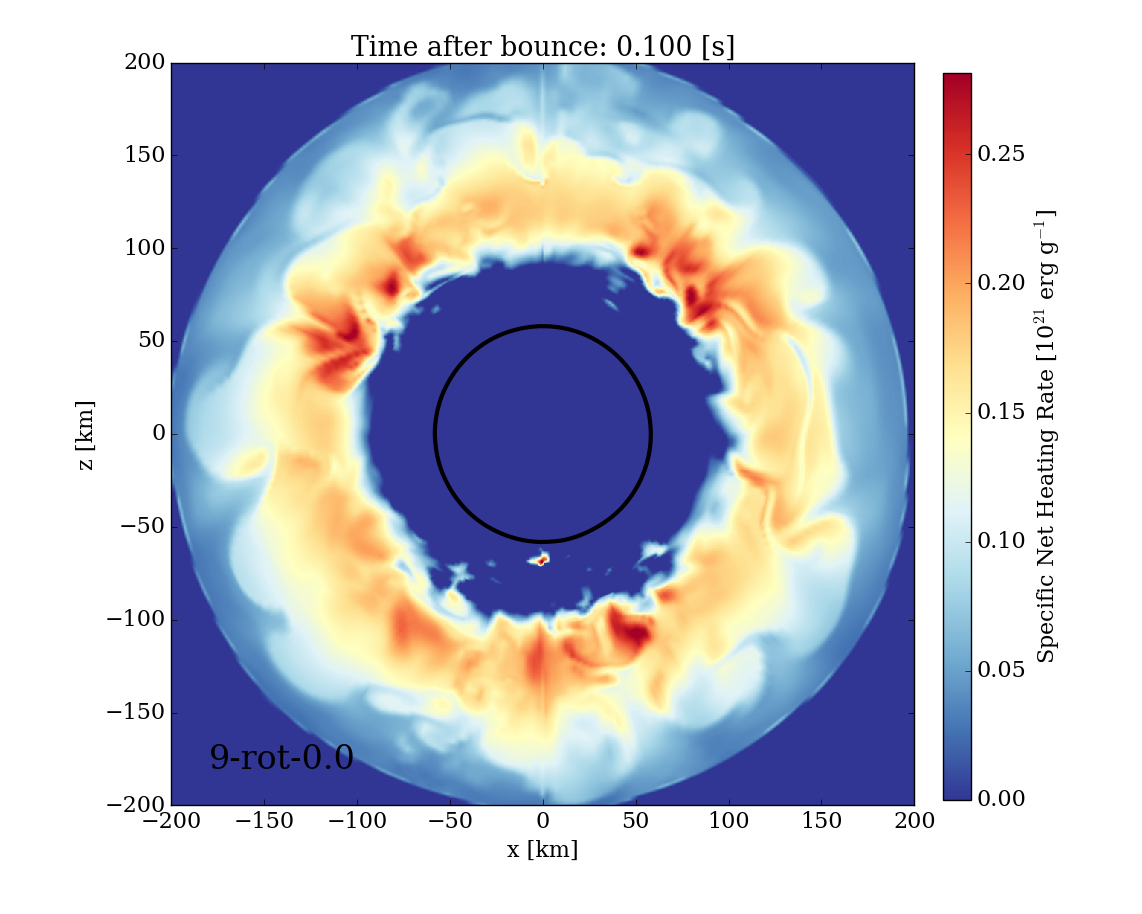}
    \includegraphics[width=0.45\textwidth]{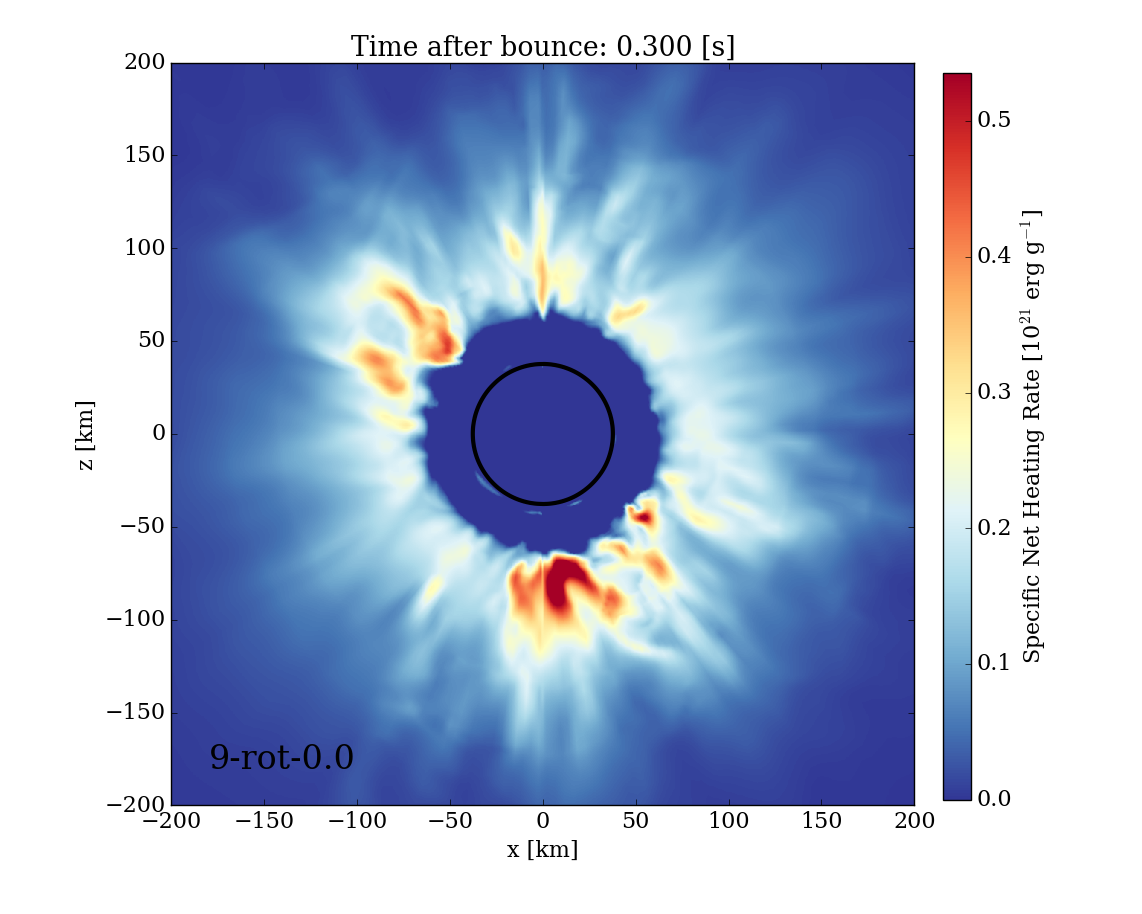}
    \includegraphics[width=0.45\textwidth]{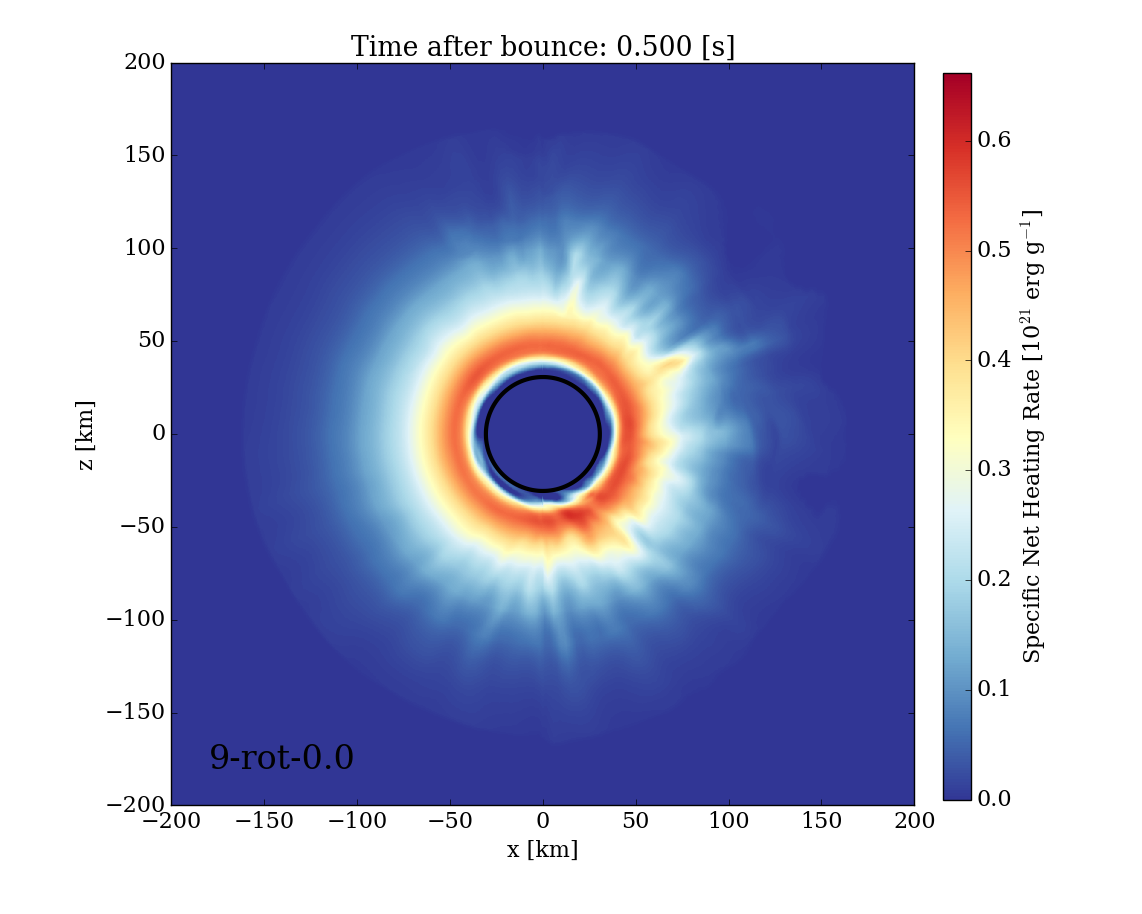}
    \includegraphics[width=0.45\textwidth]{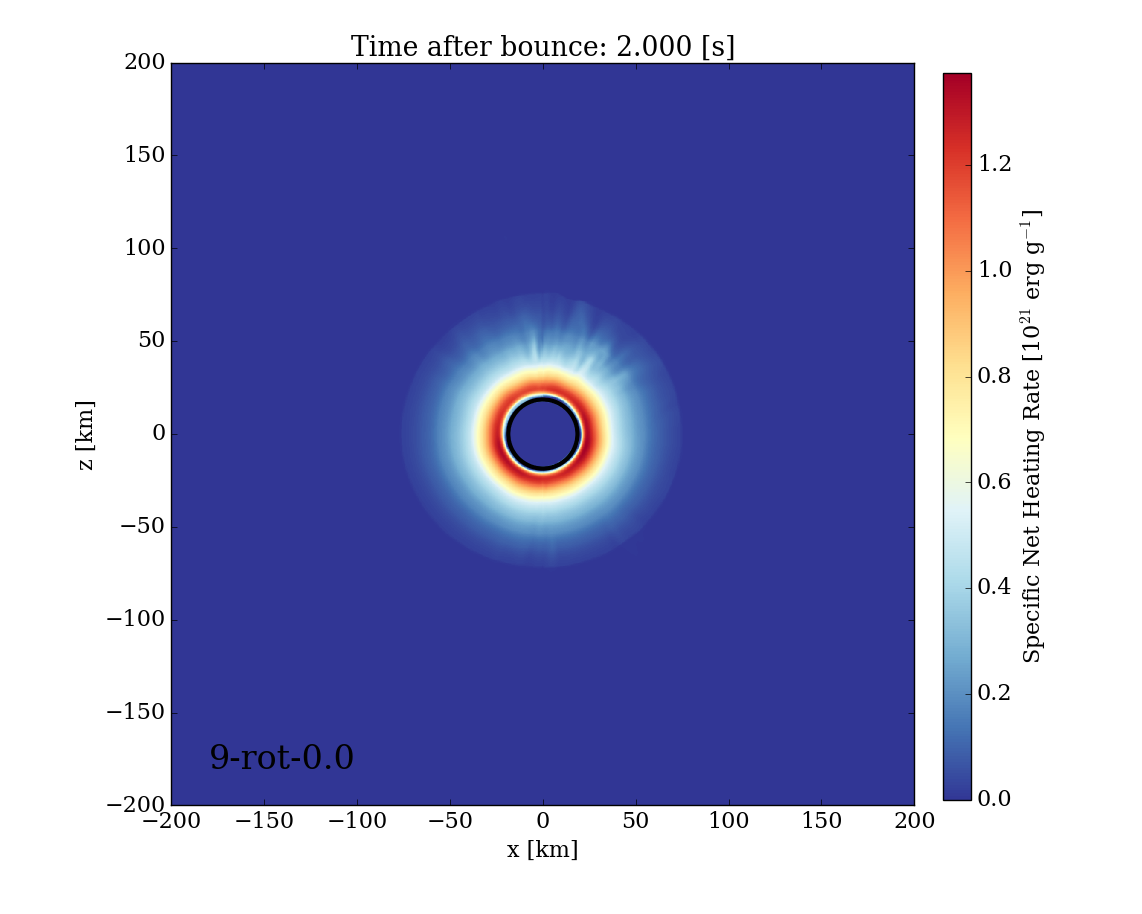}
    \caption{Specific net heating rate in the x-z plane at 0.1, 0.3, 0.5, and 2.0 seconds post-bounce of the 9-rot-0.0 model. Despite transient variations with angle of the energy deposition rate in the turbulent heating rate exterior to the inner PNS, there is no pole-equator dichotomy. }
    \label{fig:Qdot1}      
\end{figure*}

\begin{figure*}[htbp!]
    \centering
    \includegraphics[width=0.45\textwidth]{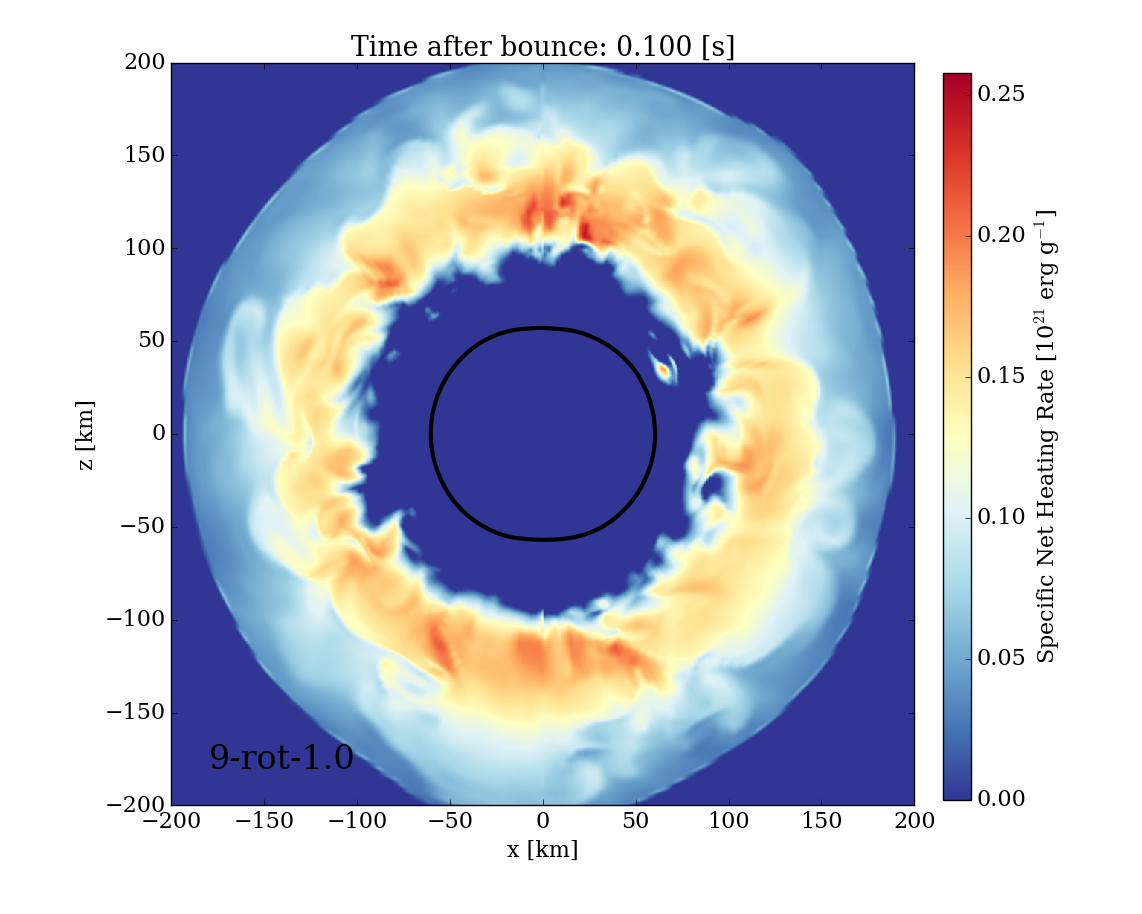}
    \includegraphics[width=0.45\textwidth]{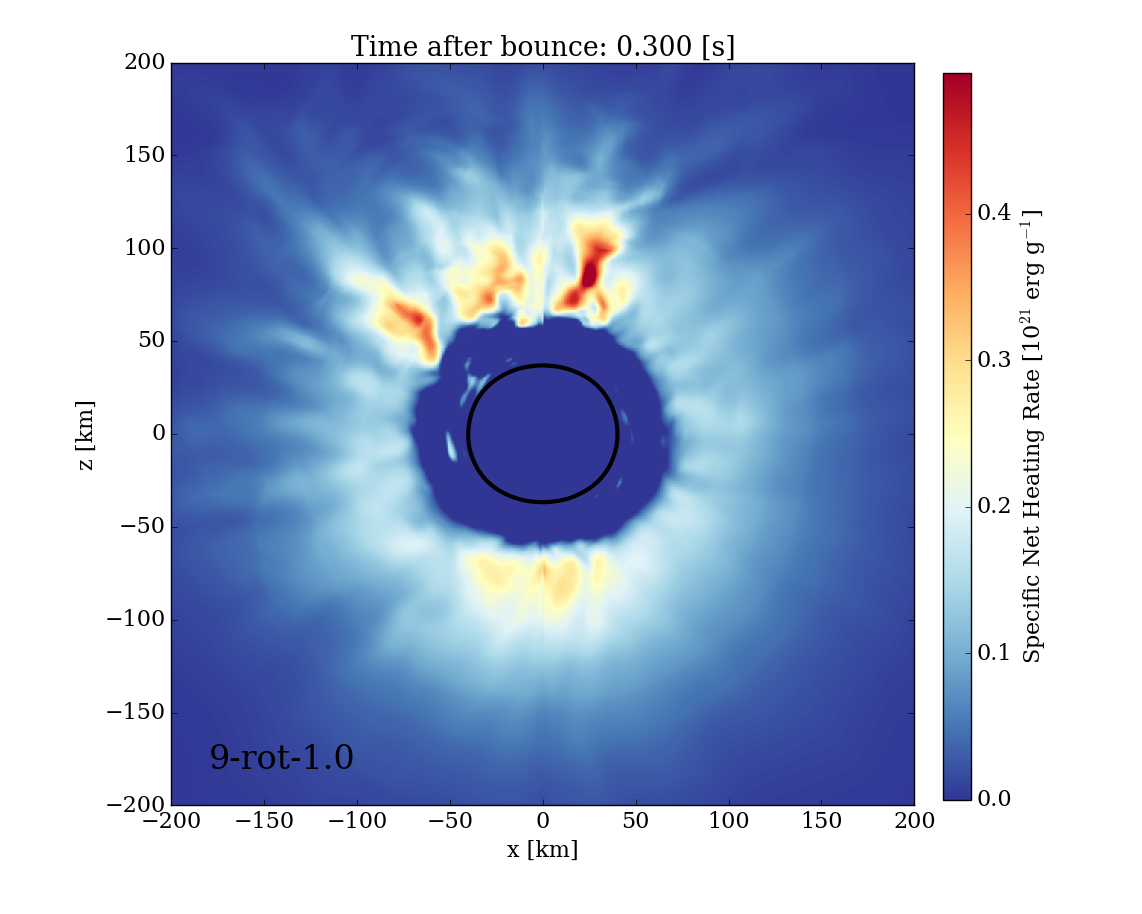}
    \includegraphics[width=0.45\textwidth]{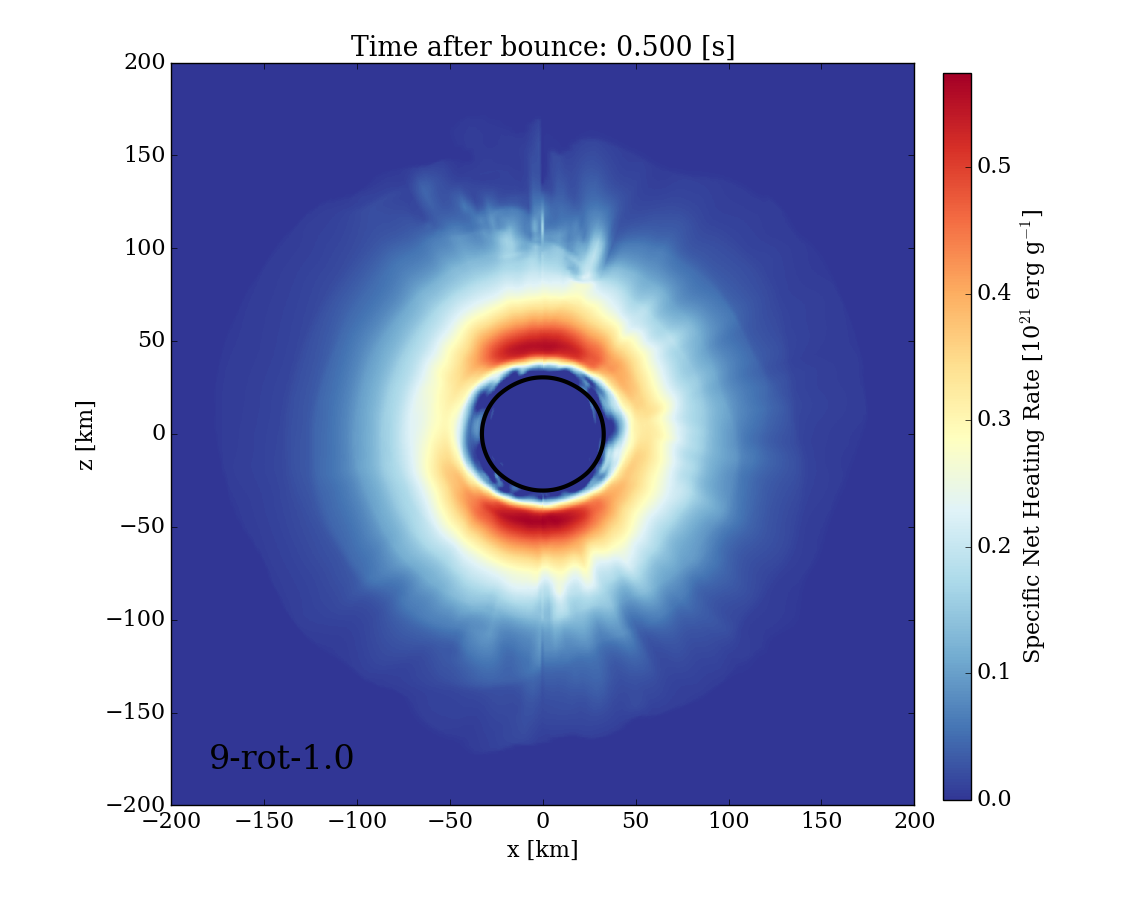}
    \includegraphics[width=0.45\textwidth]{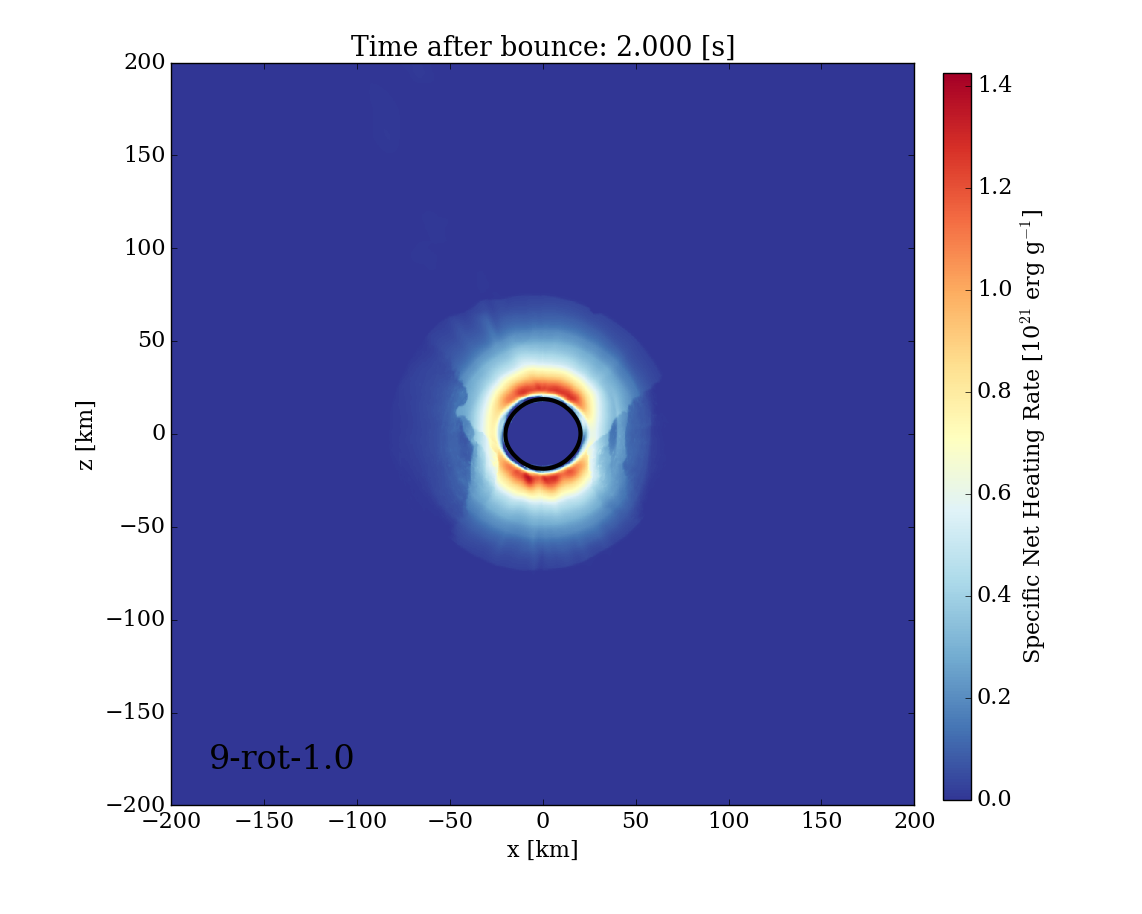}
    \caption{Specific net heating rate in the x-z plane at 0.1, 0.3, 0.5, and 2.0 seconds post-bounce of the 9-rot-1.0 model. Unlike for model 9-rot-0.0 depicted in Figure \ref{fig:Qdot1}, a persistent pole-equator variation in the neutrino energy deposition rate clearly emerges. However, and importantly, this pole-equator variation is rather mute at early times, when the core is still significantly extended by rotation, the explosion itself is launched, and most of the driving power is being deposited. See text for a discussion.}
    \label{fig:Qdot2}      
\end{figure*}

Figure \ref{fig:Qdot1} depicts the heating rate as a function of angle for the non-rotating model 9-rot-0.0 for various times after bounce.  As seen on these panels, there is a stochastic variation in the angle-dependence of the neutrino heating rate, but at late times it becomes approximately spherical, if weak. This is as expected. However, as the panels in Figure \ref{fig:Qdot2} indicate, at late times the pole-equator deviations in the specific neutrino heating rates for the 9-rot-1.0 model are significant. This is a manifestation of the Von Zeipel effect in the ``effective temperature" and energy flux as a function of angle expected and seen in rapidly rotating stars \citep{Maeder1999,Ott2008}. However, as the top left panel of Figure \ref{fig:Qdot2} indicates, at early times even the 9-rot-1.0 model has not yet spun up sufficiently due to neutrino losses during its Kelvin-Helmholtz contraction to result a significant oblateness in the core. Importantly, it is during this earlier phase that this 9.0-$M_{\odot}$ progenitor explodes. Hence, the effect of very rapid rotation on heating anisotropy for models that explode early is muted. One result is that the angular asymmetry of the explosion when the explosion is early is rather small.  This is seen on the right panel of Figure \ref{fig:rshock}.  The shock dipoles of the models at late times are not large, even for rapidly-rotating model 9-rot-1.0 ($\sim$8\%). However, they are monotonic with initial rotation rate. What the trends might be for more massive progenitors is yet to be determined.

\section{Recoil Kicks}
\label{kicks}

\begin{figure*}[htbp!]
    \centering
    \includegraphics[width=0.45\textwidth]{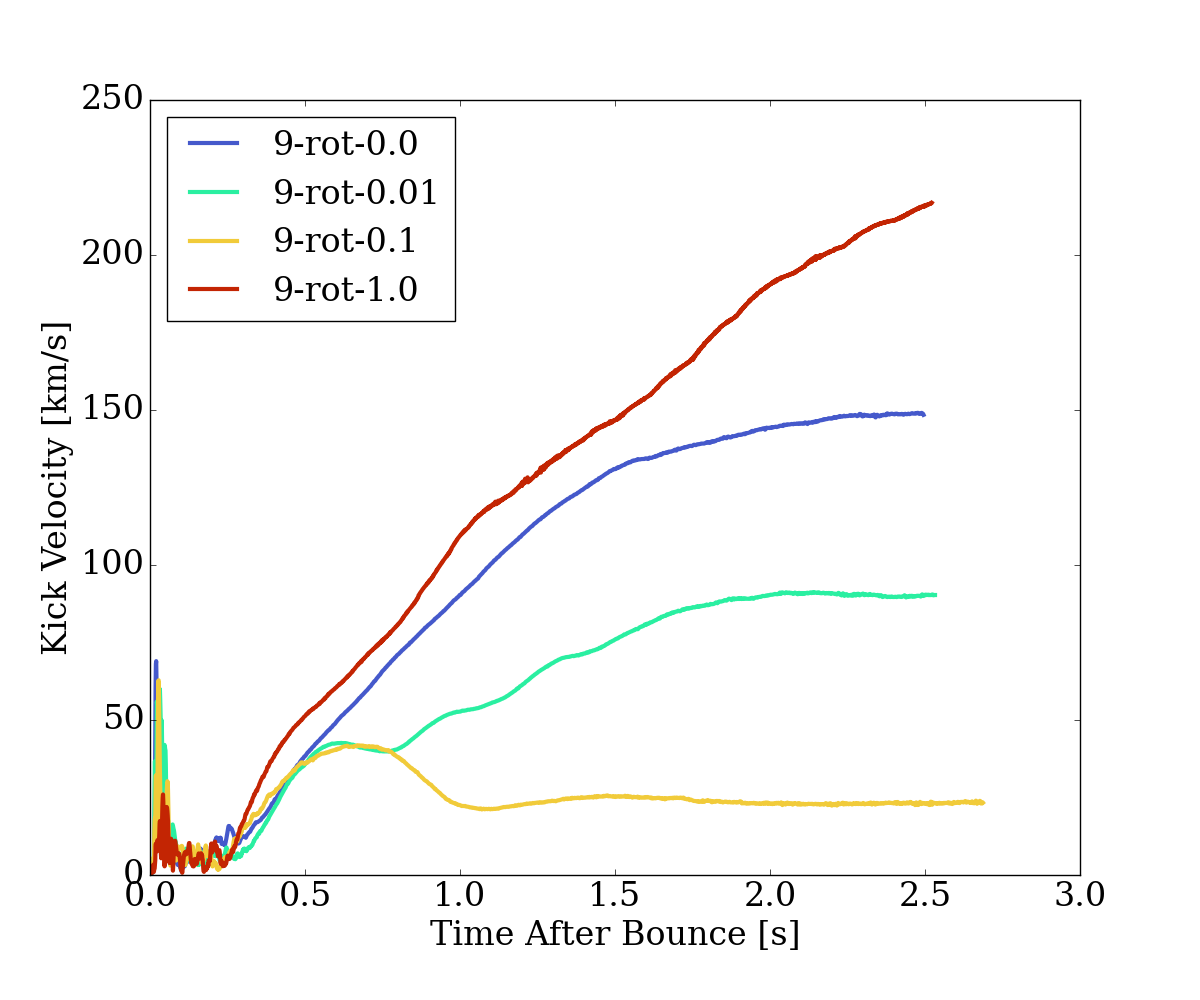}
    \includegraphics[width=0.45\textwidth]{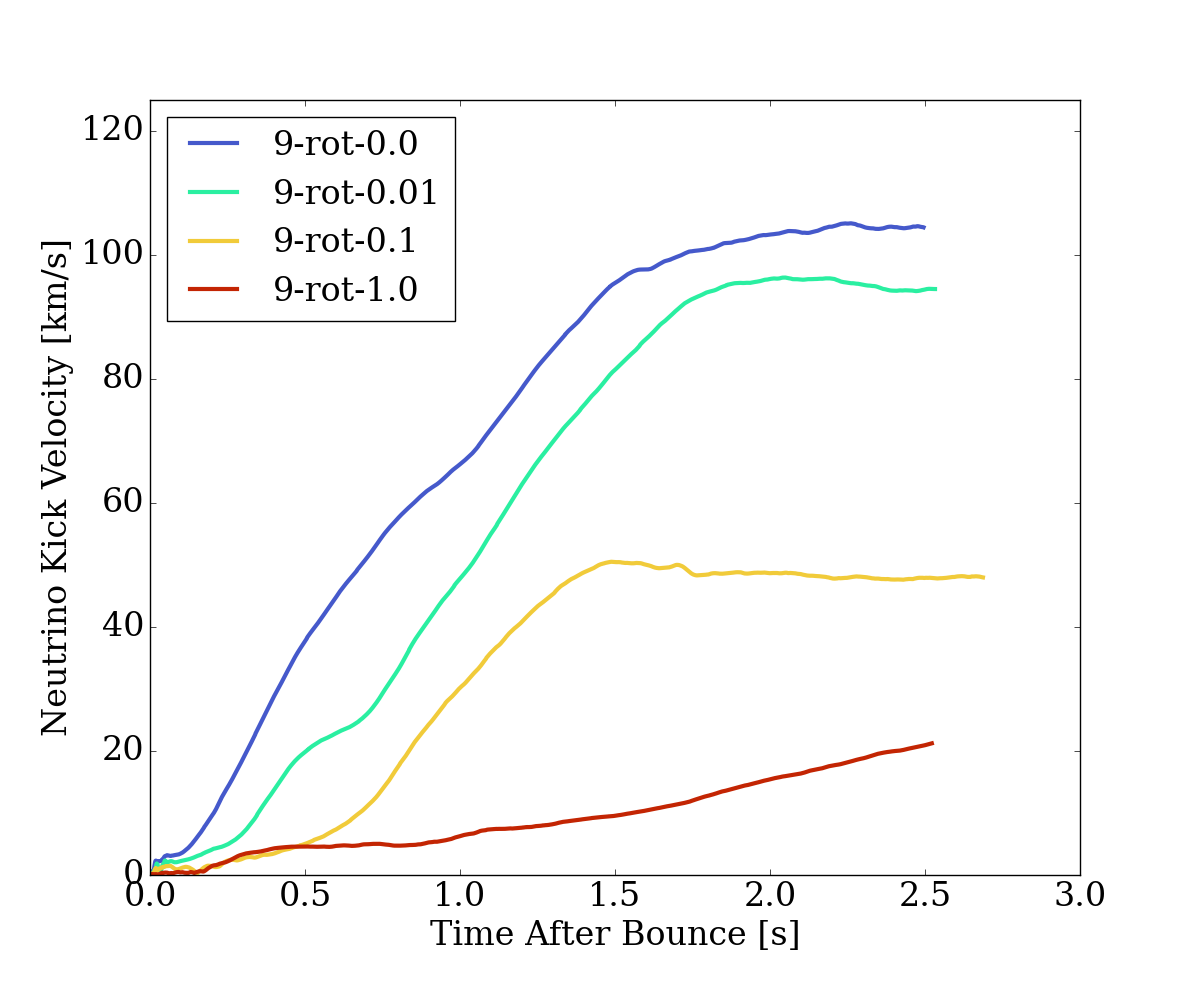}
    \caption{Total (left) and neutrino (right) recoil kicks in km s$^{-1}$ versus time after bounce (in seconds). We see a non-monotonic dependence upon initial spin rate, with the intermediate spin rates manifesting the lowest overall kicks. However, we are not confident this is not a result of the real physical chaos in the flow and would counsel caution in deriving any permanent conclusions from the trends we see. Nevertheless, it is interesting that the most rapidly rotating model shows the weakest neutrino-driven component, yet the largest total kick and that the neutrino kick magnitudes monotonically decrease with spin rate. Note that the total kick for the 9-rot-1.0 model is still increasing. Importantly, we find that the initial spin and the kick are co-linear only for the most rapidly rotating model. See text for a discussion.}
    \label{fig:kick}      
\end{figure*}

Figure \ref{fig:kick} shows the evolution of the magnitude of the total recoil kick due to the vector sum of both the matter and the neutrino recoils (left panel) and the magnitude of the contribution due to neutrinos alone (right panel). In general the matter and the neutrino contributions are not in the same direction.  We see that the total kick is not a monotonic function of initial spin, but that the magnitude of the neutrino component itself does decrease monotonically with increasing initial spin. As the right panel of Figure \ref{fig:rshock} indicates, the magnitude of the shock dipole asymmetry increases slightly with spin. However, the fact that the trend in the neutrino recoil seen in Figure \ref{fig:kick} is opposite to this suggests that the neutrino emissions become more top/bottom symmetric with increasing spin rate. Interestingly, we find that the kick and the spin vectors align for the most rapidly rotating model (9-rot-1.0), for which the dot product of the angular momentum and kick unit vectors at the end of the simulation is pegged at $\sim$1.0 from early in its evolution. We do not see this behavior for the slowest models. For models 9-rot-0.0, 9-rot-0.01, and 9-rot-0.1 respectively, the dot product of the angular momentum and kick unit vectors at the end of the simulations is $\sim$0.8, 0.76, and $\sim$0.0, and it varies significantly with time during their evolution. Therefore, we see no obvious strong alignment trend for the slower spin rates. This topic may also have interesting implications for late-time correlations between kicks and ejecta distributions \citep{vartanyan2025_blast,2025arXiv250916314V}. We note that, given the coarseness of our grid in initial spin, we have not yet determined where the transition from ``random" to aligned may occur. In summary, we see an intriguing context for potential spin/kick alignment, but (see also \citet{burrows_kick_2024}) we have yet to explain the spin-kick correlation seen for some pulsars \citep{Ng2007,noutsos2013,yao2021,biryukov2025}, except if the initial generic spin at collapse is larger than generally thought.

\section{Protoneutron Star Convection}
\label{pns}

\begin{figure*}[htbp!]
    \centering
    \includegraphics[width=0.45\textwidth]{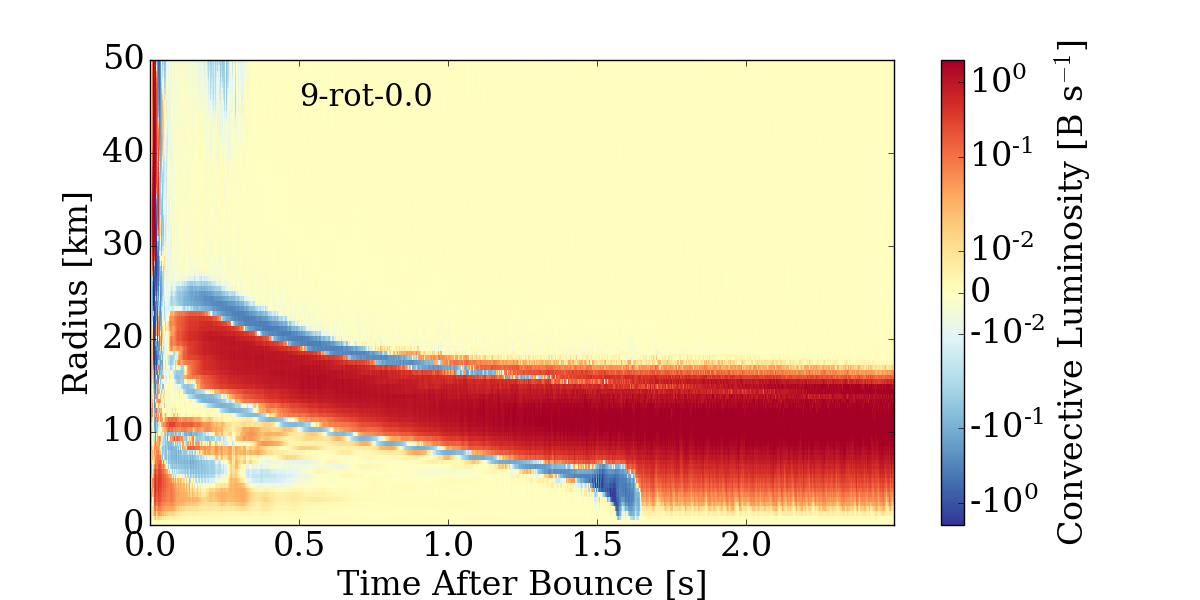}
    \includegraphics[width=0.45\textwidth]{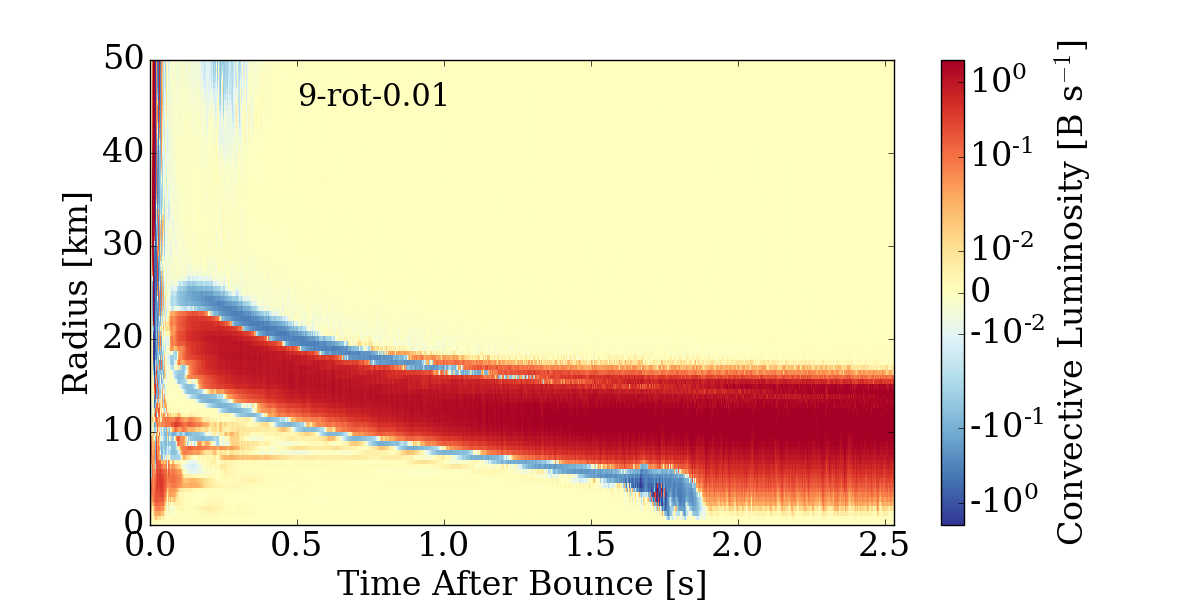}
    \includegraphics[width=0.45\textwidth]{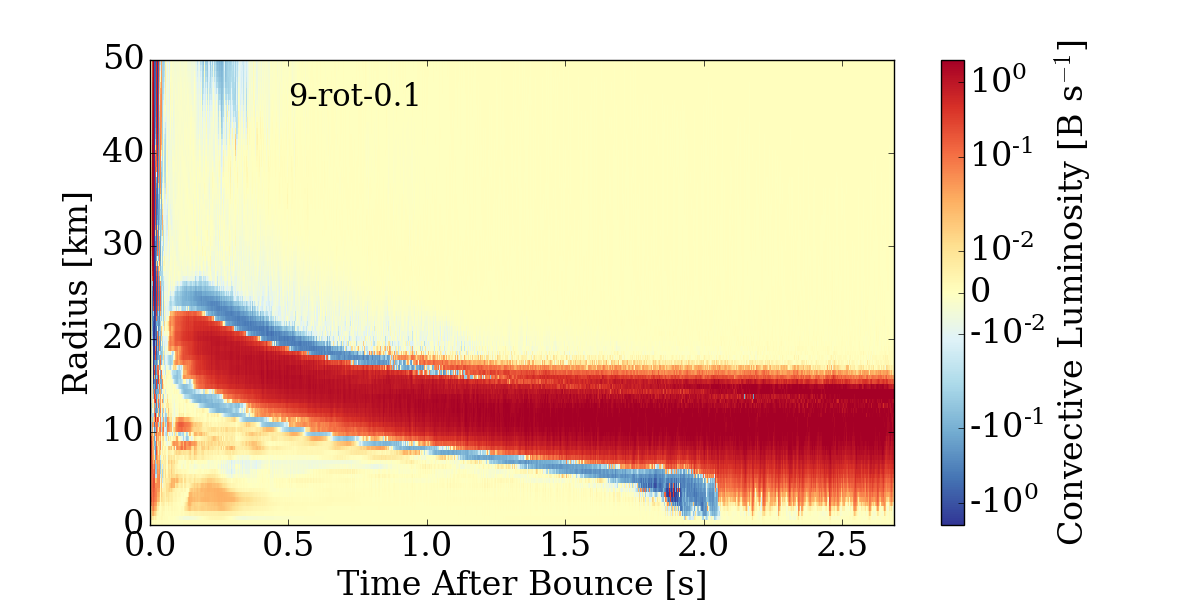}
    \includegraphics[width=0.45\textwidth]{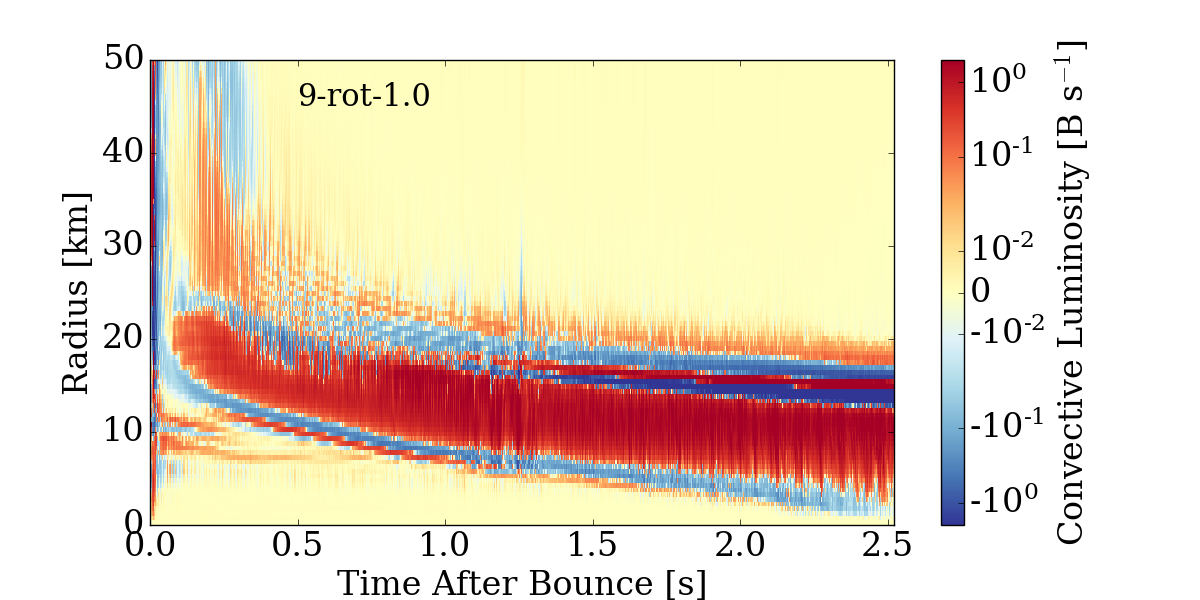}
    \caption{The angle-integrated convective luminosity profile in the protoneutron star versus time after bounce in seconds. The red regions depict where lepton-driven convection in the core is found. It begins in a shell, whose inner boundary slowly moves inward until it reaches the center. For all but the fastest rotator, this occurs near $\sim$1.7 seconds of bounce. However, for the most rapidly rotating model, it takes an extra second to achieve the center. In addition, as the bottom right panel shows,  the upper and lower boundaries of the convective region are smeared on this angle-averaged diagram. This is in part a consequence of the angular momentum's effect on the convective condition.  See Figure \ref{fig:turb_Ma} for another angle on this phenomenon and the text for a discussion.}
    \label{fig:coleman}      
\end{figure*}

Figure \ref{fig:coleman} depicts the evolution of the angle-averaged convective luminosity profile in the protoneutron star for the four models of this paper. This figure recapitulates many similar figures we have published in the past \citep[e.g.,][]{nagakura_pns} and traces the regions in the PNS core experiencing lepton-driven convection. Such convection persists in all CCSN models and achieves the center, encompassing much of the nascent neutron star, in a time that is a monotonic function of the residual neutron star mass. For the non-rotating 9.0-$M_{\odot}$ models of \citet{swbj16} that we have calculated in the past \citep{burrows_correlations_2024} this has occurred at roughly $\sim$1.7 seconds after bounce.  We see in Figure \ref{fig:coleman} the same behavior for all but the 9-rot-1.0 model.  For that model, as Figure \ref{fig:coleman} shows, achieving the center requires almost one more second. In addition, the convective power for this model is slightly diminished and the upper and lower boundaries of the convective region are smeared. Both these effects are due to the weaker effective gravity experienced in this core due to its rapid rotation and the effects along its cylindrical periphery of the Solberg-H\o{i}land stability condition. \tianshu{We note that many other simulations in the literature are not long enough to witness convection achieving the center, with the 1D simulations of \citet{roberts2012_thesis} an exception, though many works have shown shrinking inner and outer radii of the convection region (e.g., \citet{2006A&A...447.1049B}). }

It is curious though that such a rapid spin rate is required to make much of a difference in the character of PNS convection. This emphasizes yet again the secondary role rotation seems to play for this 9-$M_{\odot}$ model for most spin rates.  Whether this trend and behavior continues for more mass progenitors remains to be seen.

\begin{figure*}[htbp!]
    \centering
    \includegraphics[width=0.45\textwidth]{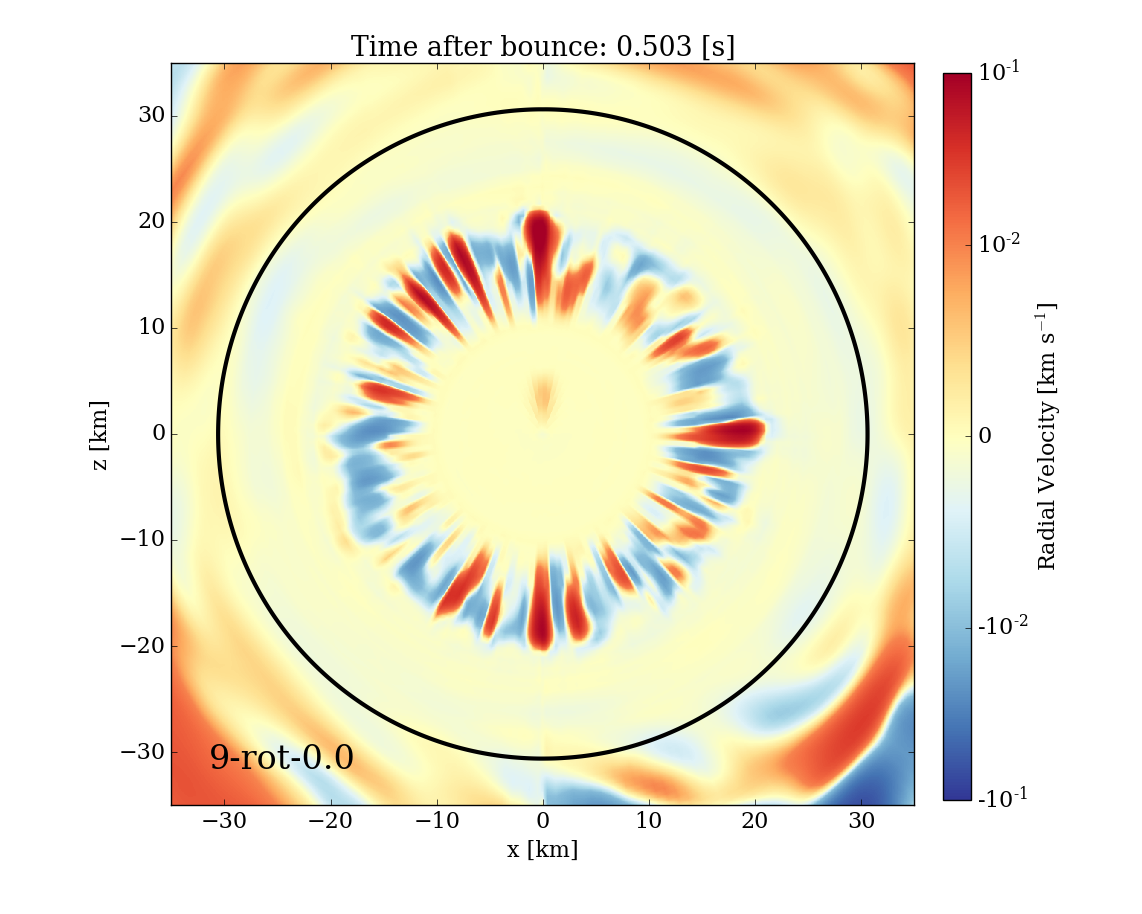}
    \includegraphics[width=0.45\textwidth]{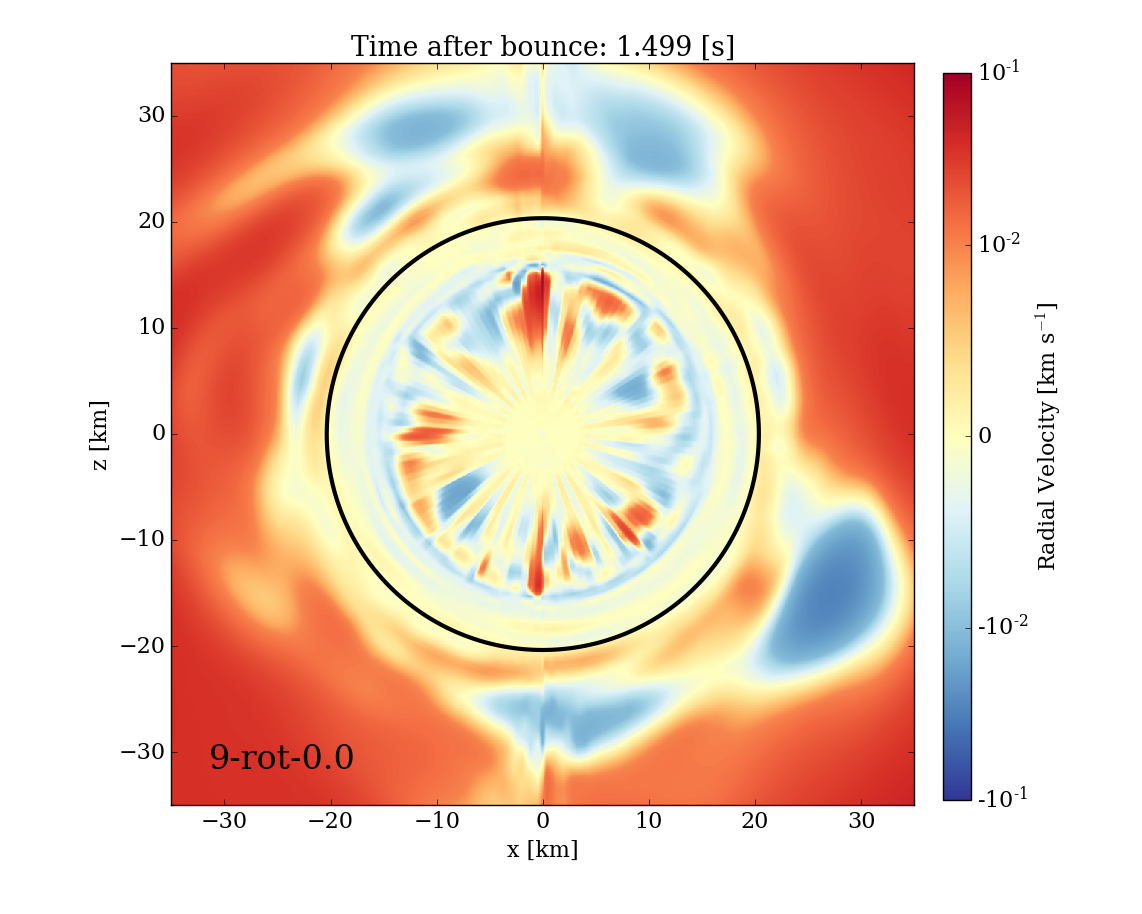}
    \includegraphics[width=0.45\textwidth]{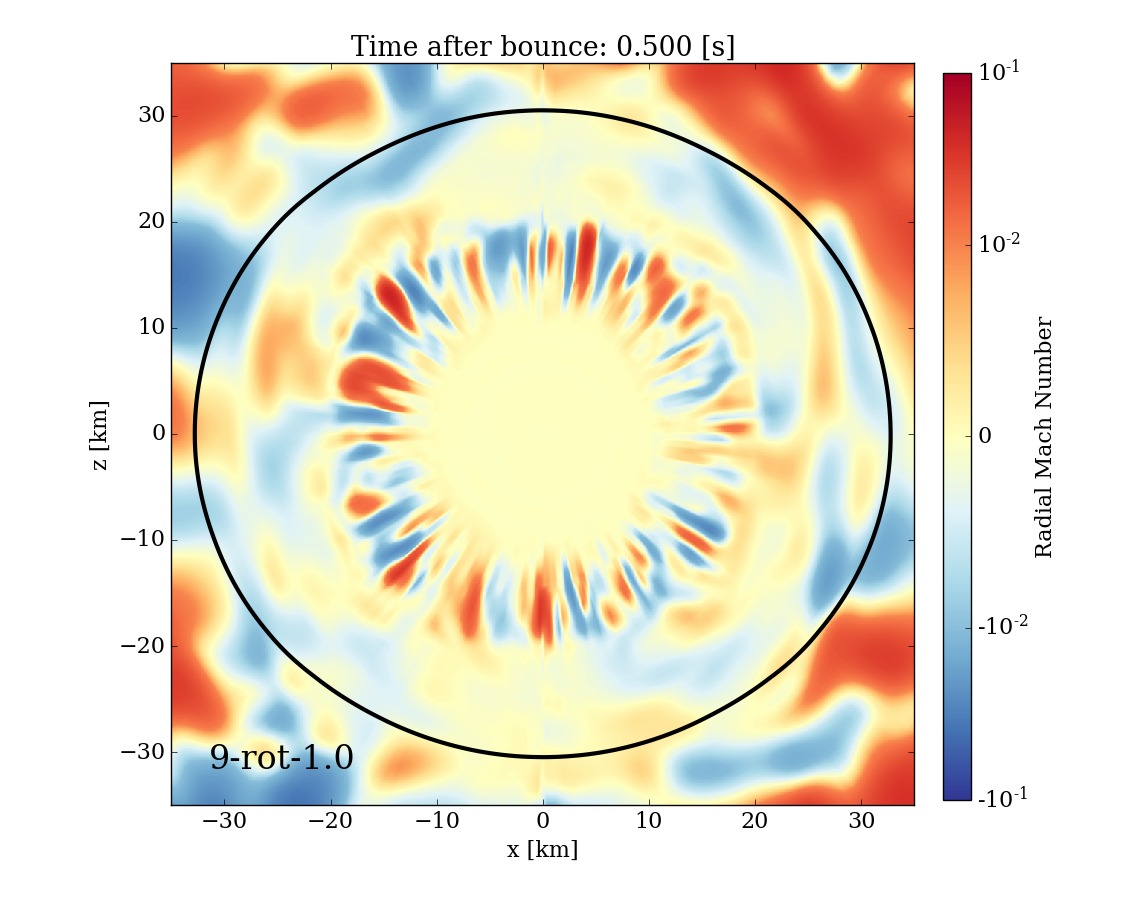}
    \includegraphics[width=0.45\textwidth]{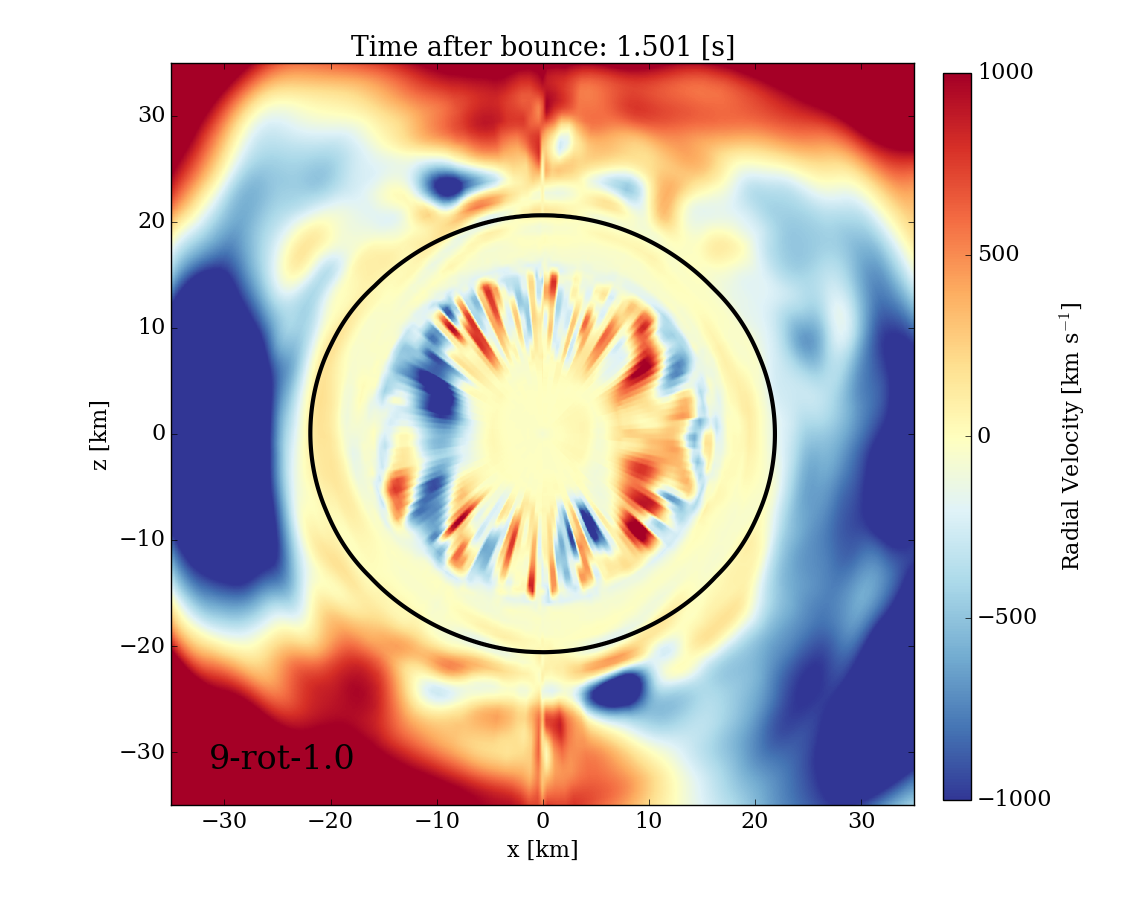}
    \caption{Radial Mach number in the x-z plane at $\sim$0.5 and $\sim$1.5 seconds post-bounce of the 9-rot-0.0 (top) and 9-rot-1.0 (bottom) models. From the radial dominance (even at late times) and roughly alternating sign of the mean convective velocity, the convective zone in the initially non-rotating model is seen to maintain its shellular structure. However, at later times for the fast rotating model there appears a slight red (positive)/blue (negative) variation from left to right and the emergence of roughly cylindrical convective structures. This is a clear and expected difference in the structure of PNS convection for rapidly rotating cores.}
    \label{fig:turb_Ma}      
\end{figure*}

Figure \ref{fig:turb_Ma} depicts the radial Mach number in this inner PNS region for the non-rotating and the most rapidly rotating models of this study at two different times after bounce. For the initially non-rotating model, the structures seem radial and the sign of the convective flow velocity alternates with angle $\theta$ in a standard way. Hence, convection in this model maintains its shellular structure. However, at later times the fast rotating model manifests a slight positive/negative variation from left to right and the emergence of roughly cylindrical convective structures. This is a clear and expected difference in the structure of PNS convection for rapidly rotating cores. However, such behavior is not much in evidence in the 9-rot-0.01 and 9-rot-0.1 models (not shown). Again, we find that very rapid spin rates are required for the hydrodynamics to differ much from the general behavior of the non-rotating progenitor.

\section{Nucleosynthesis}
\label{nucleo}
   
\begin{figure*}[htbp!]
    \centering
    \includegraphics[width=0.85\textwidth]{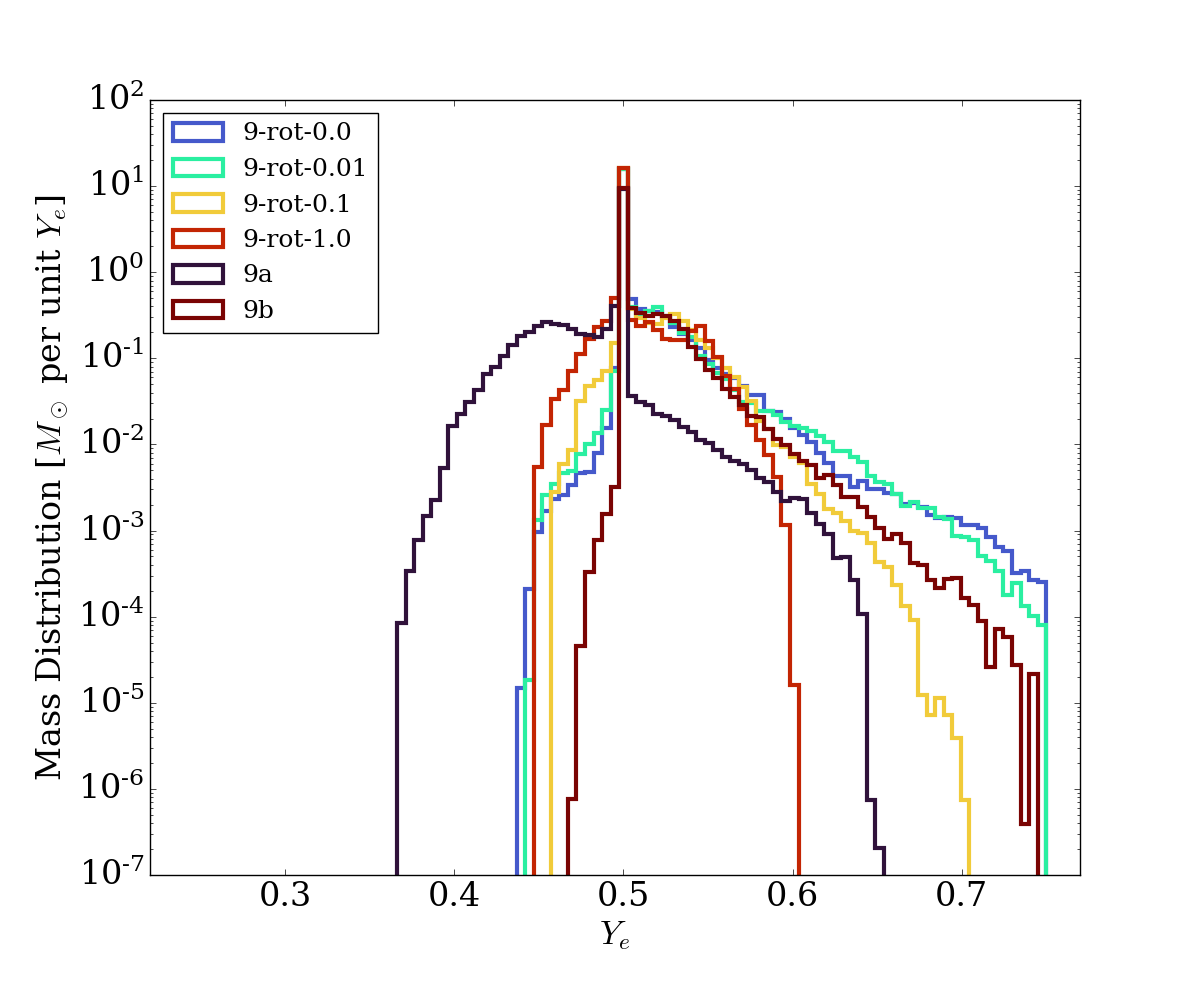}
    \caption{Final ejecta $Y_e$ distributions for all four 9.0-$M_{\odot}$ models simulated for this paper, along with those for the 9a and 9b models from \citet{burrows_correlations_2024}, \citet{wang_low_2024}, and \citet{wang_nucleo_2024}. The latter differ only in the fact that the 9a model accreted an envelope perturbed due to artificial turbulence and that the infall stage before bounce was done in 1D, mapping to 3D 10 milliseconds after bounce. All four of the models run for this paper were collapsed in 3D. From this figure we see that the neutron-richness of the ejecta increases with initial rotation rate, but that initial perturbations result in the ejection of even more neutron-rich matter. Note that the y-axis is logarithmic, so that most of the ejecta is still near 0.5. See text and the caption for Figure \ref{fig:ye-S} for a discussion.}
    \label{fig:ye}      
\end{figure*}

Nucleosynthesis in the ejecta is a key product and observable in core-collapse explosions and we seek discriminants of rotation in this sector.
Figure \ref{fig:ye} shows the $Y_e$ distribution of the ejecta for all models. Two previously published models \citep{burrows_correlations_2024,wang_nucleo_2024,Rusakov2026}, 9a and 9b, are included for comparison. Unlike the new models simulated in this work, the collapse and bounce phases of 9a and 9b were done in spherical-symmetry (1D) until 10 ms post-bounce. Therefore, the development of convection and turbulence is a bit delayed in those models compared to the new models whose collapse and bounce are done in full 3D. At 10 ms post-bounce, a 100 km/s $n=4$ and $l=m=10$ radial velocity perturbation between 200 and 1000 kilometers was added to the 9a model \citep{muller_janka_pert}. This perturbative velocity field leads to a slightly more rapid explosion and slightly higher explosion energy than the non-perturbed model 9b. As a result, the ejecta are more neutron-rich, since the effective duration over which neutrino-matter interactions can increase the ejecta $Y_e$ is  decreased. A similar effect is seen for the rotating models: for progressively increasing initial spins, the ejecta are more and more neutron-rich. 

\begin{figure*}[htbp!]
    \centering
    \includegraphics[width=0.45\textwidth]{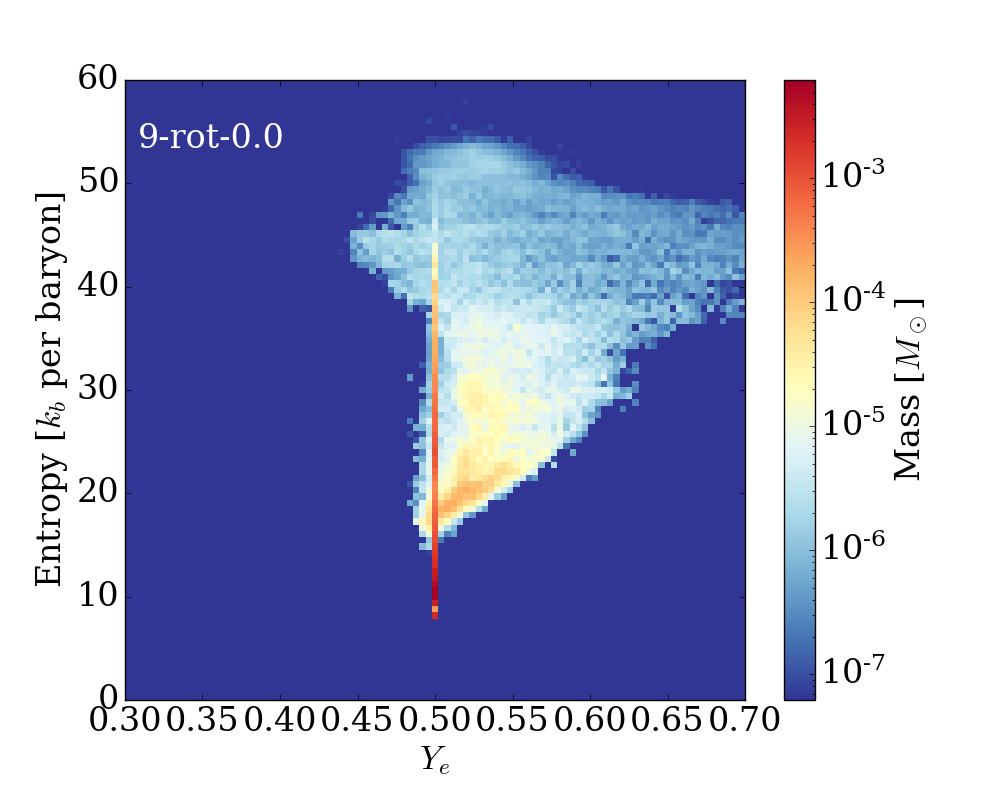}
    \includegraphics[width=0.45\textwidth]{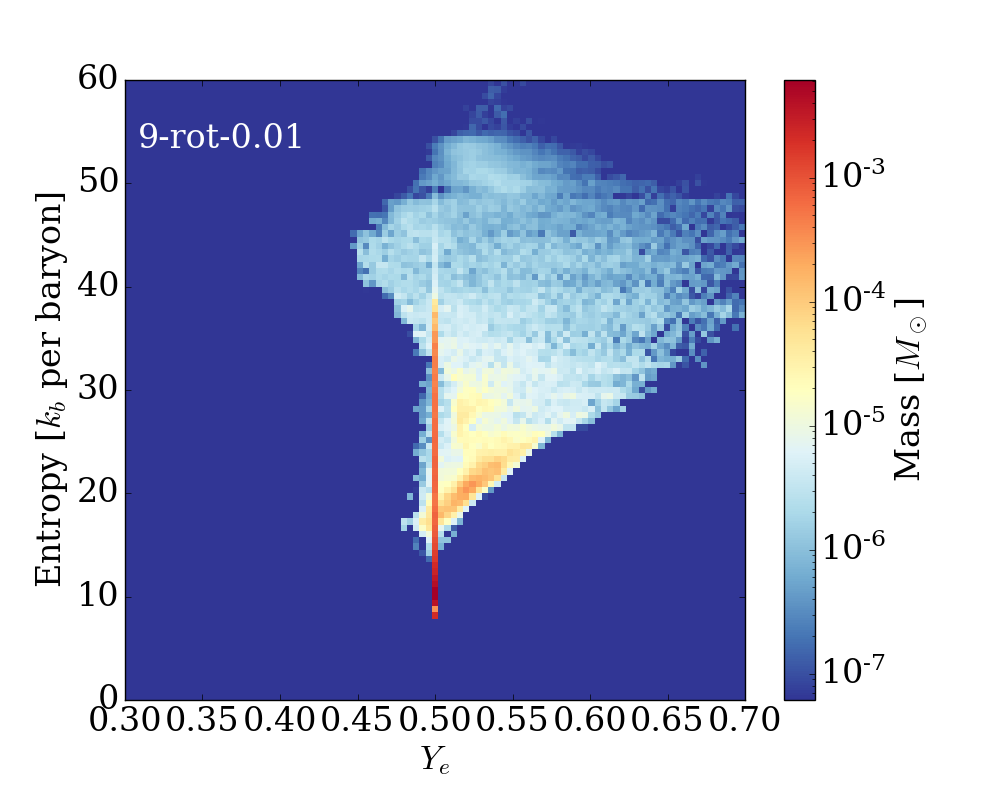}
    \includegraphics[width=0.45\textwidth]{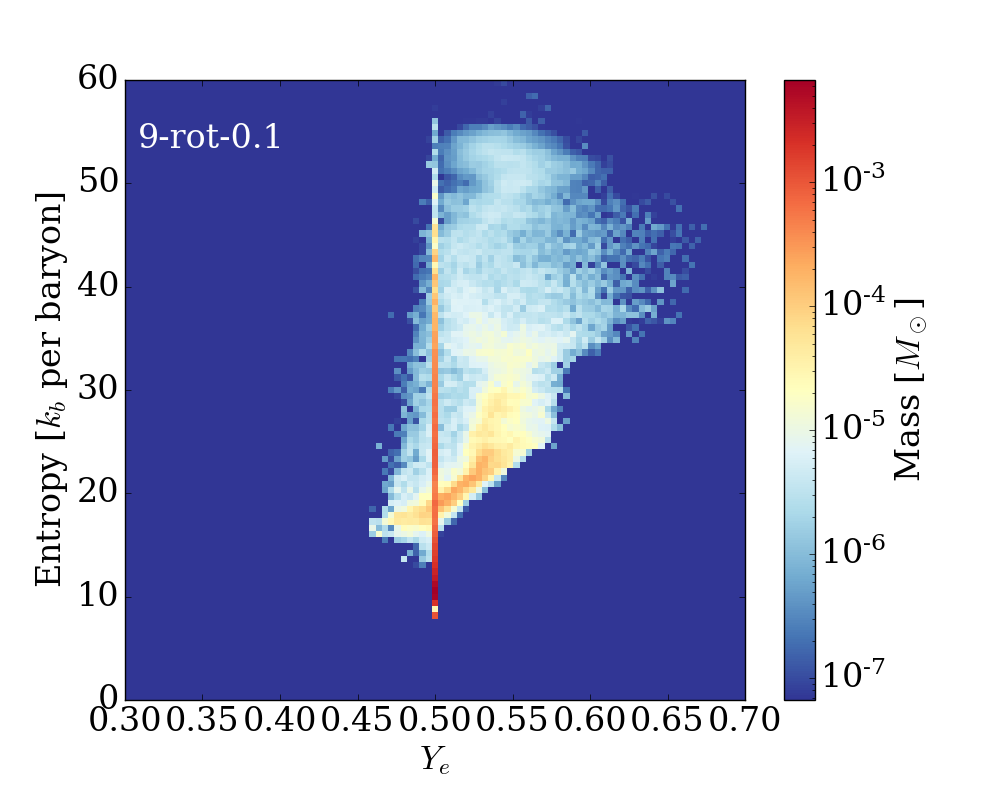}    
    \includegraphics[width=0.45\textwidth]{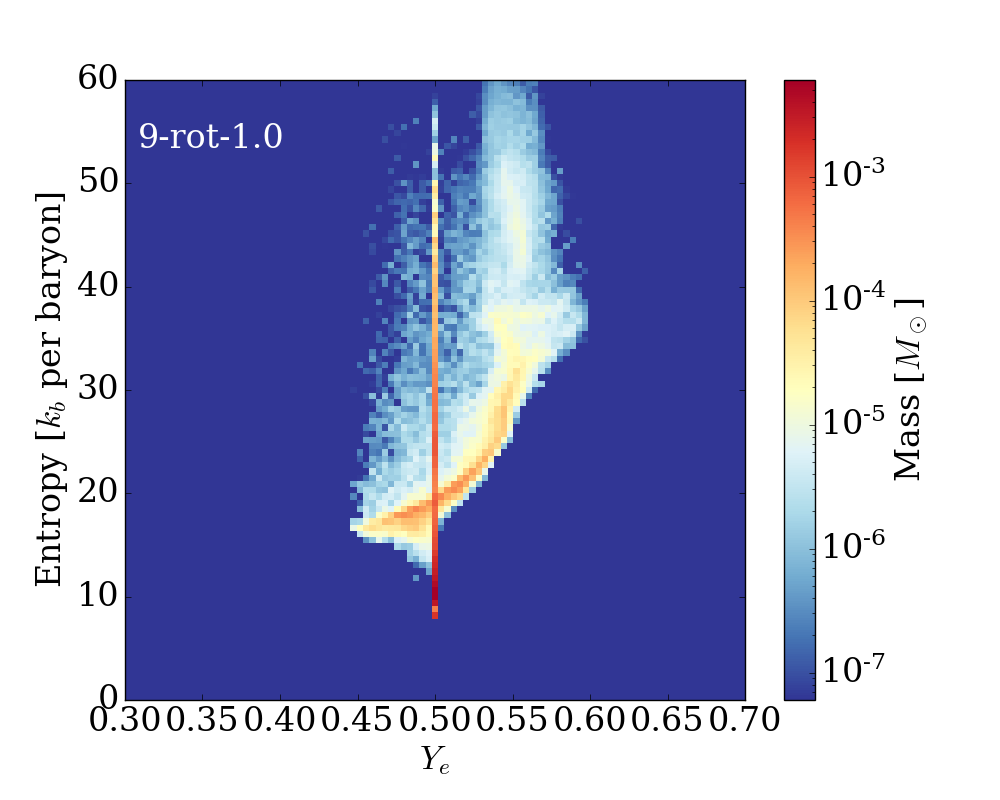}
    \includegraphics[width=0.45\textwidth]{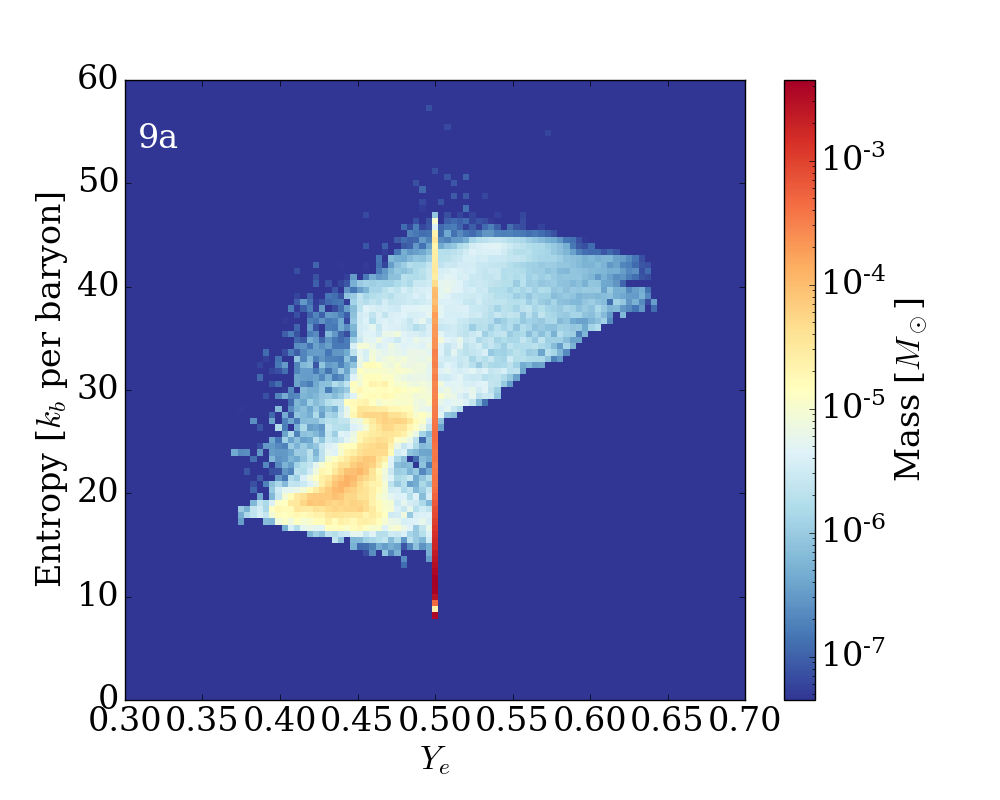}
    \includegraphics[width=0.45\textwidth]{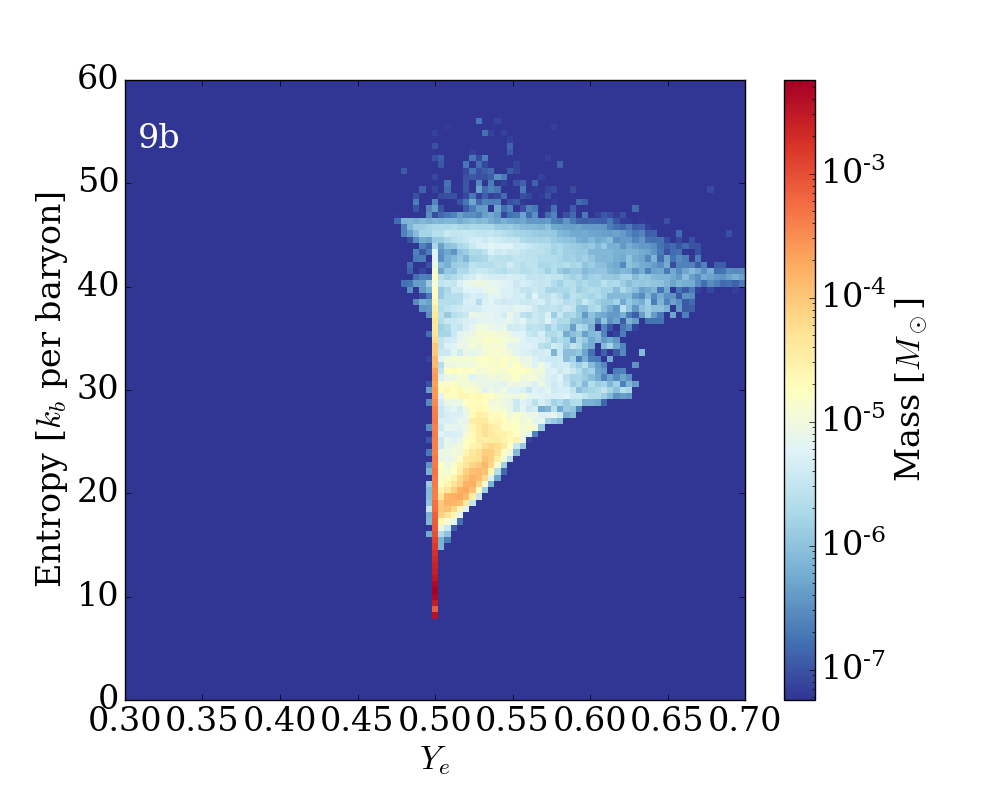}
    \caption{Ejecta $Y_e$ and entropy distribution for all models. The two 9 $M_\odot$ models published before, 9a and 9b (with/without initial perturbation), are plotted for comparison. With increasing initial spin rate, the ejecta become more and more neutron-rich, which is consistent with the trend in explosion time $-$ more rapid explosions lead to more neutron-rich ejecta. The behaviors of all the new models of this work lie between those of models 9a and 9b. Instead of collapsing in 3D  as we did in the new models, both 9a and 9b collapse and bounce in 1D until 10 ms post-bounce. A 100 km s$^{-1}$ $n=4$, $l=m=10$ perturbative velocity field between 200 and 1000 km is then added to the 9a model, while the 9b doesn't include any extra perturbation. Since 9b collapses and bounces in 1D, the development of convection is delayed compared to 9a and the new models. This makes it the last to explode, with the least neutron-rich ejecta. The initial perturbation added to 9a, on the other hand, is very efficient at aiding explosion, thus making it the fastest exploding and thereby the most neutron-rich. The impact of initial spin on ejecta $Y_e$ is less significant than that of initial perturbation, at least for this 9.0-$M_{\odot}$ model suite.}
    \label{fig:ye-S}      
\end{figure*}

Figure \ref{fig:ye-S} shows the joint distributions of the $Y_e$ and entropy of the ejecta. Since entropy scales very roughly as $T^3/\rho$, matter ejected later tends to have higher entropy. Therefore, such joint distribution plots provide a clear way to depict indirectly the evolution of $Y_e$ as a function of time. We can see that the new models and 9a start with different amounts of neutron-rich ejecta, and as time evolves, the $Y_e$ of the matter ejected later gradually increases. 

\begin{figure*}[htbp!]
    \centering
    \includegraphics[width=0.45\textwidth]{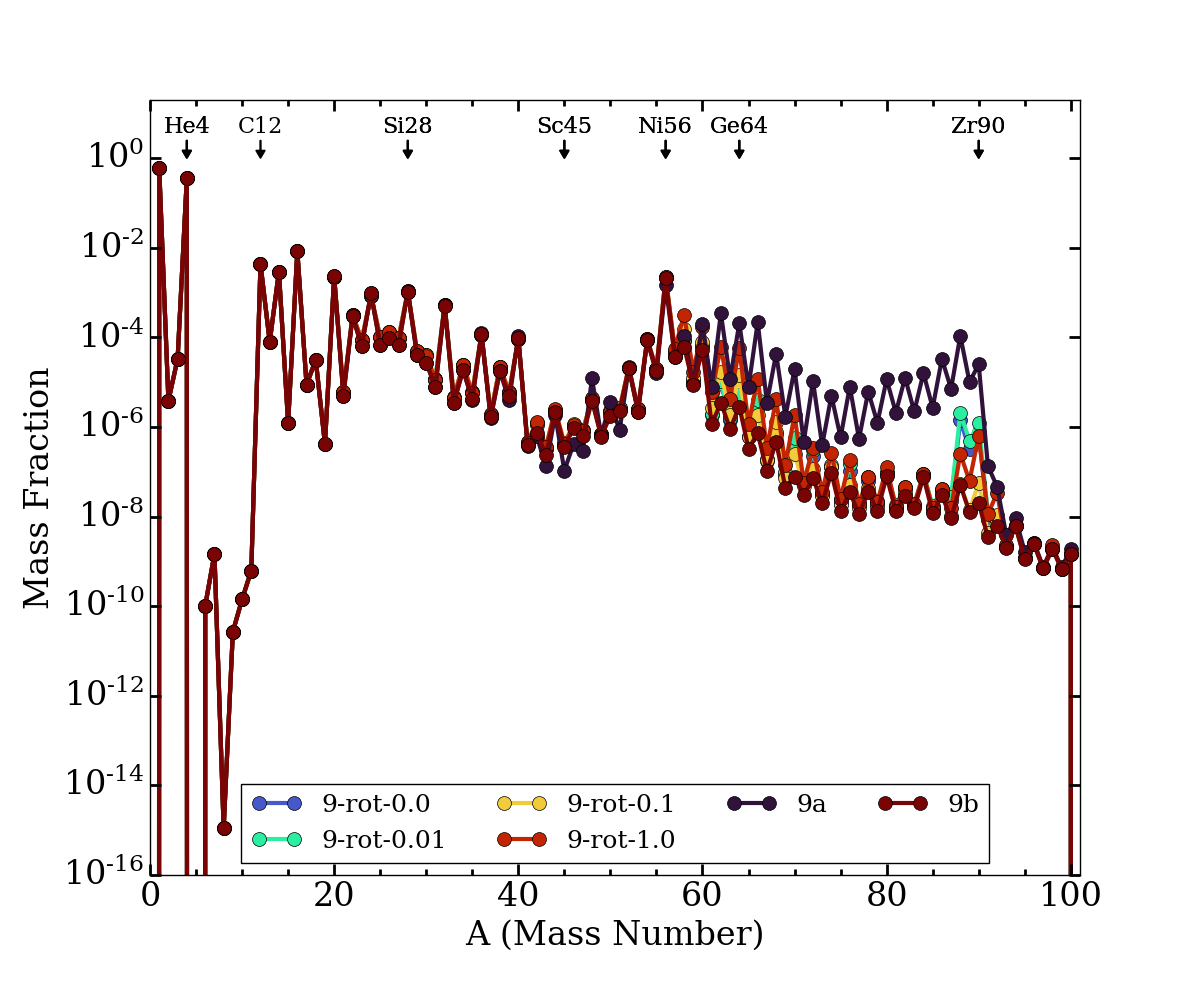}
    \includegraphics[width=0.45\textwidth]{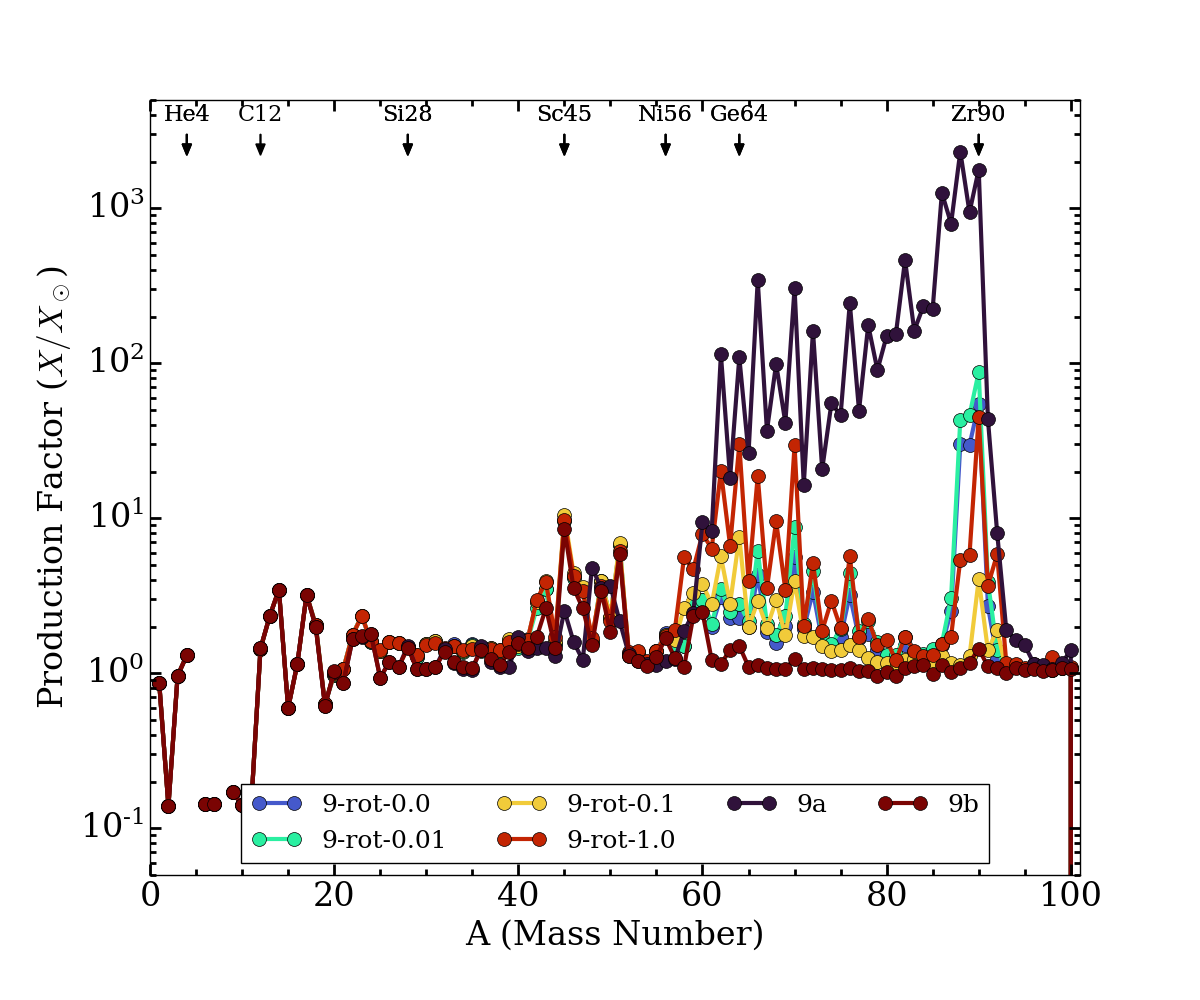}
    \caption{Nucleosynthetic yields (left) and production factors (right) of all six of our 9.0-$M_{\odot}$models. The two 9 $M_\odot$ models published before, 9a and 9b (with/without initial perturbation), are plotted for comparison. As discussed in Figure \ref{fig:ye-S}, increasing initial spin rates leads to more neutron-rich ejecta, which creates more elements heavier than iron. This effect can be more clearly seen in the production-factor plot. However, also as shown in Figure \ref{fig:ye-S}, the neutron-rich ejecta have quite low entropy and are unable to produce r-process isotopes except for some weak r-process peak nuclei near $^{90}$Zr. Here too model 9a with perturbations is much more prolific in producing isotopes beyond the iron peak.}
    \label{fig:yield}      
\end{figure*}

Figure \ref{fig:yield} shows the nucleosynthetic yields and production factors of all models. The trend in neutron-richness is clearly converted into a hierarchy of heavy element production. The yields of isotopes heavier than iron increase with faster initial spins, but initial perturbation shows a more significant impact. \tianshu{As indicated in Figure \ref{fig:ye-S}, the ejecta in the rotating models are at most modestly neutron-rich with low entropy ($Y_e>0.45$ and $S<20$ $k_b$ per baryon). Such ejecta conditions are mostly represented by the G0 group ($Y_e$ near 0.5 with low entropy, create no r-process) and a small amount of high-$Y_e$ end of the G1 group ($Y_e$ between 0.3 and 0.45 with low entropy, create only the first r-process peak) in \citet{kuske2025}, and thus are unable to create a significant amount of r-process elements, even the first peak.} From this analysis we find that rotation without magnetic fields has a small effect on the nucleosynthetic byproducts of explosion, and likely less of an effect than initial perturbations, at least for the models upon which we have focused in this paper.

\section{Gravitational Radiation}
\label{GW}

There is a large literature on the effects of rotation on gravitational-wave signatures in the context of core collapse that we will not summarize here. The reader is referred to the pioneering work of \citet{muller1982} and to salient representatives of that large literature \citep{ott_2005_onearmed,2009CQGra..26f3001O,ott2012_rapid,kuroda:14,fuller_klion2015,TaKoSu16,richers:17,muller2017b,tk18,pajkos2019,shibagaki2020,Shibagaki2021}.

\begin{figure*}[htbp!]
   \centering
   \includegraphics[width=0.45\textwidth]{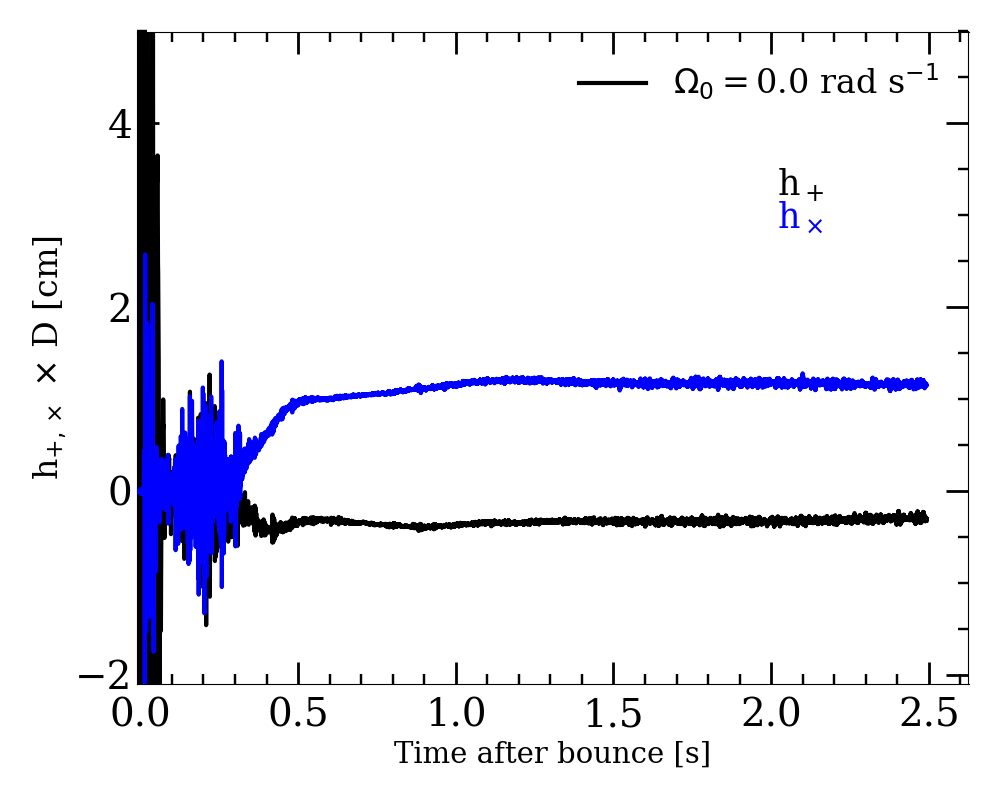}
   \includegraphics[width=0.45\textwidth]{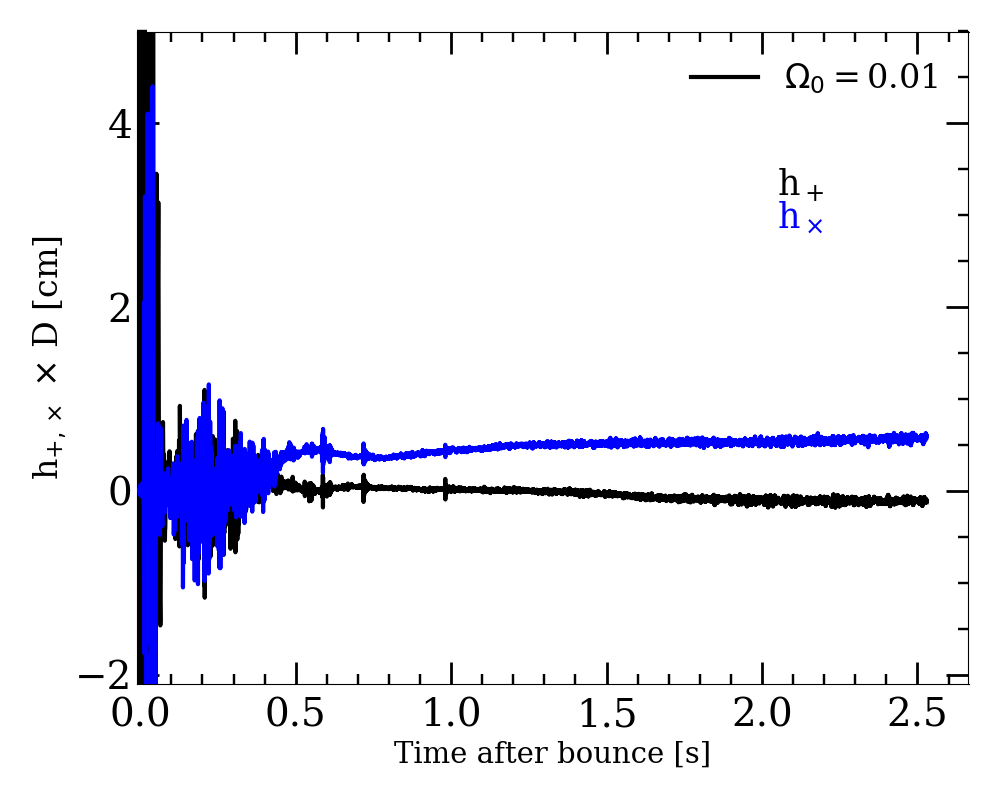}
   \includegraphics[width=0.45\textwidth]{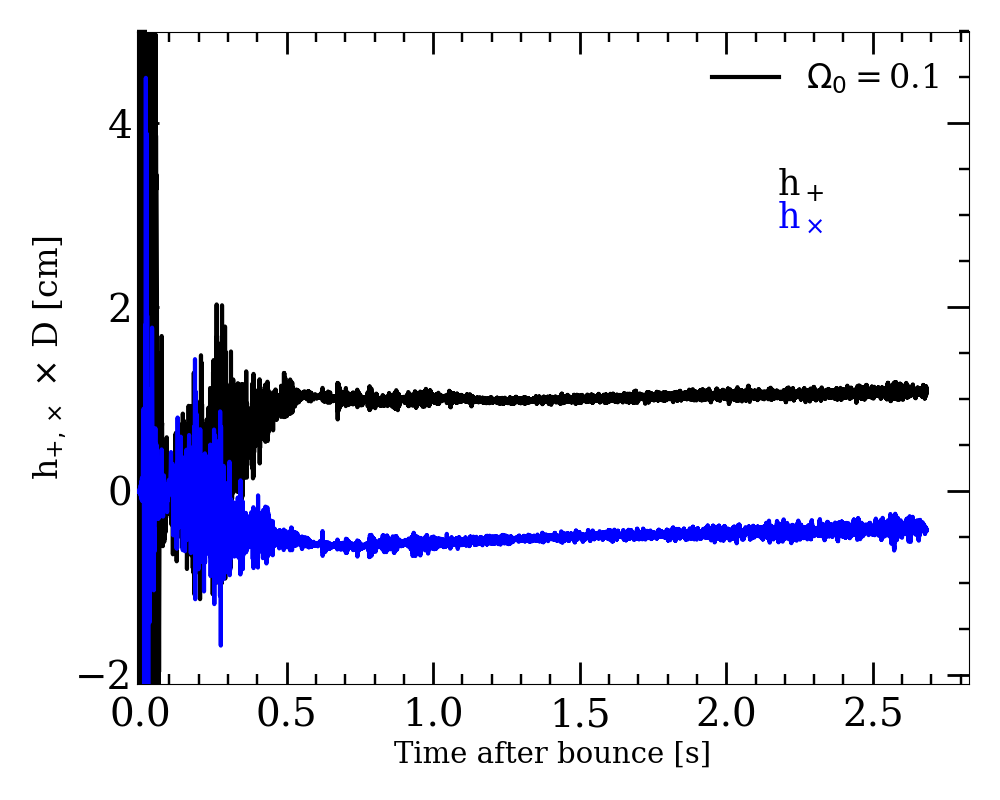}
   \includegraphics[width=0.45\textwidth]{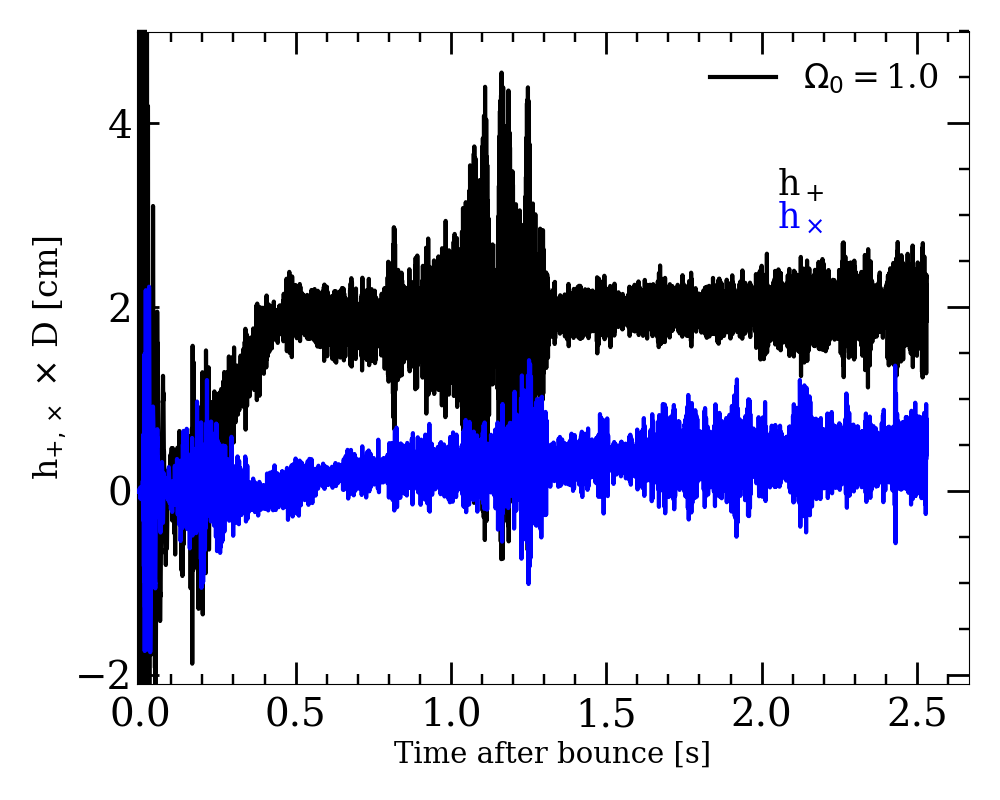}
   \caption{Matter GW strain (cm) versus time after bounce (in seconds) for the sequence of rotating models. All signals provided here are in the direction of the x-axis (recall that the z-axis is the rotation axis). See the text for details and a discussion.}
   \label{fig:gwstrain_matter}      
\end{figure*}

Figure \ref{fig:gwstrain_matter} depicts the time evolution of metric strain (times distance in cm) due to matter motions in the $x$ direction for the two polarizations for all four models. We see that for the slowest three models the signals due to matter motions are quite similar, lasting only about $\sim$0.5 seconds, after which a memory signal due to the asymmetric blast evolution on long timescales of the explosive debris emerges and roughly flattens. As has been discussed elsewhere \citep[e.g.,][]{vsg2018,vartanyan2023}, most of the power is due to the excitation of g/f-modes (transitioning into a pure f-mode) by accretion streams onto the PNS and a ``haze" due to the rapid deceleration of those streams. 
\david{We find that the matter memory h$_+$ along the equatorial direction becomes stronger with higher rotation rates than the memory along the polar direction. However, we do not see a similar directional trend at higher rotation rates for h$_\times$. The memory for both polarizations reflects the late-time ejecta morphology, and the stronger h$_+$ memory carries the imprint from the pole/equator asymmetry, whereas the h$_\times$ polarization is due to stochastic nonaxisymmetric ejecta morphology. We use the convectional spherical polarization basis defined in, e.g., \cite{1997PThPS.128..183O}, and our rotation axis here is aligned with the z-axis. For this low-mass progenitor (and more generally, \citealt{vartanyan2020}), the matter memory asymptotes in the first seconds, before our ejecta leave the grid boundary.}

\begin{figure*}[htbp!]
    \centering
    \includegraphics[width=0.45\textwidth]{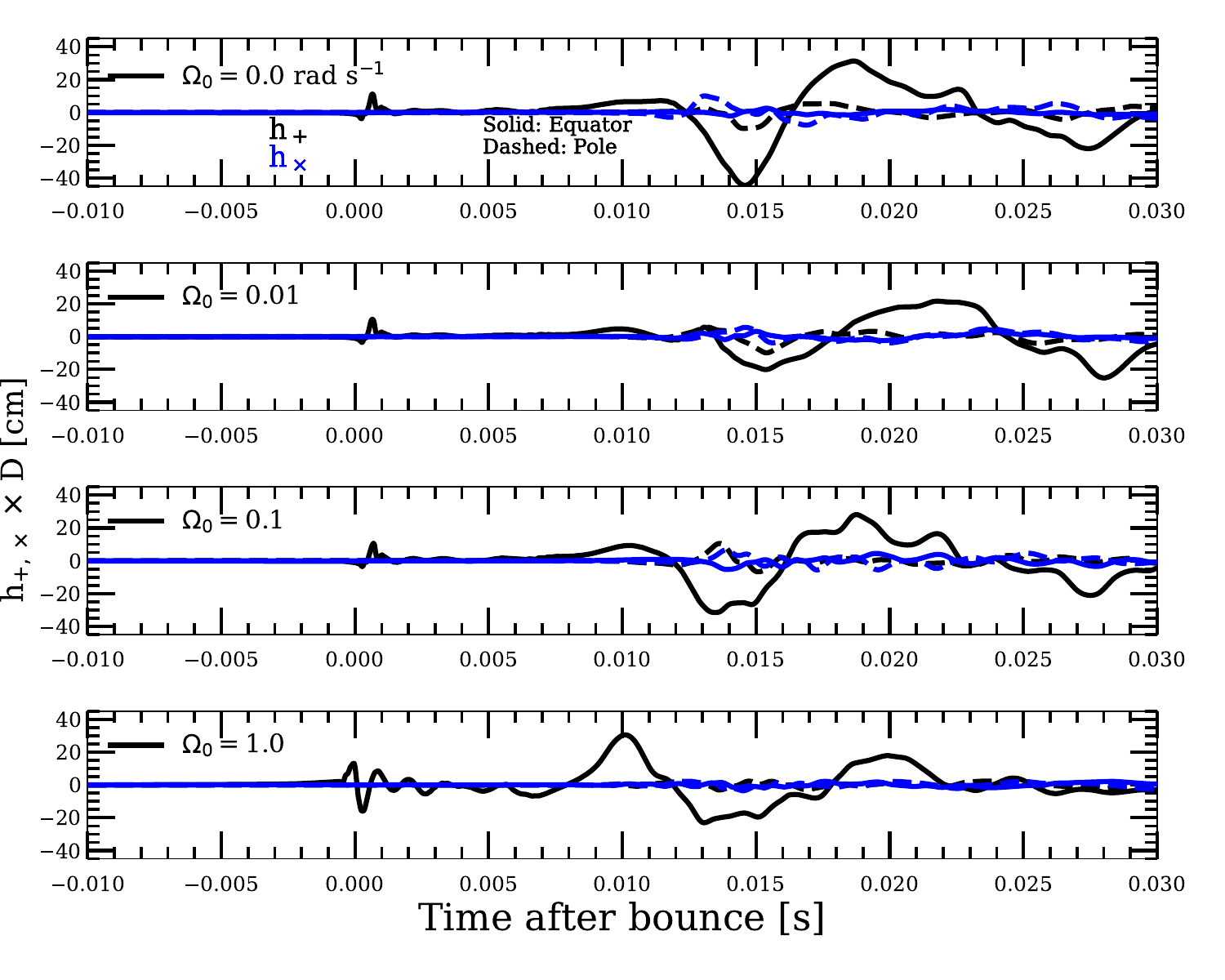}
    \caption{Matter GW strain (cm) versus time after bounce (in seconds) for the sequence of rotating models in the first 30 ms after core bounce for both polarizations and along the pole (dashed) and equator (solid). The bounce signature for the 9-M$_{\odot}$ model is only weakly sensitive to the initial rotation frequency. All signals provided here are in the direction of the x-axis. See text for a discussion.}
    \label{fig:gwstrain_matter_early}      
\end{figure*}

Figure \ref{fig:gwstrain_matter_early} depicts the earliest phase (obscured in Figure \ref{fig:gwstrain_matter}) for this sequence of four models during the first $\sim$30 ms after core bounce for the $h_+$ and $h_x$  polarizations along the pole and equator. This bounce signature is a combination of that due to prompt convection and the time-changing oblateness due to rotation. The associated frequency range in the radiated power is near $\sim$100 Hz and we discuss this below (see Figure \ref{fig:egw2}). Curiously, this component is only weakly sensitive to the initial rotation rate. Due to its faster rotation, the rapidly rotating model bounces more slowly and this attenuates its bounce and ringdown signal strain. As shown by \citet{ott2006_spin}, there is a rotational speed ``sweet spot, above and below which the rotational bounce signal is weaker. Generally, the plus and equatorial components are stronger than the cross and polar components for all models.

However, as depicted in the bottom right panel of Figure \ref{fig:gwstrain_matter}, the more rapidly rotating model maintains significant relative strengths in both polarizations for the duration of the simulation. This is a clear manifestation of the excitation of low-$T/|W|$ modes for this rapidly rotating model. As indicated on this figure, the strength of these modes in gravitational waves gradually increases from $\sim$300 ms to $\sim$1.0 second, at which time the amplitude jumps, signaling even more vigorous excitation of the associated modes. This highly-excited phase lasts for another $\sim$250 ms, after which time it persists to the end of the simulation in both polarizations, but at a reduced amplitude.

\begin{figure*}[htbp!]
    \centering
    \includegraphics[width=0.45\textwidth]{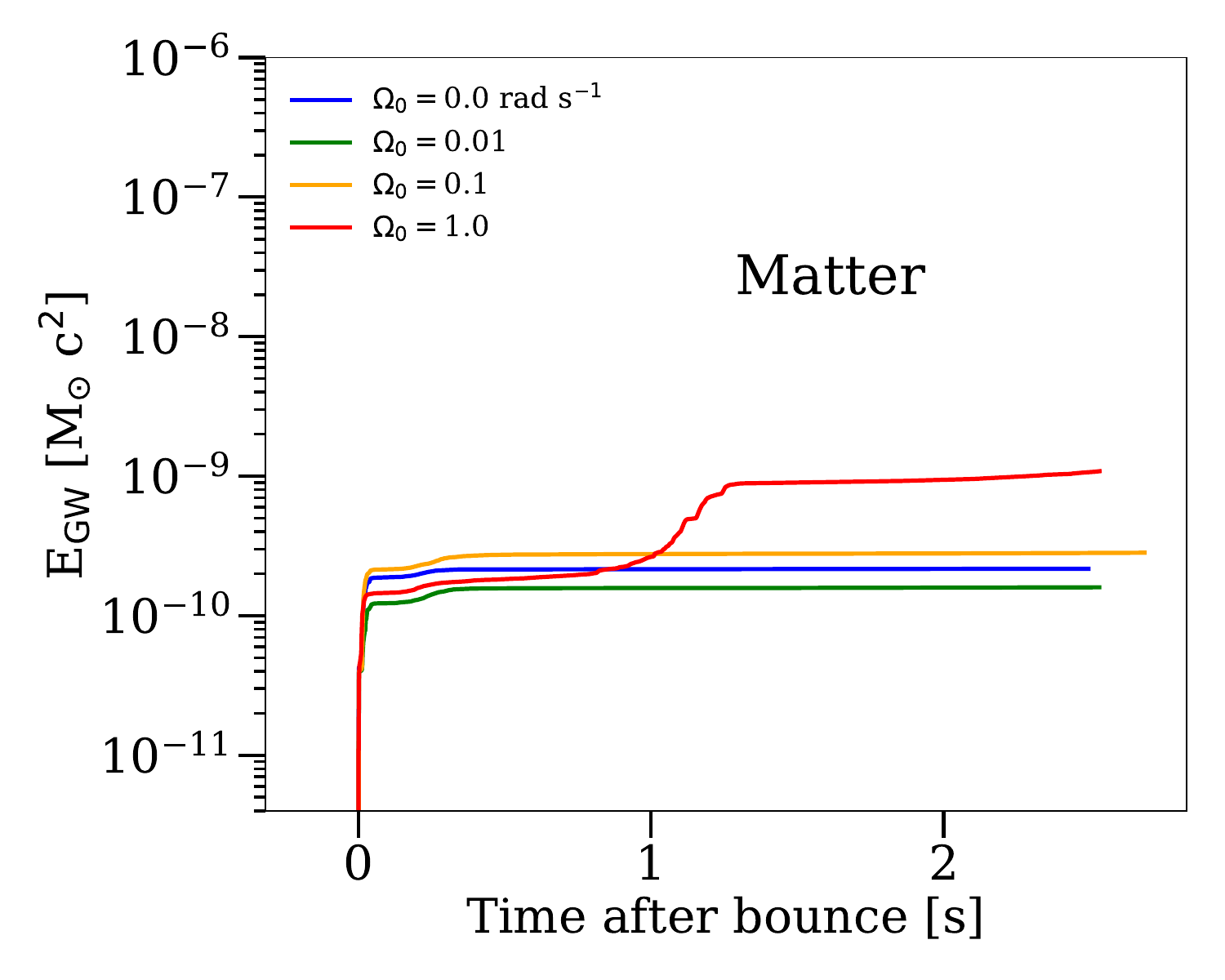}
    \includegraphics[width=0.45\textwidth]{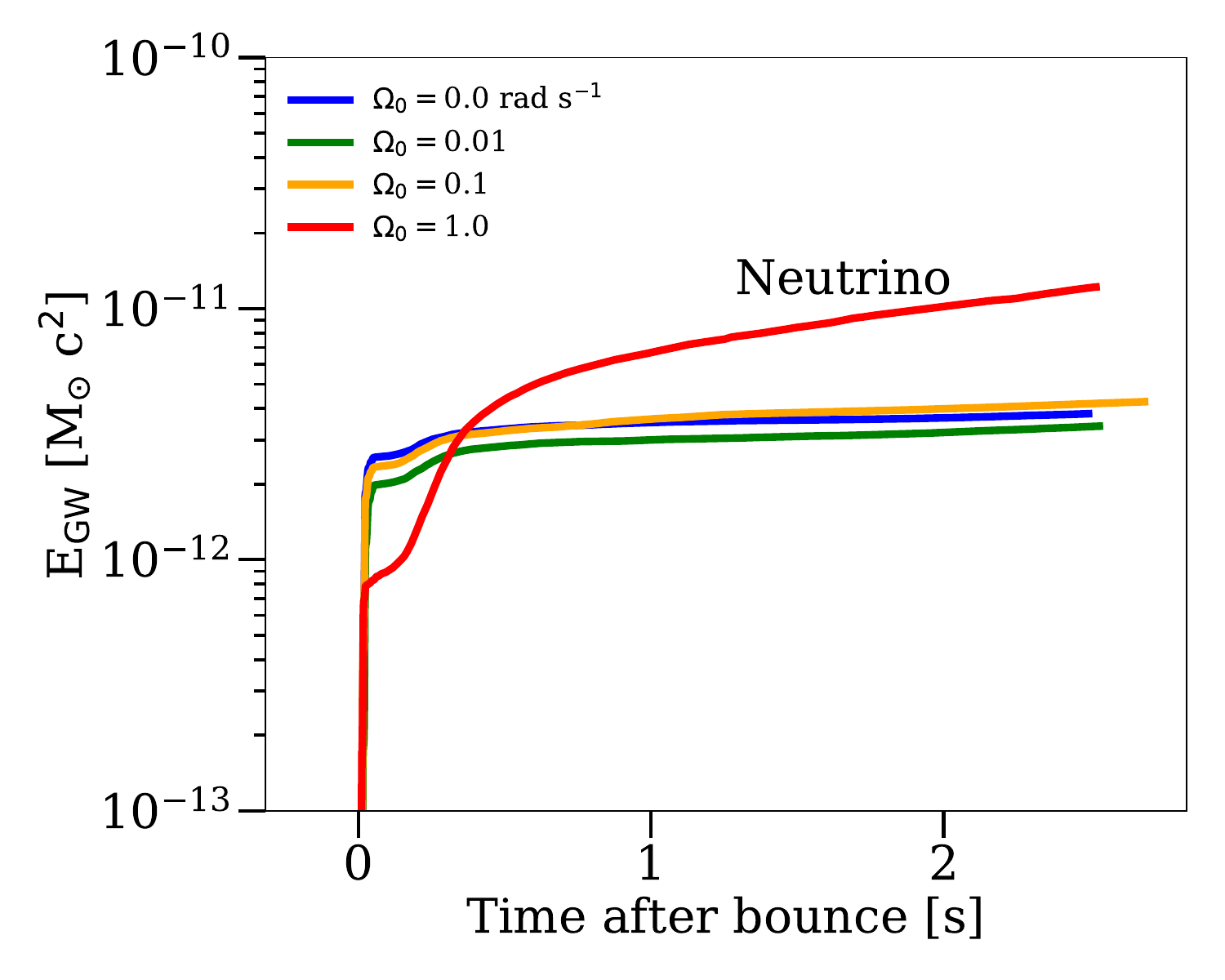}
    \caption{Matter (left) and neutrino memory (right) total GW energy emitted versus time after bounce.  We see a quasi-monotonic increase in the asymptotic gravitational wave energy with increasing rotation for both the matter and neutrino GW contributions, with the exception that the non-rotating model has a stronger signal than that of the weakly rotating model with $\Omega_0$=0.01 rad s$^{-1}$. We see that the GW energy of the rapidly-rotating model surpasses the more slowly-rotating models near 1 second for the matter contribution, and near $\sim$300 ms for the neutrino contribution.}
    \label{fig:egw4}      
\end{figure*}

Figure \ref{fig:egw4} shows the total energy radiated in gravitational waves for both the matter (left) and neutrino (right) components, the latter due to ``neutrino memory" sourced from the time-changing quadrupole moment of the anisotropically-radiated neutrinos \citep{epstein1978,burrows1996,emuller,vartanyan2020,choi2024}.
The jump for model 9-rot-1.0 in both these components due to the onset of the various low-$T/|W|$ modes is clearly seen. As a result, the total radiated gravitational wave energy for this model is many factors bigger than that for the other models.

\begin{figure*}[htbp!]
    \centering
    \includegraphics[width=0.45\textwidth]{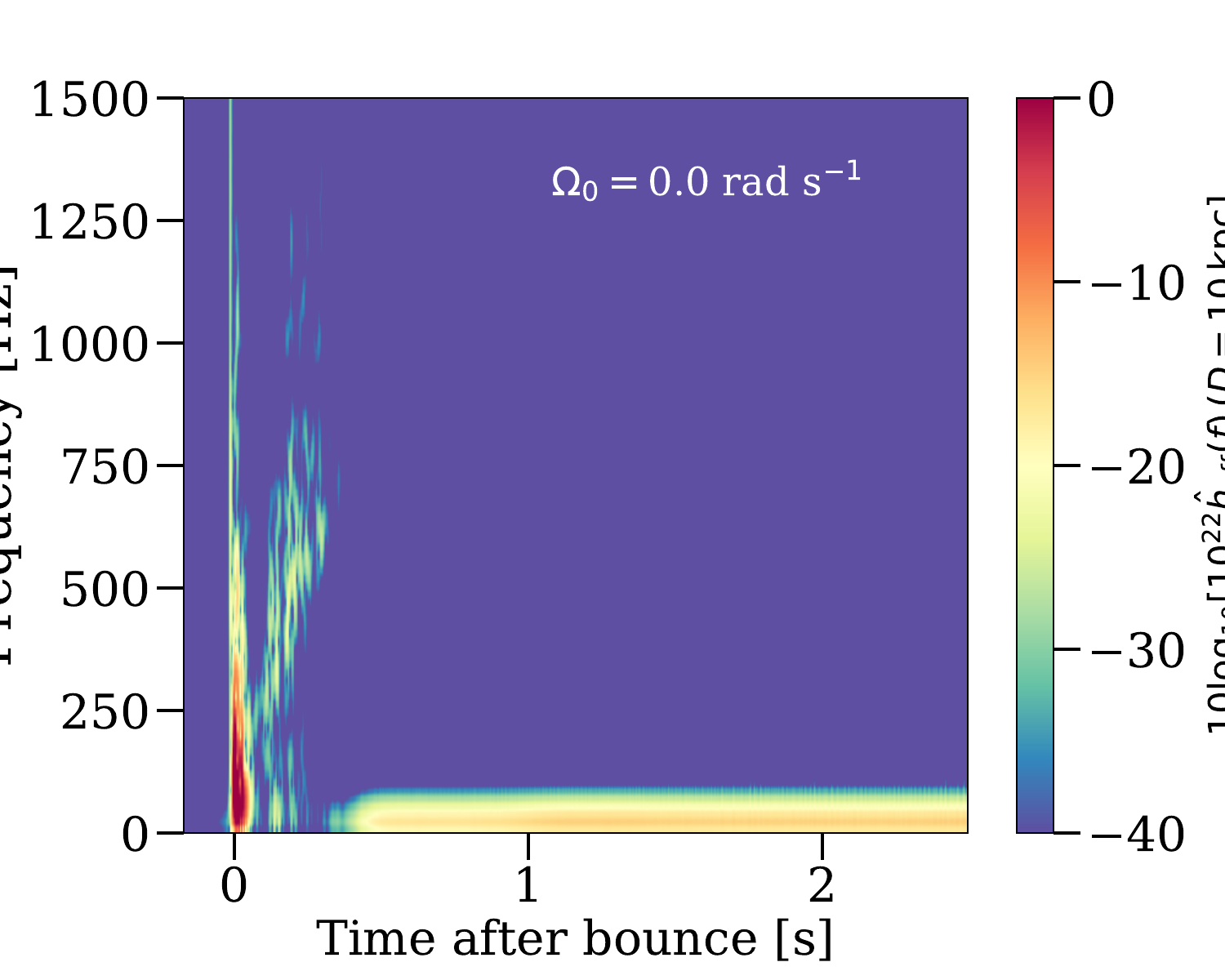}
    \includegraphics[width=0.45\textwidth]{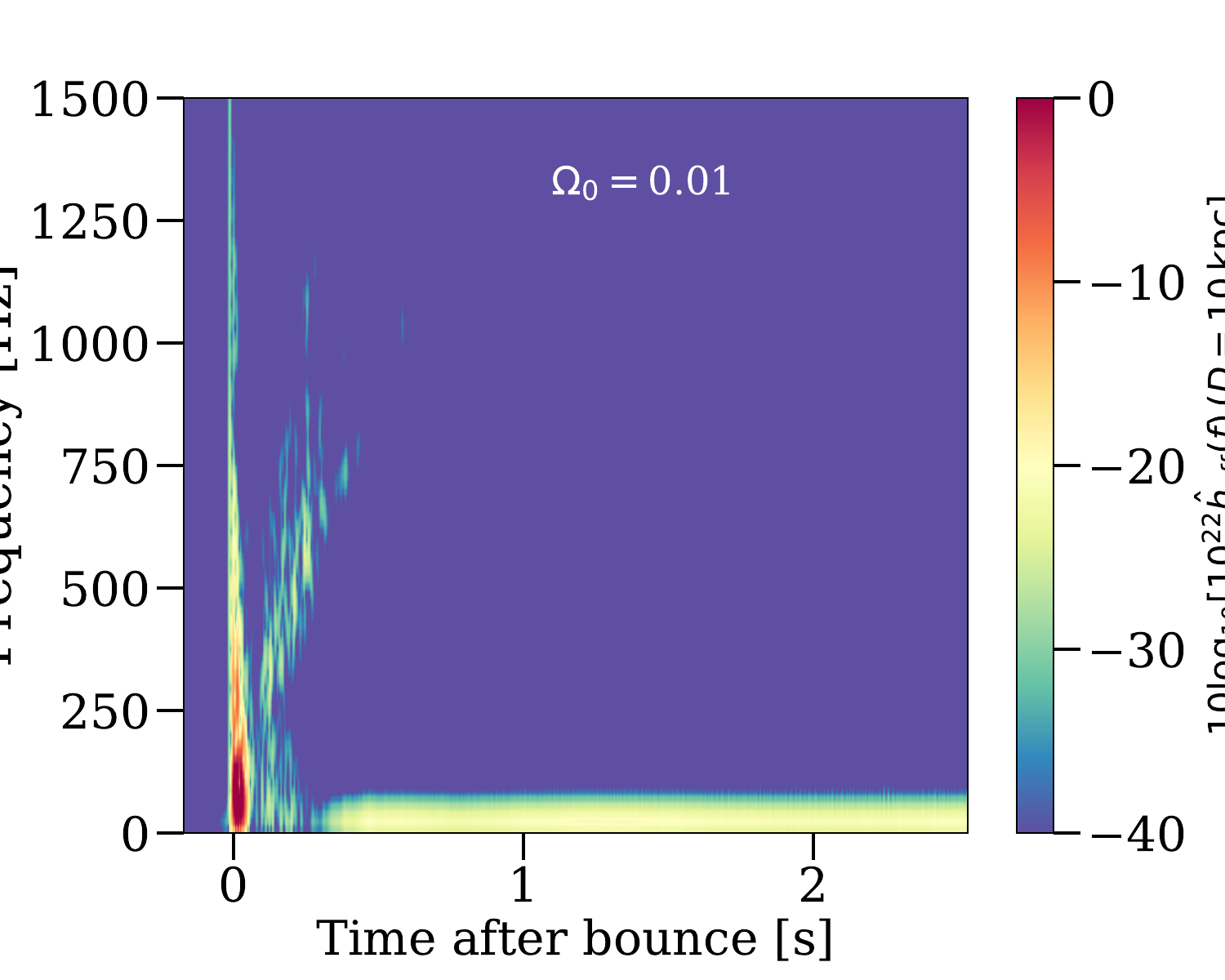}
    \includegraphics[width=0.45\textwidth]{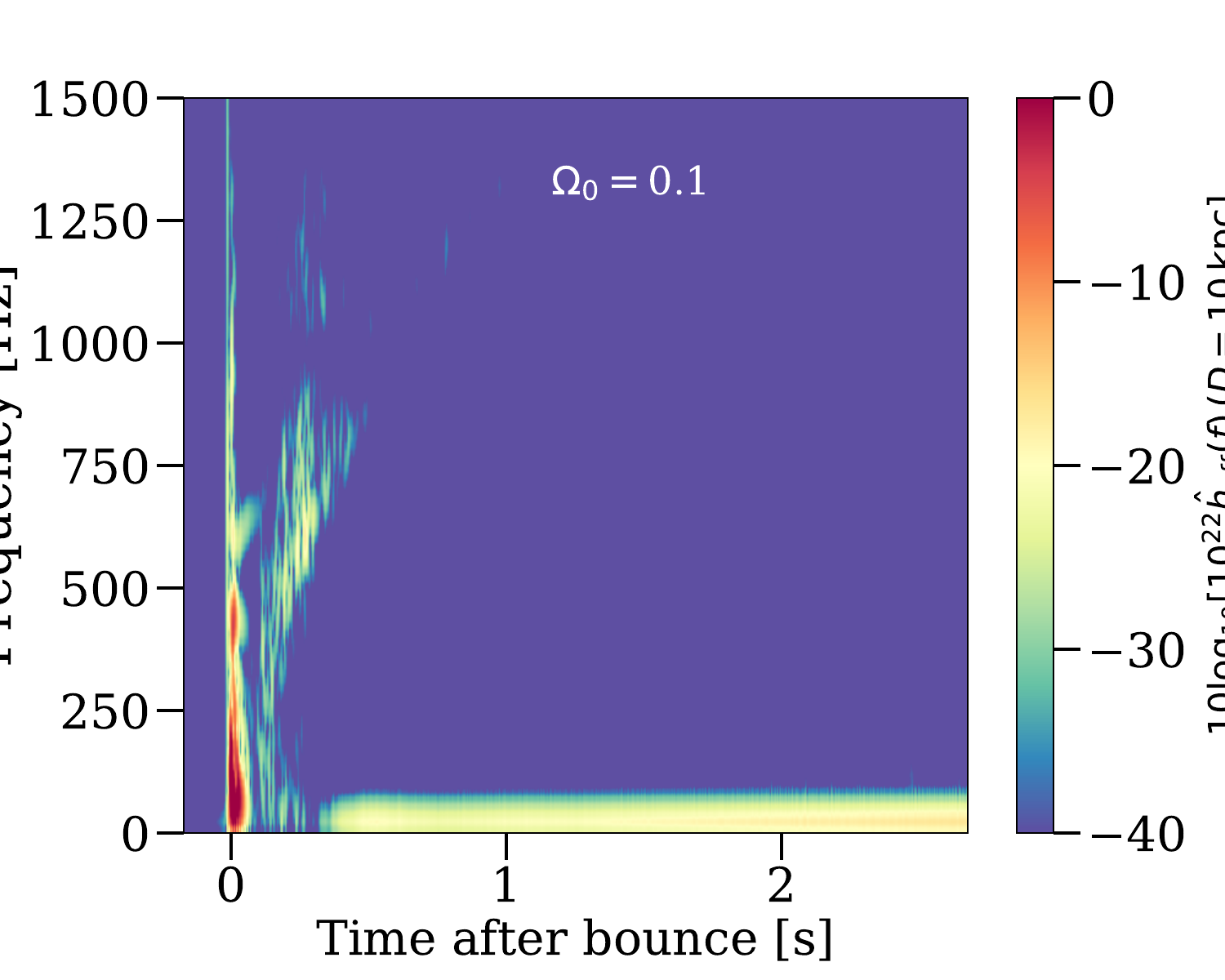}
    \includegraphics[width=0.45\textwidth]{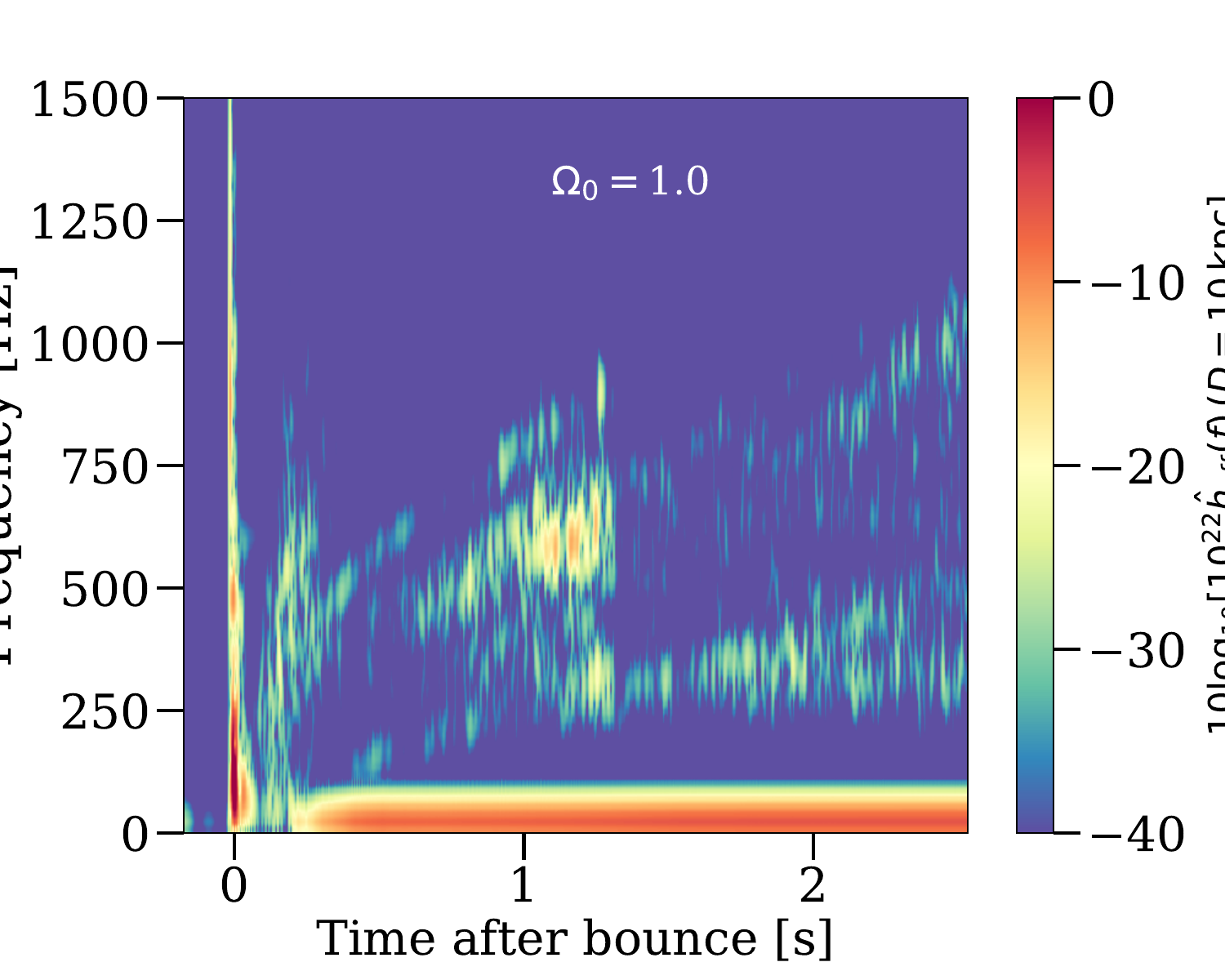}
    \caption{Matter-driven gravitational wave strain spectrogram plotted against time after bounce (in seconds). We see prominent prompt convection and/or rotational bounce signatures for all models in the very early phase, followed by weak early g/f-mode signals, slightly suppressed for the rapid-rotating model. The overall weakness of the these early signals is consistent with the low ``compactness" of this low-mass progenitor \citep{vartanyan2023}. The rapidly-rotating model evinces a significantly stronger matter memory signal (low-frequency band at the bottom). See text for a discussion.}
    \label{fig:egw}      
\end{figure*}

\begin{figure*}[htbp!]
    \centering
    \includegraphics[width=0.45\textwidth]{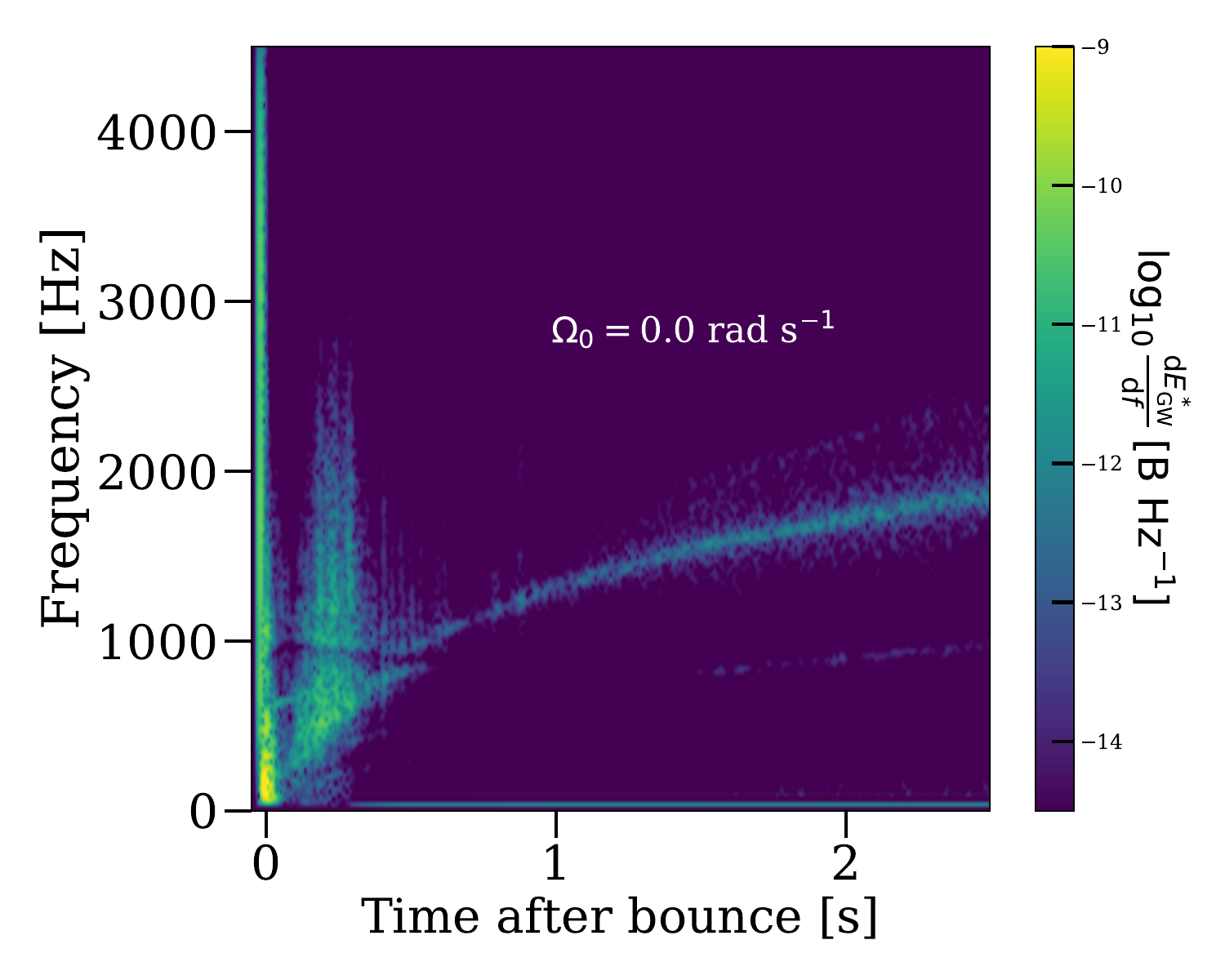}
    \includegraphics[width=0.45\textwidth]{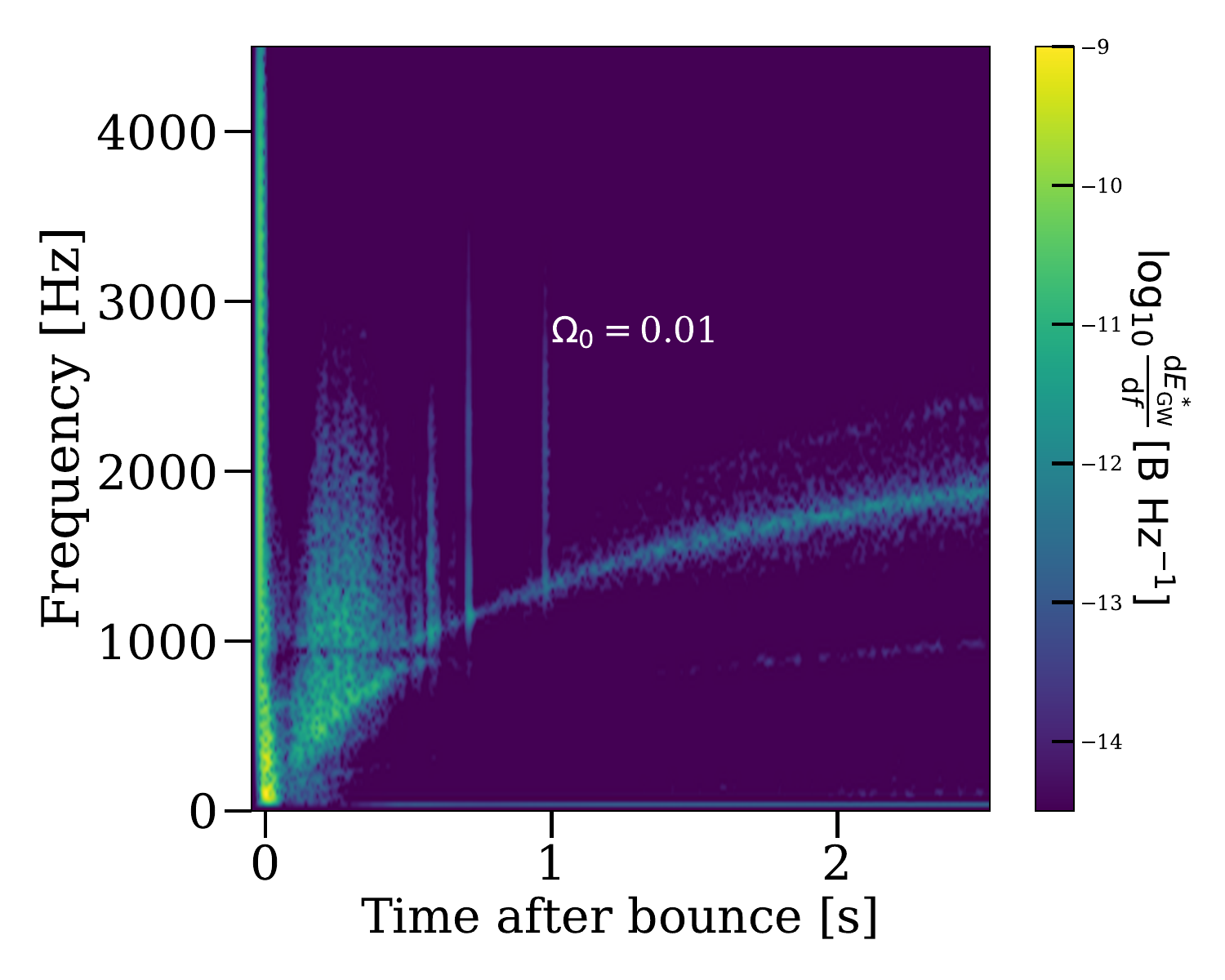}
    \includegraphics[width=0.45\textwidth]{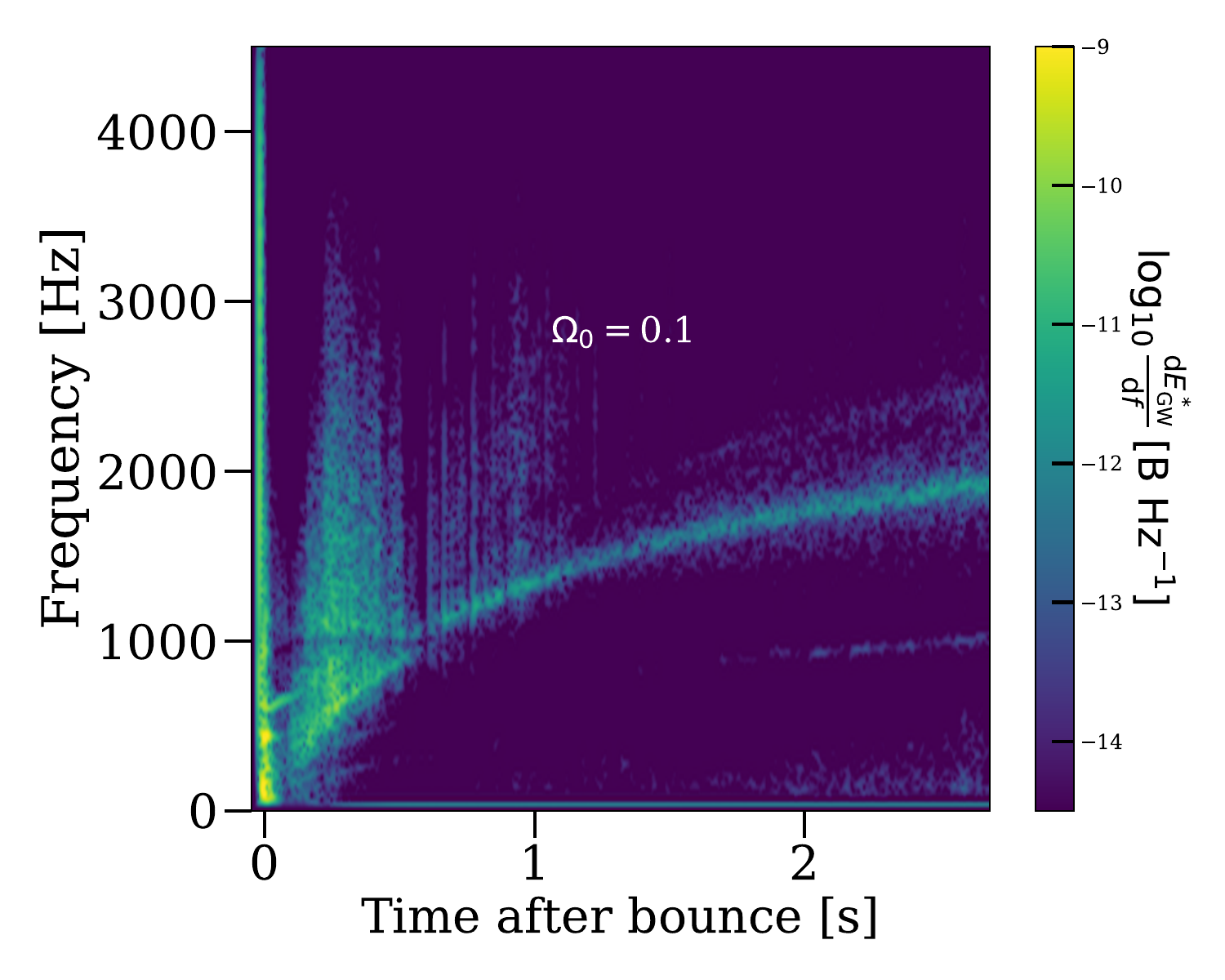}
    \includegraphics[width=0.45\textwidth]{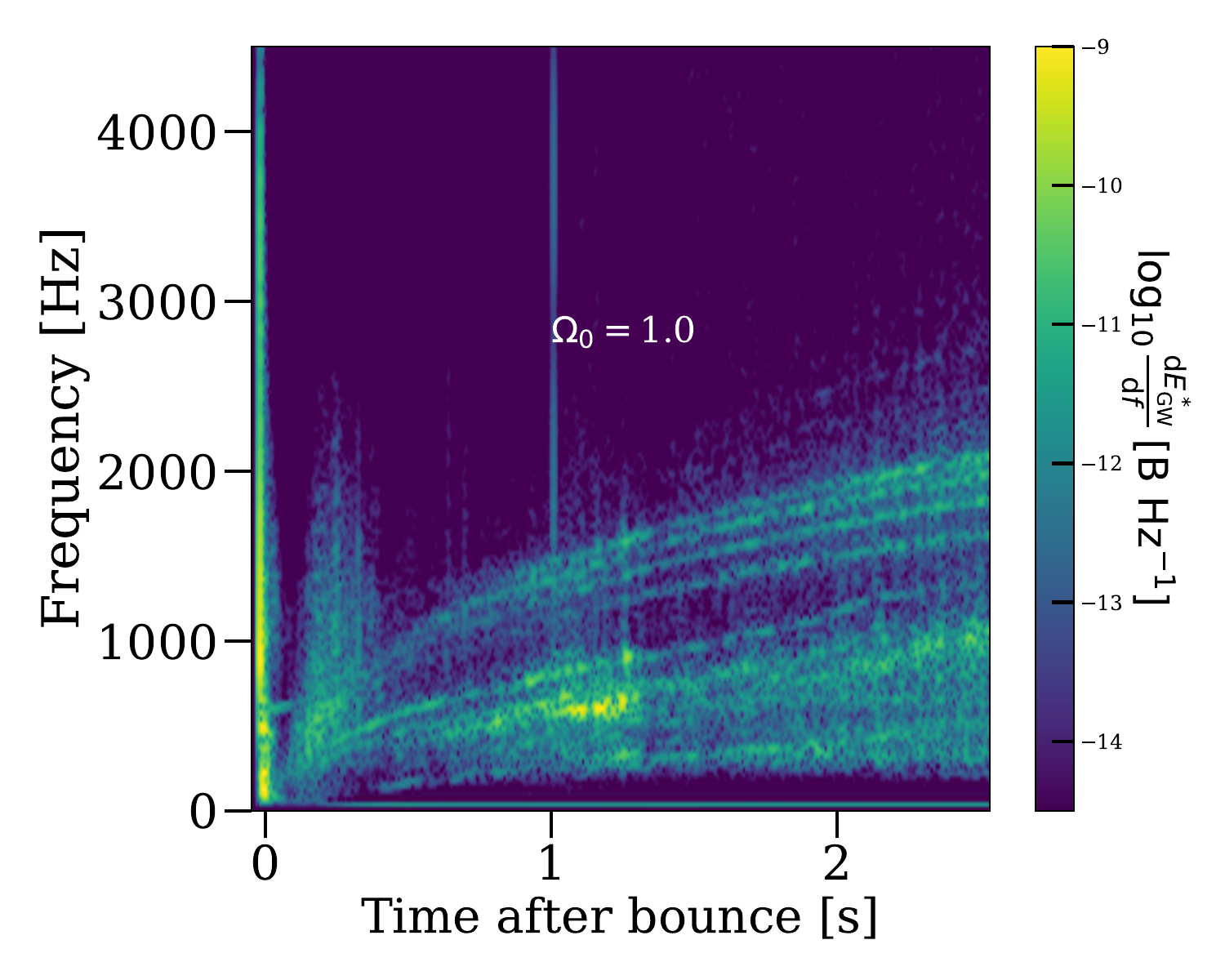}
    \caption{Matter-driven gravitational wave energy spectrograph. All models except the rapid-rotating model show a power gap near 1000 Hz. The 9-M$_{\odot}$ model rotating with an initial angular frequency of 0.1 rad s$^{-1}$ shows a stronger `fan' of GW radiation powered by accretion in the first second (extending to $\sim$3200 Hz), compared to the most rapid rotating model which has a shorter-lasting `fan' and at lower frequencies (extending to $\sim$2200 Hz). The rapidly rotating model shows a cascade of low-$T/|W|$ modes with a strong signal near $\sim$1 second at $\sim$600 Hz. See text for a discussion.}
    \label{fig:egw2}      
\end{figure*}

Figure \ref{fig:egw} depicts the frequency/time spectrogram of the strain for the four models.  As expected from Figure \ref{fig:gwstrain_matter}, the slowest three models show similar behavior, with the early prompt and rotational components and the g/f-mode component during the early hundreds of milliseconds after bounce clearly in evidence. However, the onset and persistence of the low-$T/|W|$ modes for the 9-rot-1.0 model, and the enhanced vigor around $\sim$1.0 seconds,  distinguish it clearly. We note that the rapidly-rotating model evinces a significantly stronger matter memory signal (low-frequency band at the bottom). 

However, when we generate and plot the energy spectrogram, interesting features emerge\footnote{Note that this quantity is derived from the square of the strain, and as a result the frequencies are double those in Figure \ref{fig:egw}.}. Figure \ref{fig:egw2} clearly shows the similarities for models 9-rot-0.0, 9-rot-0.01, and 9-rot-0.1, with the haze, g/f-mode, and ``power gap" \citep{eggenberger2021,Rusakov2026} seen near $\sim$1000 Hz during the first $\sim$0.5 seconds. However, the bottom right panel of Figure \ref{fig:egw2}, portraying the 9-rot-1.0 model, shows distinct modes that persist and grow in frequency, covering a frequency range from $\sim$200 to 2000 Hz.  The bounding frequency curve is clearly the f-mode seen in all CCSN models at late times \citep{murphy:09,vsg2018,vartanyan2023,choi2024}. Moreover, the period of enhanced amplitude near and around $\sim$1.0 second after bounce for this model clearly pops out.

The multitude of distinct bands shown in the bottom right panel of Figure \ref{fig:egw2} suggests (but does not prove) we are seeing multiple distinct low-$T/|W|$ modes, in addition to the f-mode. Are these inertial modes contingent on Coriolis effects? Are these resonances with core g-modes? Is there a relationship with the f-mode? What is the role of the evolving PNS convective zone (\S\ref{pns})? The rich set of features in this plot motivate a much more rigorous analysis that we must postpone to a later paper, but which suggests many fruitful directions for future research.

\begin{figure*}[htbp!]
    \centering
    \includegraphics[width=0.45\textwidth]{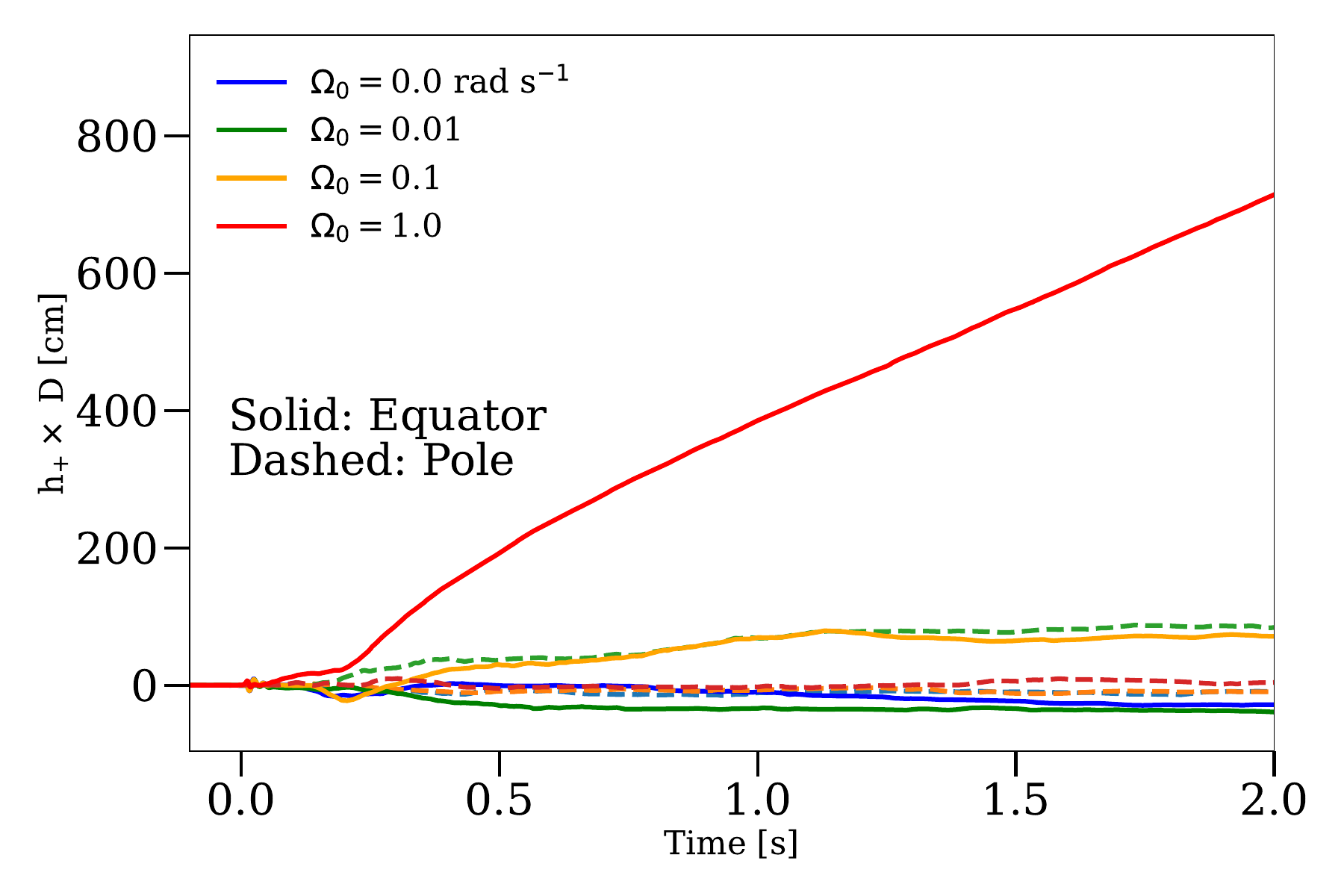}
    \includegraphics[width=0.45\textwidth]{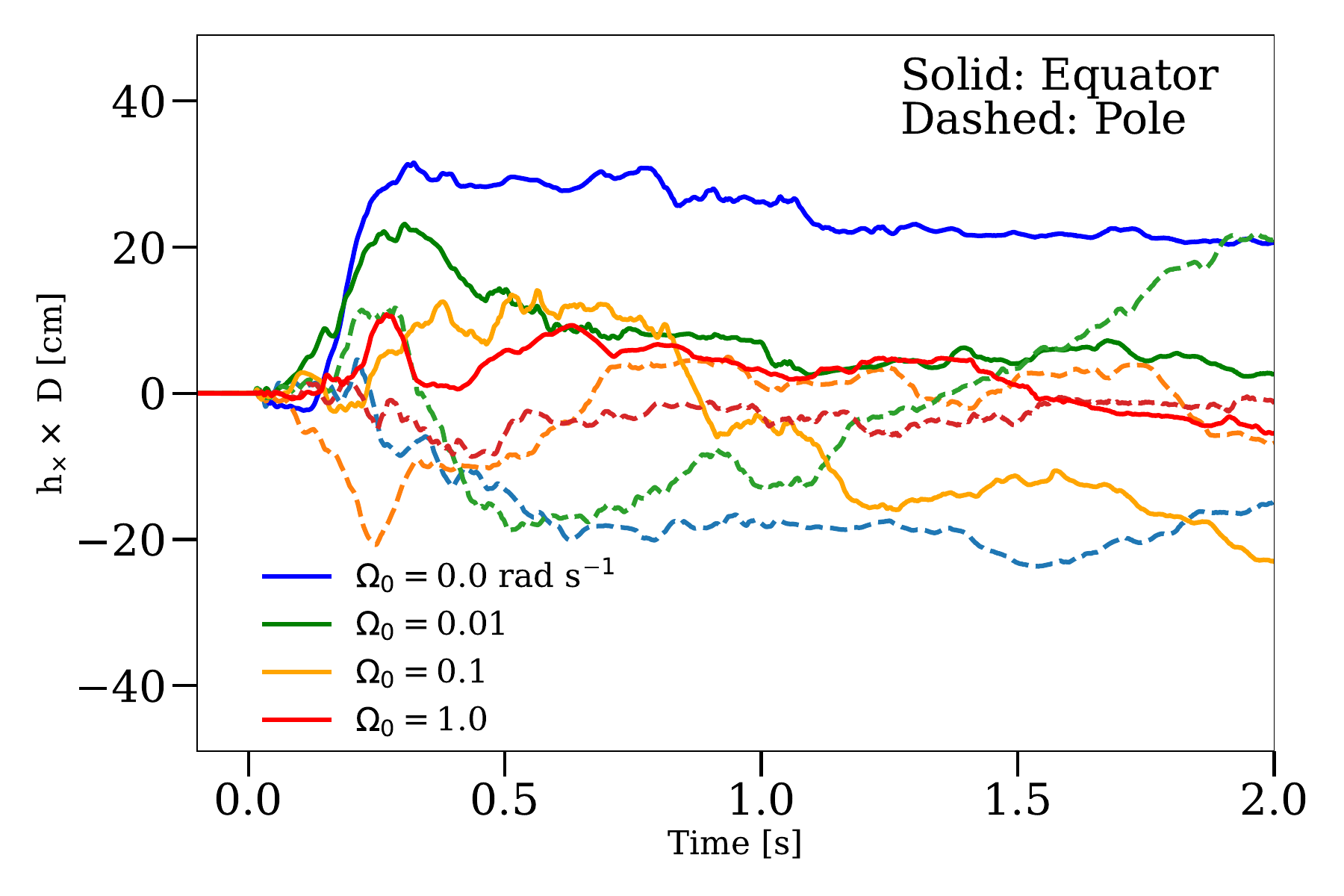}   
    \caption{Neutrino-driven GW strain versus time after bounce (in seconds) (\textbf{left:} h$_+$; \textbf{right:} h$_\times$) comparing the pole (dashed) versus equator (solid) emission along the positive x-axis. Only for the rapidly-rotating 9-rot-1.0 model do we see a significant pole/equator strain asymmetry visible in the plus-polarization, which has strains more than  $\sim$20 times stronger than the cross-polarization at late times and $\sim$100 times stronger than for the weakly and non-rotating models.}
    \label{fig:gwstrain_nu_angle}      
\end{figure*}

Figure \ref{fig:gwstrain_nu_angle} depicts the neutrino memory strain versus time after bounce for the $h_+$ and $h_x$ polarizations at the pole and equator (along the x-axis). The rapidly-rotating model with $\Omega_0$=1.0 rad s$^{-1}$ has the only significant pole/equator asymmetry, and this in the plus-polarization. Note, however that the amplitude of the neutrino memory for all models is generally greater than that of the matter, though the latter has much more power at high frequencies. This is clearly seen in Figure \ref{fig:egw4}, which depicts the time-integrated power spectrum of all our models for the matter (top left), neutrinos (top right), and sum (bottom). The signals near $\sim$100 Hz depict the prompt convection and rotation/bounce components.  Note that the prompt component for the 9-rot-0.0 model is not small, reflecting the effect of perturbations on infall on the bounce component. The relative strength of the 9-rot-1.0 model around and below $\sim$1000 Hz captures the low-$T/|W|$ mode complex. 
Note again the large pole/equator difference for the neutrino memory for this model. When summed, the total power spectrum emphasizes the large relative contribution of neutrino memory below $\sim$50-60 Hz, a feature seen elsewhere \citep{choi2024}
that is emerging to be important for detection if the low-frequency sensitivity of the emerging flotilla of gravitational wave detectors can reach it.


\section{Discussion and Conclusions}
\label{conclusions}

In this paper, we have explored the dependence upon initial rotation rate (without considering magnetic fields) of various supernova observables for the explosion of a 9-$M_{\odot}$ supernova progenitor simulated to or near the asymptotic state of explosion launch.  We have found that the explosion energy is non-monotonic with, and weakly dependent upon, spin across a broad range of initial spins. It is only for spin rates which leave pulsars with spins faster than $\sim$15 milliseconds that a boost in explosion energy near $\sim$20\% is seen. Slower initial spin rates, in the range expected to leave pulsars with measured and inferred spin, do not much effect the explosion energy. The dipole asymmetry of the blast is weakly dependent upon spin, with faster spins leading to slightly greater asymmetries. However, there is no significant pole-equator neutrino heating asymmetry during explosion.  This is due to the fact that explosion is early for such progenitors, while the core is still extended before Kelvin-Helmholtz contraction leads to spin up and an oblate protoneutron star. Moreover, it is only for the fastest rotator that a pole-equator variation of significance is ever seen.

We find that rotation only marginally decreases the angle-integrated neutrino luminosities, perhaps at most by only $\sim$10\%.  Correspondingly, and only for the fastest rotator, the mean neutrino energies are diminished by up to $\sim$0.5 MeV. However, these differences diminish with time after bounce.
Moreover, only for the fastest rotator does the resulting neutrino heating rate in the gain region diminish noticeably, and this by no more than $\sim$15\% during the first $\sim$200 milliseconds after bounce. 

We find that the recoil kick and spin are clearly aligned only for the most rapidly rotating model. This leaves unresolved the explanation for the spin/kick correlation inferred observationally. A major result we obtain is that for the fastest rotator and within $\sim$300 milliseconds of bounce spiral arm modes emerge due to strong $T/|W|$ resonances, with angular momentum and energy fluxes predominantly in the plane perpendicular to the spin access. This is most manifest in the gravitational radiation signal. For the slowest three models, the gravitational wave signatures are quite similar, with the standard components of a rotational bounce and prompt convection, followed by a brief period of g/f-mode excitation and a ``haze"  powered by stochastic accretion streams, accompanied by a ``power gap" near $\sim$1000 Hz. However, for our fastest model and within a few hundred milliseconds, what seem to be numerous $T$/$|W|$ corotation resonances emerge, with discrete frequencies in the power spectrum (Figure \ref{fig:egw2}) from $\sim$200 Hz to the late-time $\sim$2000 Hz f-mode. This, along with the classic rotation/bounce signature of rapidly-rotating models, is the most distinctive gravitational wave feature to emerge from our study.

We find that the nucleosynthetic yields depend little upon the rotation rate, with the ejecta being only slightly more neutron-rich with increasing initial spin rate and producing only slightly more isotopes heavier than the iron peak. Comparing with our previous study for the same progenitor, but with initial perturbations in the collapsing mantle, we find that such perturbations are more consequential for heavy element production than even rapid rotation.

The goal of this paper was to determine the effect of rotation on all salient aspects of supernova dynamics. We found that most effects can be small, even for rapid rotation.  
Since we also found that core spins are amplified during collapse by a factor near $\sim$4000, the initial spin periods of pulsar progenitors could be minutes. This is much slower than derived by \citet{2005ApJ...626..350H} and is qualitatively consistent with the slower interior spin rates measured asteroseismically  \citep[e.g.,][]{cantiello2014,deheuvels2012,beck2013} and suggested theoretically \citep{fuller2019,fuller2022} due to enhanced magnetic torquing and angular momentum transport out of the core. Nevertheless, even if the initial spins are larger than expected from pulsar data, we suggest that rotation per se should play little role in supernova dynamics and observables most of the time. 

However, we focused on only one low-mass progenitor and did not include magnetic fields. Coupled with rapid rotation, the effect of the latter on explosion dynamics and observables could be significant \citep{leblanc1970,burrows2007_mag,Kuroda2014,2020ApJ...896..102K,Mosta2014,mosta2015,shankar_mosta2025,Obergaulinger2018,Obergaulinger2020,Obergaulinger2021,Aloy2021,varma_muller_rot_2023,bugli2023,powell2023,kovalenko2026}, but only for rapid rotation for which the free energy in PNS core rotation can be in the supernova energy range. This requires periods at bounce faster than $\sim$10 milliseconds. Otherwise, as Table \ref{tab:main} shows, even for rotation rates likely faster than those of most pulsar progenitors the effects of rotation on supernova dynamics are likely small. 

A major caveat and limitation of this study is the absence of other progenitor models. The effect of rotation alone on the explodability, explosion dynamics, and asymptotic observables of more massive progenitors could be quite different, and this is something we are exploring. Nevertheless, at least for low-mass progenitors which explode early, we find muted consequences of rotation overall in most major particulars across a wide range of initial spins.


\section*{Data Availability}

The data presented in this paper can be made available upon reasonable request to the authors. In addition, we have  deposited in the \url{https://zenodo.org/records/22665142}
(\url{https://zenodo.org/doi/10.5281/zenodo.22665141}) repository the data need to generate Figure 8 and two associated movies. We have updated our ongoing repository of gravitational wave strains and quadrupole moments in
\url{http://www.astro.princeton.edu/~burrows/gw.3d.2024.update/} and \url{https://dvartany.github.io/data/} with those associated with this set of rotating models.

\section*{Acknowledgments}

TW acknowledges support by the U.~S.\ Department of Energy under grant DE-SC0004658, support by the Gordon and Betty Moore Foundation through Grant GBMF5076, and support through a Simons Foundation grant (622817DK). DV acknowledges support from the NASA Hubble Fellowship Program grant HST-HF2-51520. AB acknowledges former support from the U.~S.\ Department of Energy Office of Science and the Office of Advanced Scientific Computing Research via the Scientific Discovery through Advanced Computing (SciDAC4) program and Grant DE-SC0018297 (subaward 00009650) and former support from the U.~S.\ National Science Foundation (NSF) under Grant AST-1714267. We are happy to acknowledge access to the Frontera cluster (under awards AST20020 and AST21003). This research is part of the Frontera computing project at the Texas Advanced Computing Center \citep{Stanzione2020}. Frontera is made possible by NSF award OAC-1818253. Additionally, a generous award of computer time was provided by the INCITE program, enabling this research to use resources of the Argonne Leadership Computing Facility, a DOE Office of Science User Facility supported under Contract DE-AC02-06CH11357. Finally, the authors acknowledge computational resources provided by the high-performance computer center at Princeton University, which is jointly supported by the Princeton Institute for Computational Science and Engineering (PICSciE) and the Princeton University Office of Information Technology, and our continuing allocation at the National Energy Research Scientific Computing Center (NERSC), which is supported by the Office of Science of the U.~S.\ Department of Energy under contract DE-AC03-76SF00098.

\newpage

\bibliographystyle{aasjournal}
\bibliography{References}


\begin{deluxetable*}{cccccccc}
\tablecolumns{8}
\tablewidth{0pt}

\begin{minipage}{\textwidth}
  \centering
  \textbf{Model Result Summary} \\  
\end{minipage}
    \tablehead{$\Omega_0$ [rad s$^{-1}$] & $2\pi/\Omega_0$ [s] &$\rho_c$ at 1s [g cm$^{-3}$] &$T_{\rm trans}$ at 2s [erg] & T/$|W|$ at 2s &Cold NS Mass [$M_\odot$] & Explosion Energy [B]& Final Period [ms]}
\startdata
0.0   &--   &4.45$\times10^{14}$ &1.43$\times10^{49}$ &8.53$\times10^{-5}$ &1.2371 &0.100 &  605.9\\
0.01  &628  &4.72$\times10^{14}$ &1.74$\times10^{49}$ &1.03$\times10^{-4}$ &1.2381 &0.096 &  131.9\\
0.1   &62.8 &4.84$\times10^{14}$ &1.21$\times10^{50}$ &7.11$\times10^{-4}$ &1.2365 &0.113 &  16.76\\
1.0   &6.28 &4.60$\times10^{14}$ &7.67$\times10^{51}$ &4.75$\times10^{-2}$ &1.2344 &0.120 &  1.662\\
\enddata
\caption{Some summary quantities for our suite of rotating 9-$M_{\odot}$ models. The non-rotating progenitor was taken from \citet{swbj16}. The cold neutron star moment of inertia used to derive the final period is estimated using $I_\text{NS}=(0.237+0.674\xi_{\rm NS}+4.48\xi_{\rm NS}^4)MR_{\text{NS}}^2$, where $\xi_{\rm NS}=\frac{GM}{R_{\text{NS}}c^2}$ is the compactness parameter of the neutron star itself \citep{breu2016} and $R_{\text{NS}}$ has been set to 12 km. This moment of inertia assumes a spherical neutron star so it will underestimate the final period of the rapid rotating models.}
\label{tab:main}      
\end{deluxetable*}

\begin{figure*}[htbp!]
    \centering
    \includegraphics[width=0.45\textwidth]{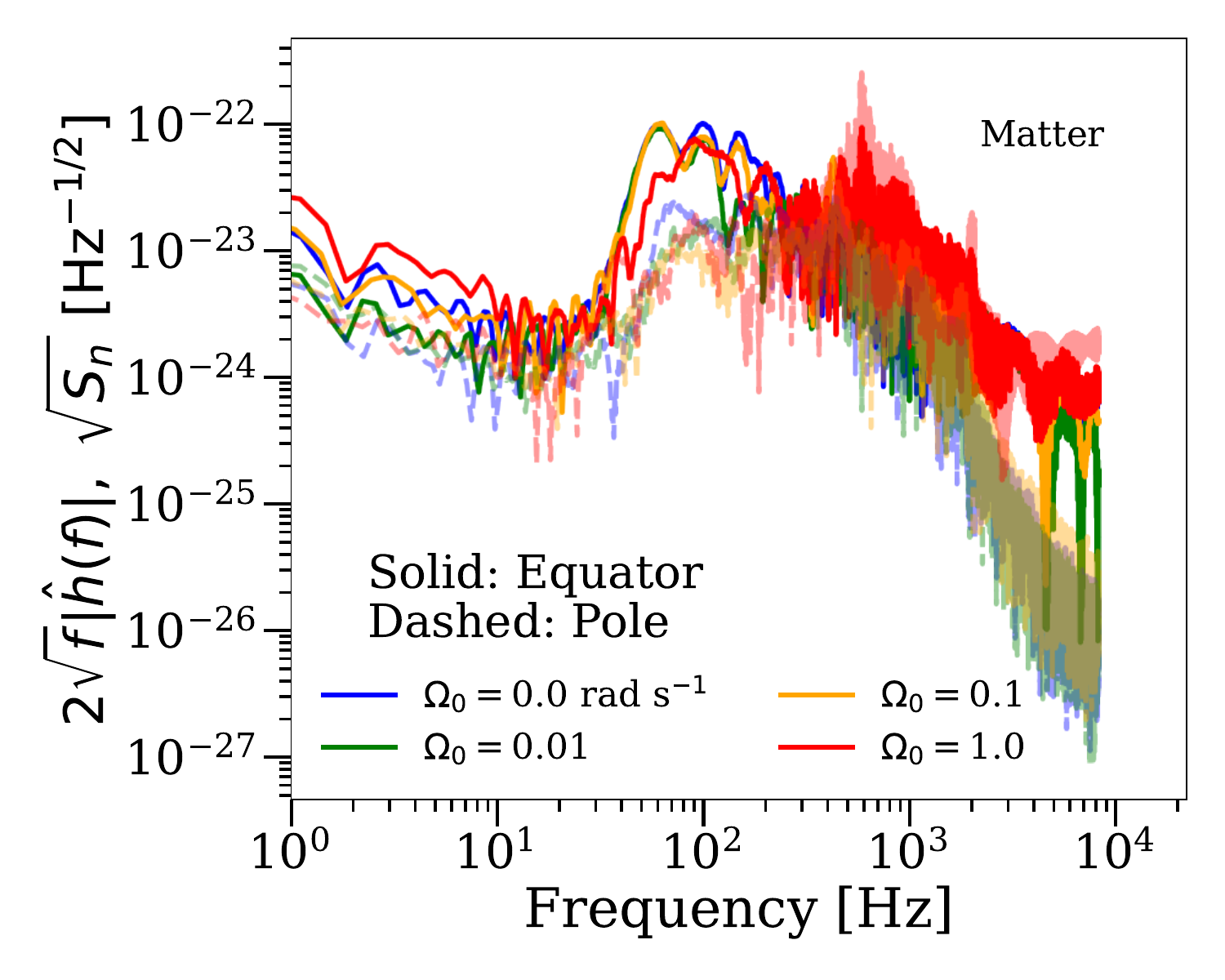}
        \includegraphics[width=0.45\textwidth]{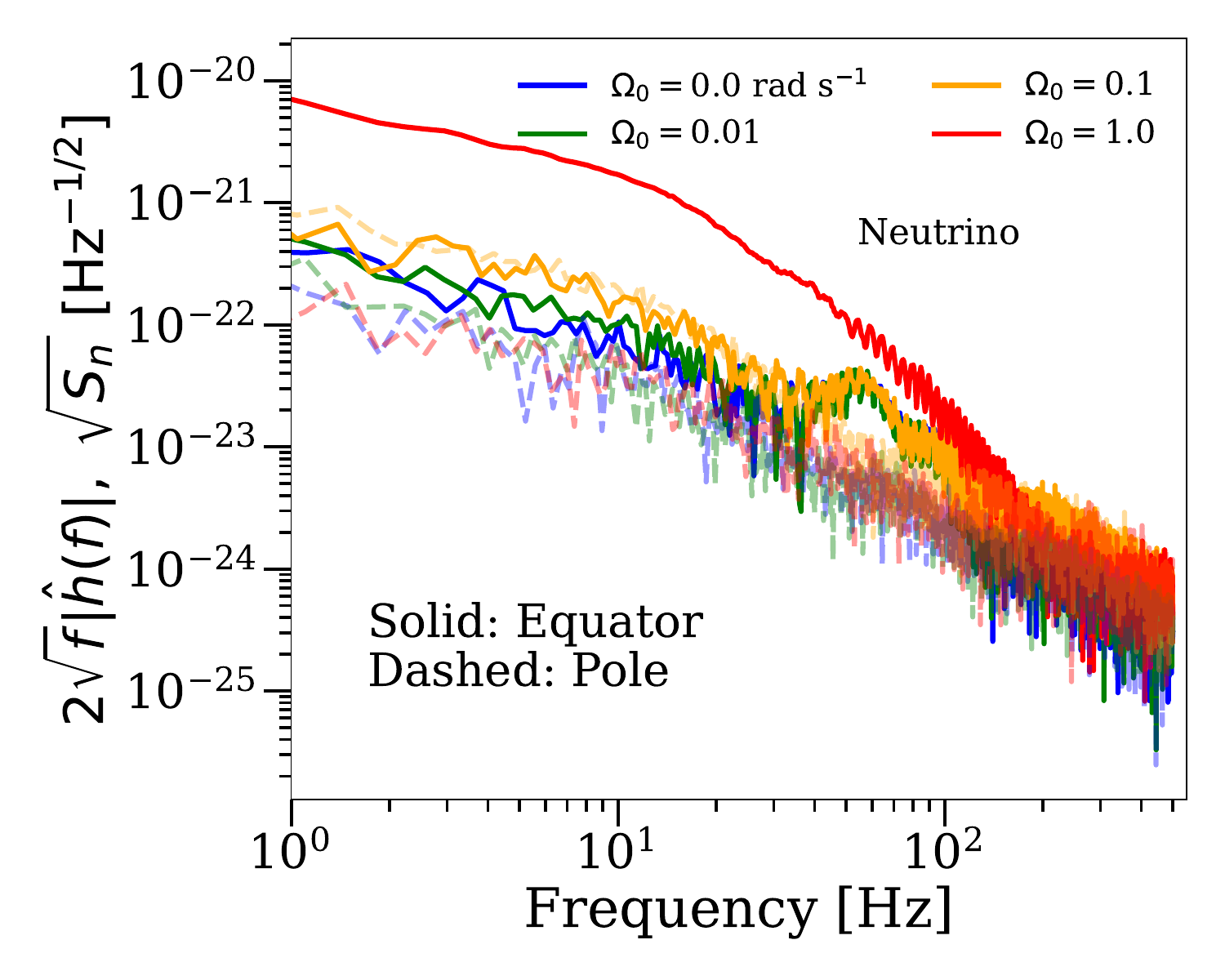}
            \includegraphics[width=0.45\textwidth]{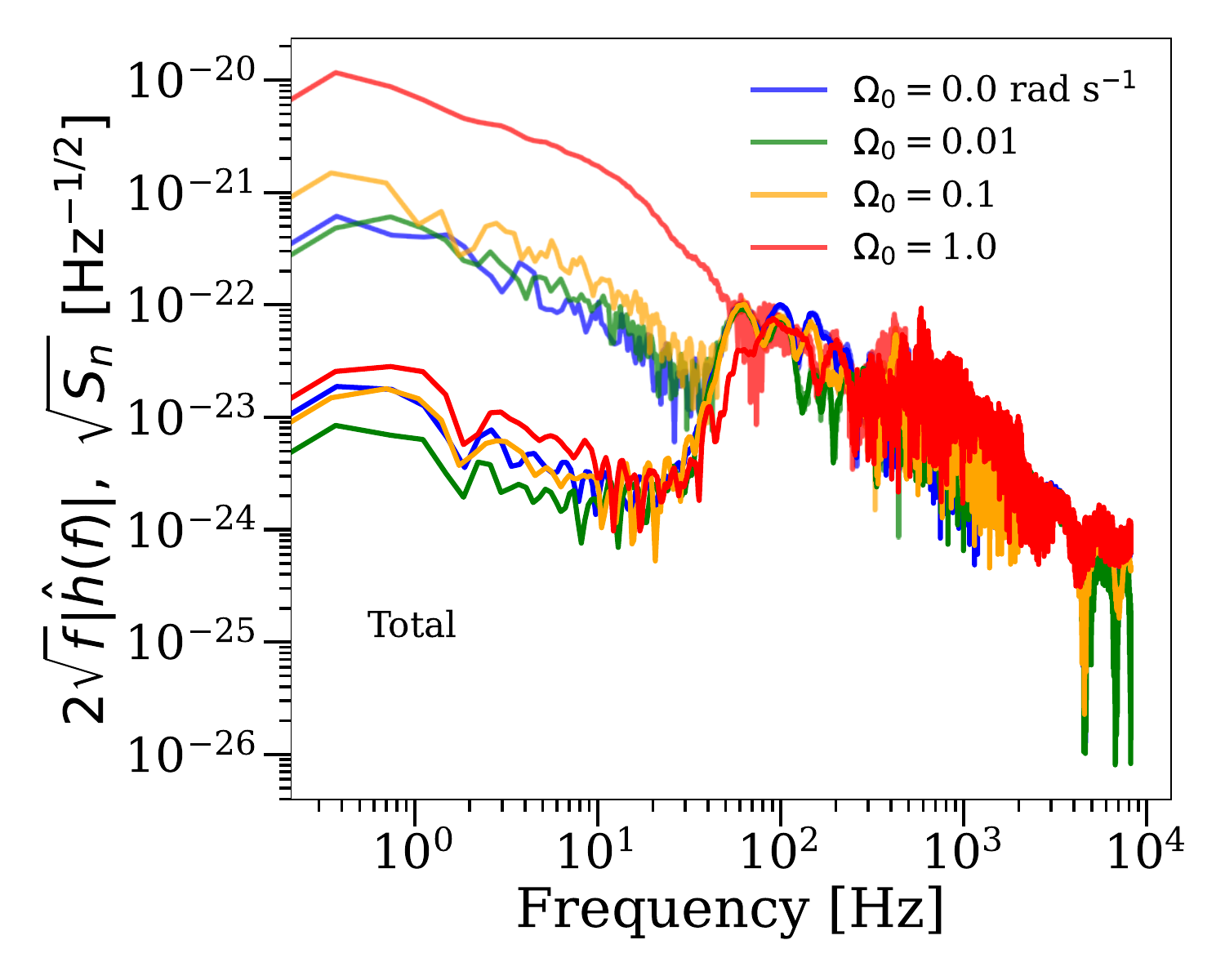}
    \caption{\textbf{Top left}: Matter-driven GW spectra as viewed along the pole (dashed, low opacity) and equator (solid) on the positive x-axis plotted against frequency [Hz], extending to 10 kHz. \textbf{Top right:} Likewise, but for the neutrino memory (extending to 500 Hz, limited by our neutrino data write cadence). Only for the rapidly rotating model do we see a significant pole/equatorial asymmetry, with the equatorial emission nearly $\sim$100 times stronger than polar GW emission at low frequencies, and similarly much stronger than the weakly/non-rotating models. The matter contribution shows a weaker pole/equator dependence, with the most rapid-rotating model showing the strongest anisotropy with a $\sim$10-fold increase in spectral strength along the pole at low frequencies. For the weakly rotating models, the equatorial matter GW emission at high frequencies is larger than the polar emission.  \textbf{Bottom:} The summed matter and strain GW spectra along the equator indicates that the neutrino memory dominates at low frequencies, below 50 Hz. See text for a discussion.}
    \label{fig:gwstrain_nu_angle2}      
\end{figure*}

\label{lastpage}

\end{document}